\documentclass[3p,times,twocolumn]{elsarticle}

\usepackage{ecrc}

\volume{XX}

\firstpage{1}

\journalname{Nuclear Physics B Proceedings Supplement}

\runauth{Mathias Butenschoen}

\jid{nuphbp}

\jnltitlelogo{Nuclear Physics B Proceedings Supplement}

\usepackage{amssymb}

\usepackage[figuresright]{rotating}

\begin{document}

\begin{frontmatter}



\dochead{}

\title{$J/\psi$ production in NRQCD: A global analysis of yield and polarization}


\author{Mathias Butenschoen}
\author{Bernd A. Kniehl}

\address{{II.} Institut f\"ur Theoretische Physik, Universit\"at Hamburg,
Luruper Chaussee 149, 22761 Hamburg, Germany}

\begin{abstract}
We present a rigorous next-to-leading order analysis of $J/\psi$ yield and polarization within the factorization theorem of nonrelativistic QCD (NRQCD). To the orders considered, this framework depends on three free parameters, the color-octet long-distance matrix elements. We extract their values in a global fit to inclusive $J/\psi$ production data from various hadroproduction, photoproduction, two-photon scattering and electron-positron annihilation experiments. We show that this fit is constrained and stable and describes all data sufficiently well. We then make predictions for $J/\psi$ polarization in photo- and hadroproduction and compare them to the currently available data. As for photoproduction, HERA data is not precise enough to draw definite conclusions. But as for hadroproduction, CDF data measured at Tevatron run~II is in strong conflict with NRQCD predictions. With early ALICE data being however compatible with NRQCD, the future, more precise polarization measurements at the LHC will thus have the potential to clearly confirm or dismiss LDME universality.
\end{abstract}

\begin{keyword}
Lepton-Nucleon Scattering \sep Nucleon-Nucleon Scattering \sep Lepton-Lepton Interactions  \sep Heavy quarkonia \sep Nonrelativistic QCD

\end{keyword}

\end{frontmatter}



\section{Introduction}

Heavy quarkonia are bound states of a heavy quark and its antiquark. There are charmonia and bottomonia. The charmonium $J/\psi$ is the most extensively studied quarkonium state, because of its experimentally clean signature due to the large branching fraction of its leptonic decay modes. Almost 40 years since its discovery, the underlying mechanisms governing heavy quarkonium production and decay are however still not understood and subject to dispute. According to the factorization theorem of nonrelativistic QCD (NRQCD) \cite{Bodwin:1994jh}, the cross section to produce a heavy quarkonium $H$ factorizes according to
\begin{displaymath}
 \sigma(ab\to H+X) = \sum_n \sigma(ab\to c\overline{c}[n] + X) \langle{\cal O}^{H}[n]\rangle,
\end{displaymath}
where the $\sigma(ab\to c\overline{c}[n] + X)$ are perturbatively calculated short distance cross sections describing the production of a heavy quark pair (here $c\overline{c}$) in an intermediate Fock state $n$, which does not have to be color neutral. The $\langle{\cal O}^{H}[n]\rangle$ are nonperturbative long distance matrix elements (LDMEs), which currently have to be extracted from experiment and which describe the transition of that intermediate $c\overline{c}$ state into the physical $H$ via soft gluon radiation. NRQCD predicts each of the LDMEs to scale with a definite power of the relative heavy quark velocity $v$, which serves as an additional expansion parameter besides $\alpha_s$: In case of $H=J/\psi$, the leading order contribution in the $v$ expansion stems from $n={^3S}_1^{[1]}$ and equals the traditional color singlet model (CSM) prediction, while the leading relativistic corrections are made up by the $^1S_0^{[8]}$, $^3S_1^{[8]}$, and $^3P_J^{[8]}$ states. The upper index ``8'' stands for color octet (CO), and these contributions are usually just called {\em the} color octet contributions. The aim of our work is to test the universality of the LDMEs and thereby to challenge NRQCD factorization.


\section{History of NLO $J/\psi$ production calculations}

Next-to-leading order (NLO) corrections to inclusive $J/\psi$ production cross sections have been calculated within the CSM for direct photoproduction \cite{Kramer:1994zi}, for hadroproduction \cite{Campbell:2007ws} and for electron-positron annihilation \cite{Ma:2008gq}.

The calculation of NLO corrections to the short distance cross sections of the intermediate CO states, especially to the $^3P_J^{[8]}$ states, have on the other hand proven to be very challenging. But up to now they have been calculated for all relevant collision processes as well. The $2\to 1$ processes for photo- and hadroproduction have already been calculated in 1998 \cite{Petrelli:1997ge}. Inclusive production in direct two-photon collisions followed in 2005 \cite{Klasen:2004az}, in direct photoproduction \cite{Butenschoen:2009zy} and
electron-positron scattering neglecting the small $^3S_1^{[8]}$ contribution \cite{Zhang:2009ym} in 2009. The hadroproduction calculations \cite{Gong:2008ft} were still missing the $^3P_J^{[8]}$ contributions. Full calculations involving all CO states followed in 2010 with two independent works \cite{Ma:2010yw,Butenschoen:2010rq}. The missing pieces of single and double resolved two photon scattering, resolved photoproduction and the $^3S_1^{[8]}$ contributions of electron-positron scattering were finally presented in 2011 \cite{Butenschoen:2011yh}.

Polarized NLO $J/\psi$ production cross sections have been evaluated within the CSM, for direct photoproduction \cite{Artoisenet:2009xh} and hadroproduction \cite{Gong:2008sn}, and for the $^1S_0^{[8]}$ and $^3S_1^{[8]}$ intermediate states in hadroproduction \cite{Gong:2008ft}. Recently, polarized NLO cross sections including all CO contributions have been calculated, namely for direct photoproduction at HERA \cite{Butenschoen:2011ks} and for hadroproduction at Tevatron and the LHC \cite{HadroPolLetter}.

\section{$J/\psi$ yield: A global fit of the CO LDMEs}

\begin{figure*}
\centering
\includegraphics[width=4cm]{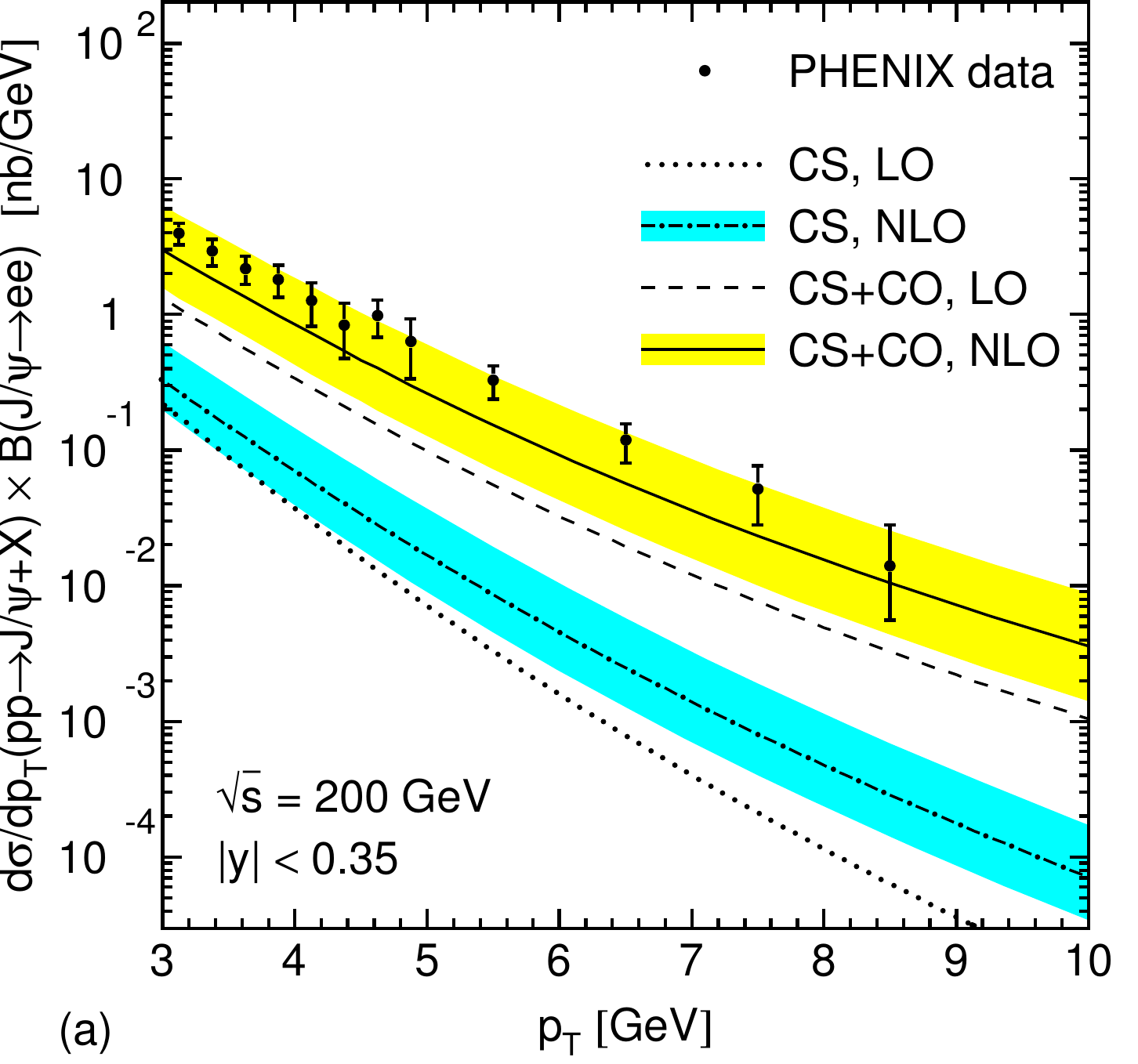}
\includegraphics[width=4cm]{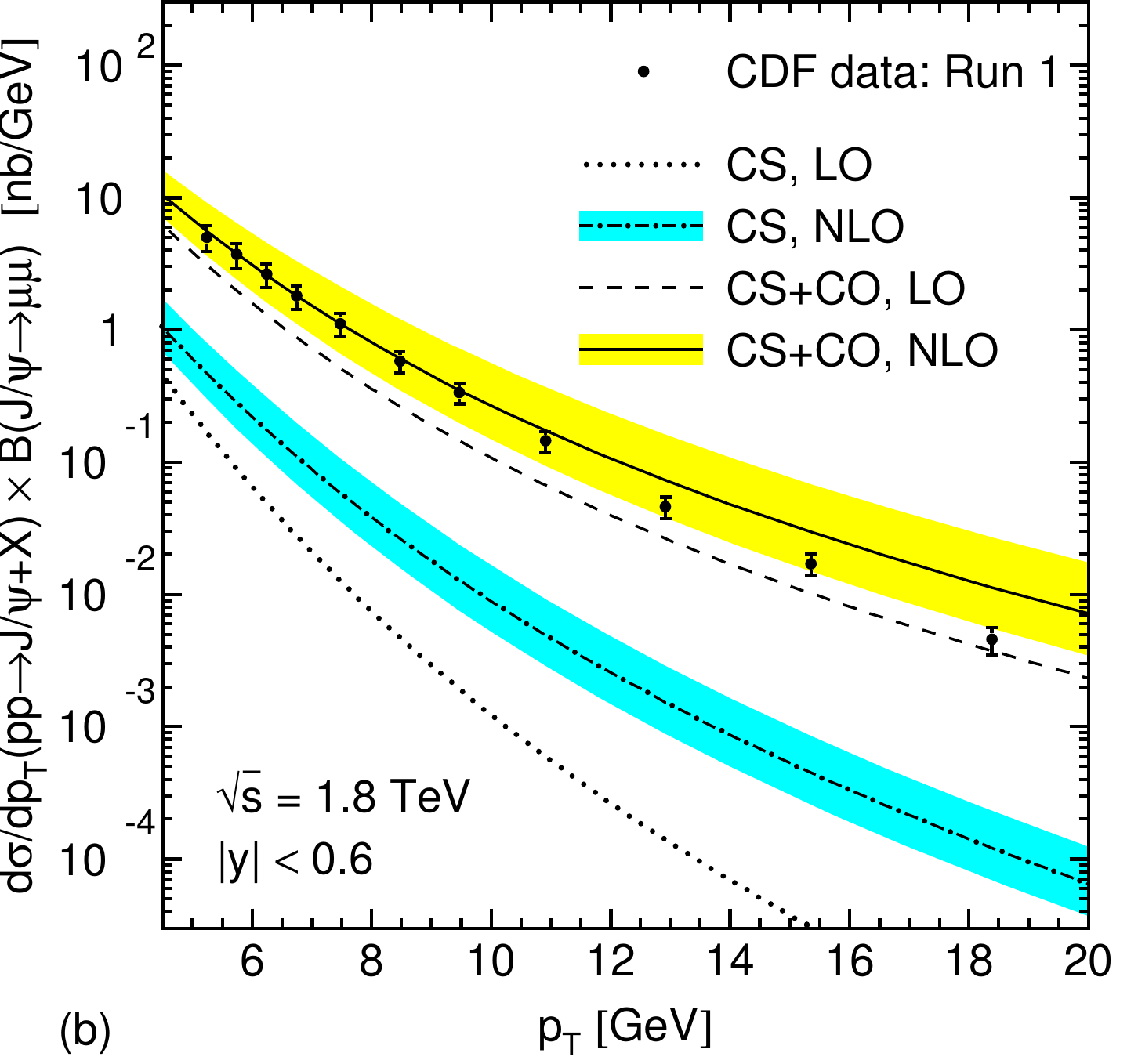}
\includegraphics[width=4cm]{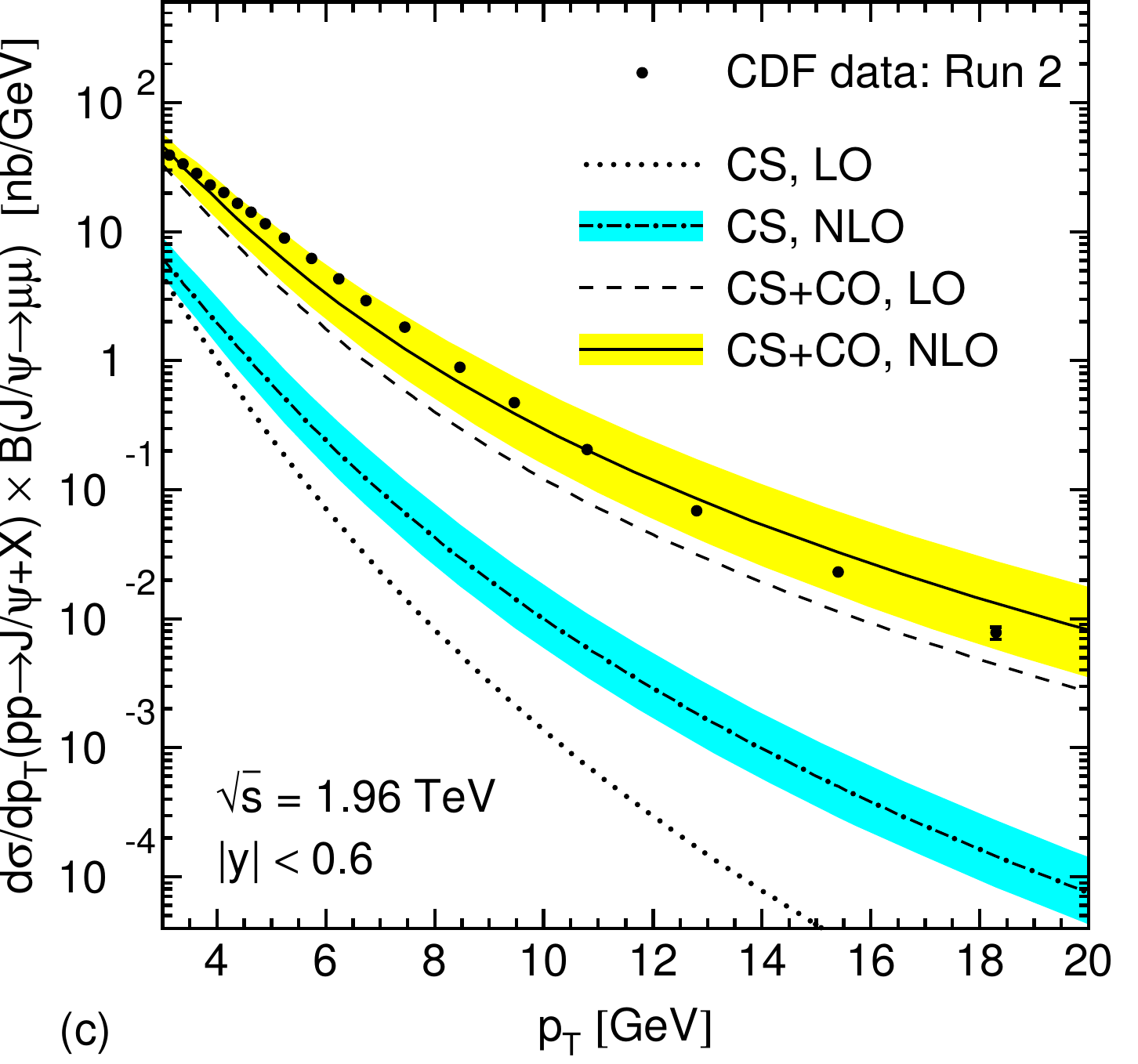}
\includegraphics[width=4cm]{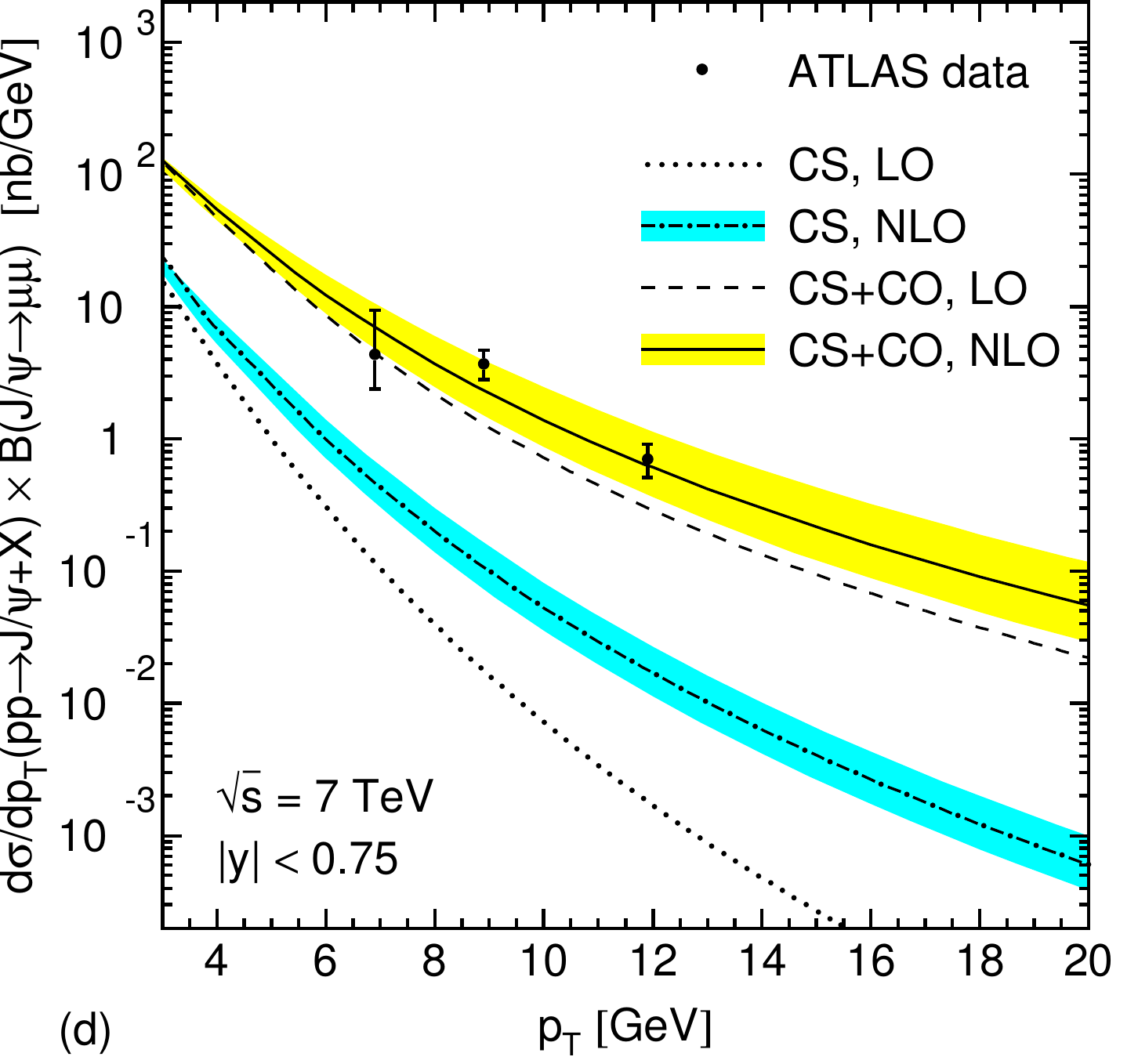}

\vspace{5pt}
\includegraphics[width=4cm]{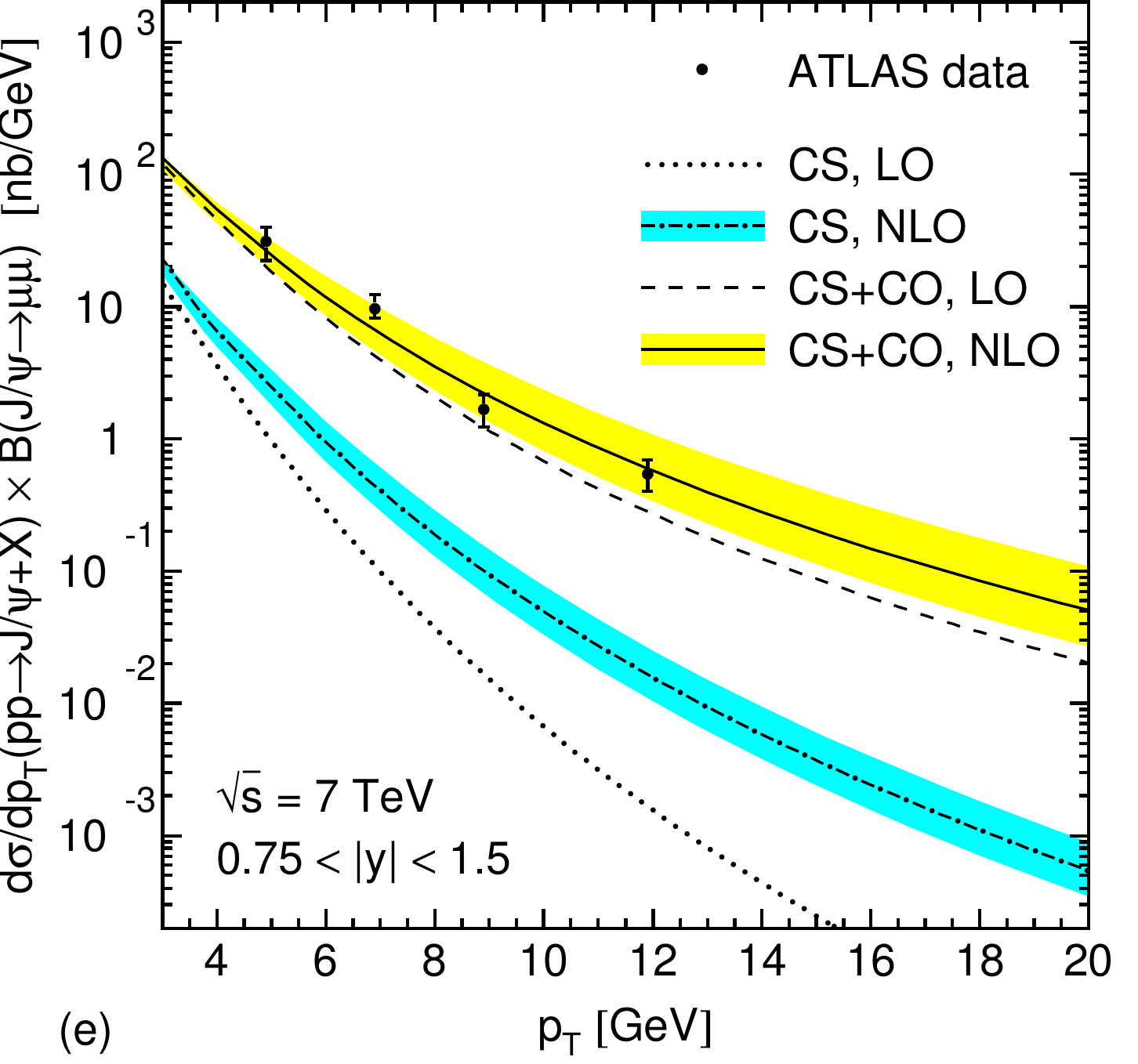}
\includegraphics[width=4cm]{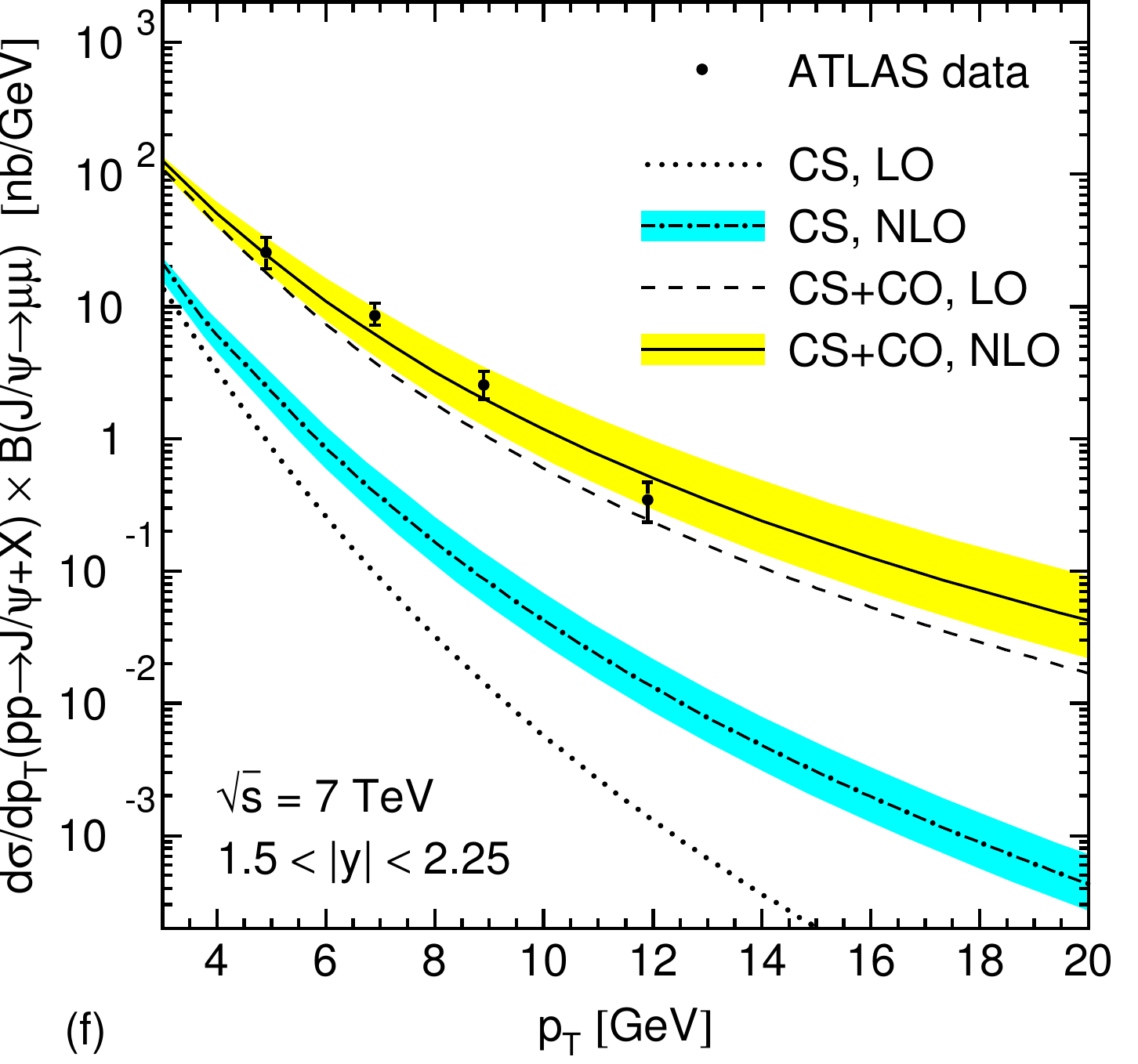}
\includegraphics[width=4cm]{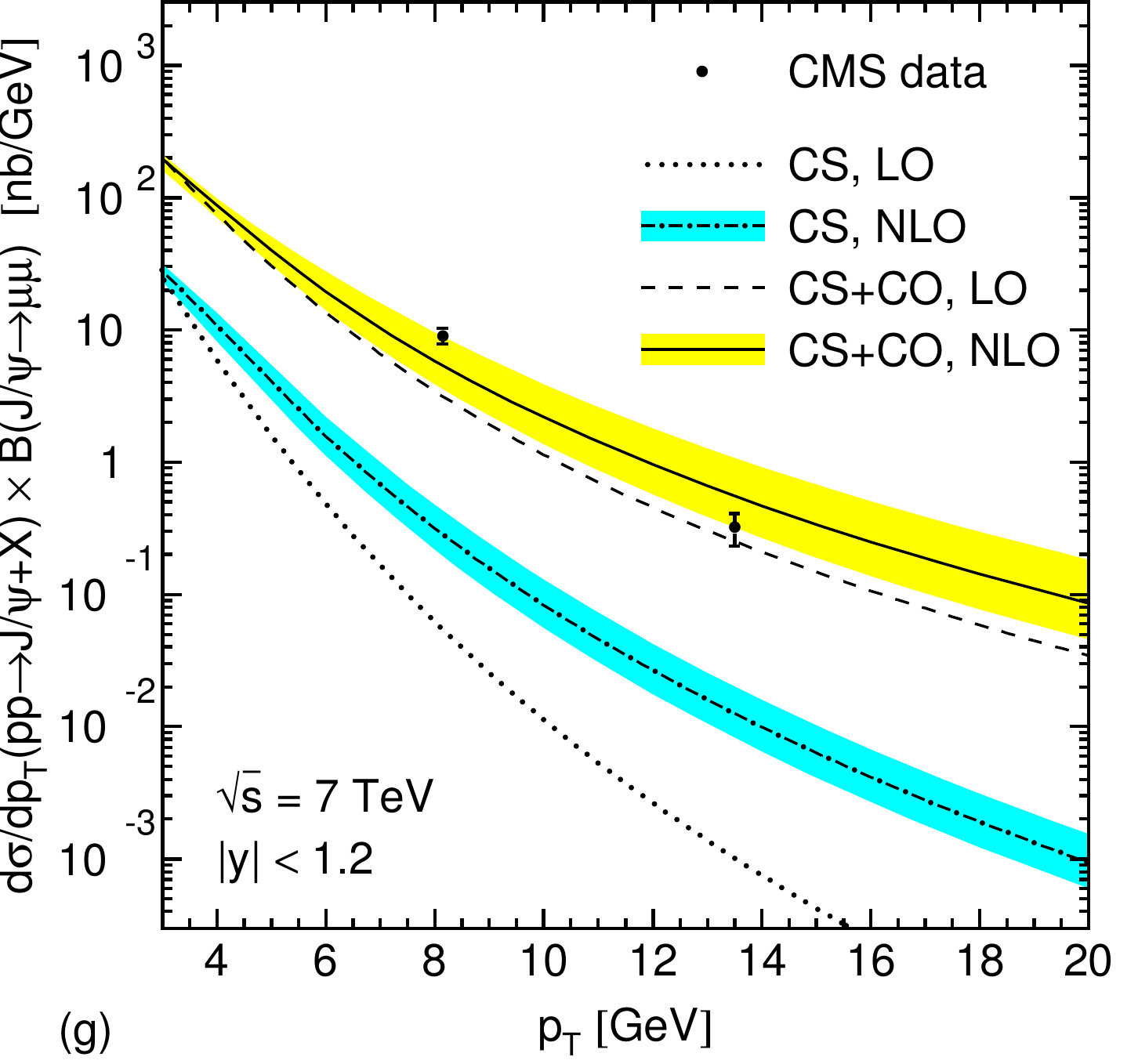}
\includegraphics[width=4cm]{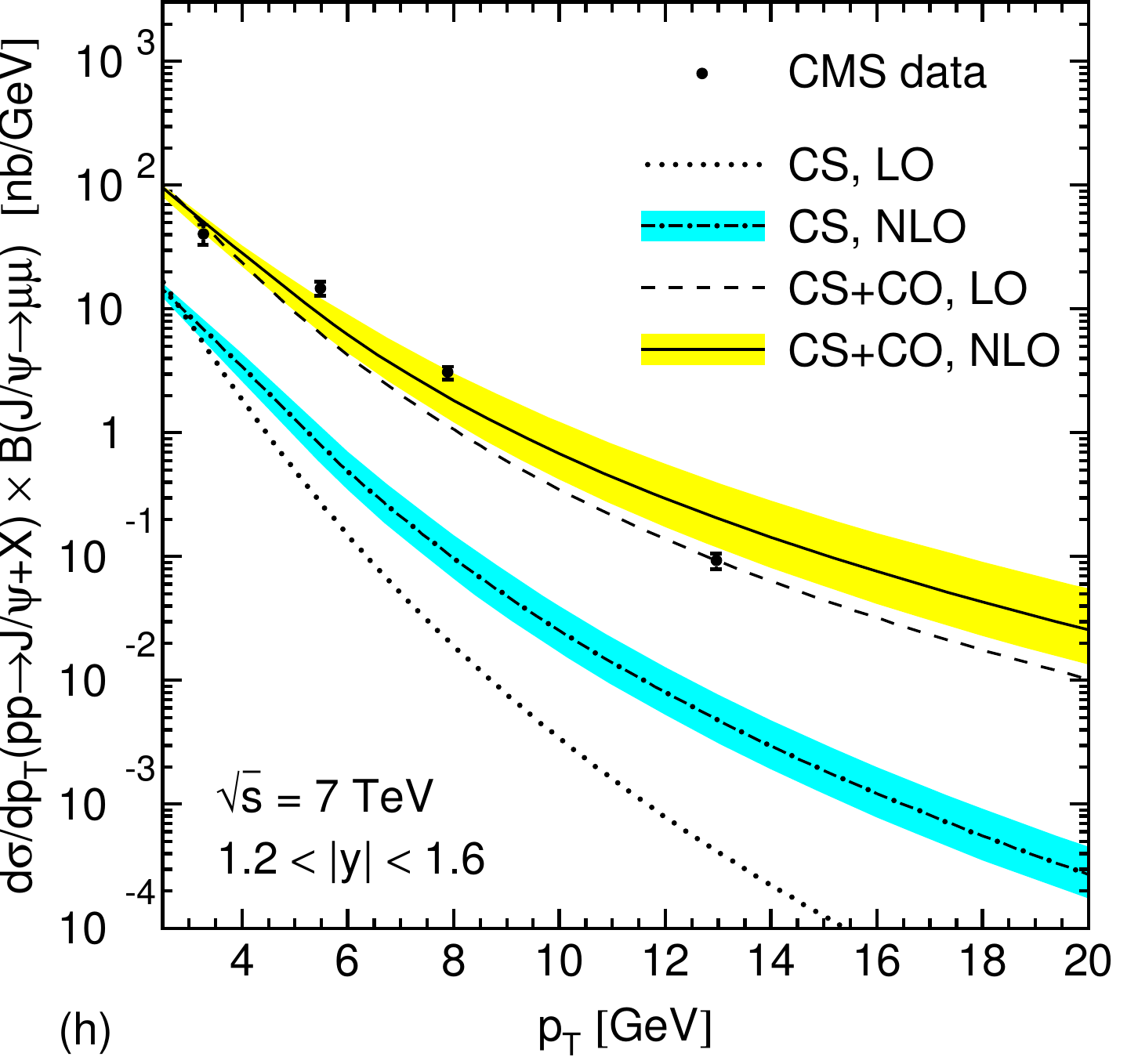}

\vspace{5pt}
\includegraphics[width=4cm]{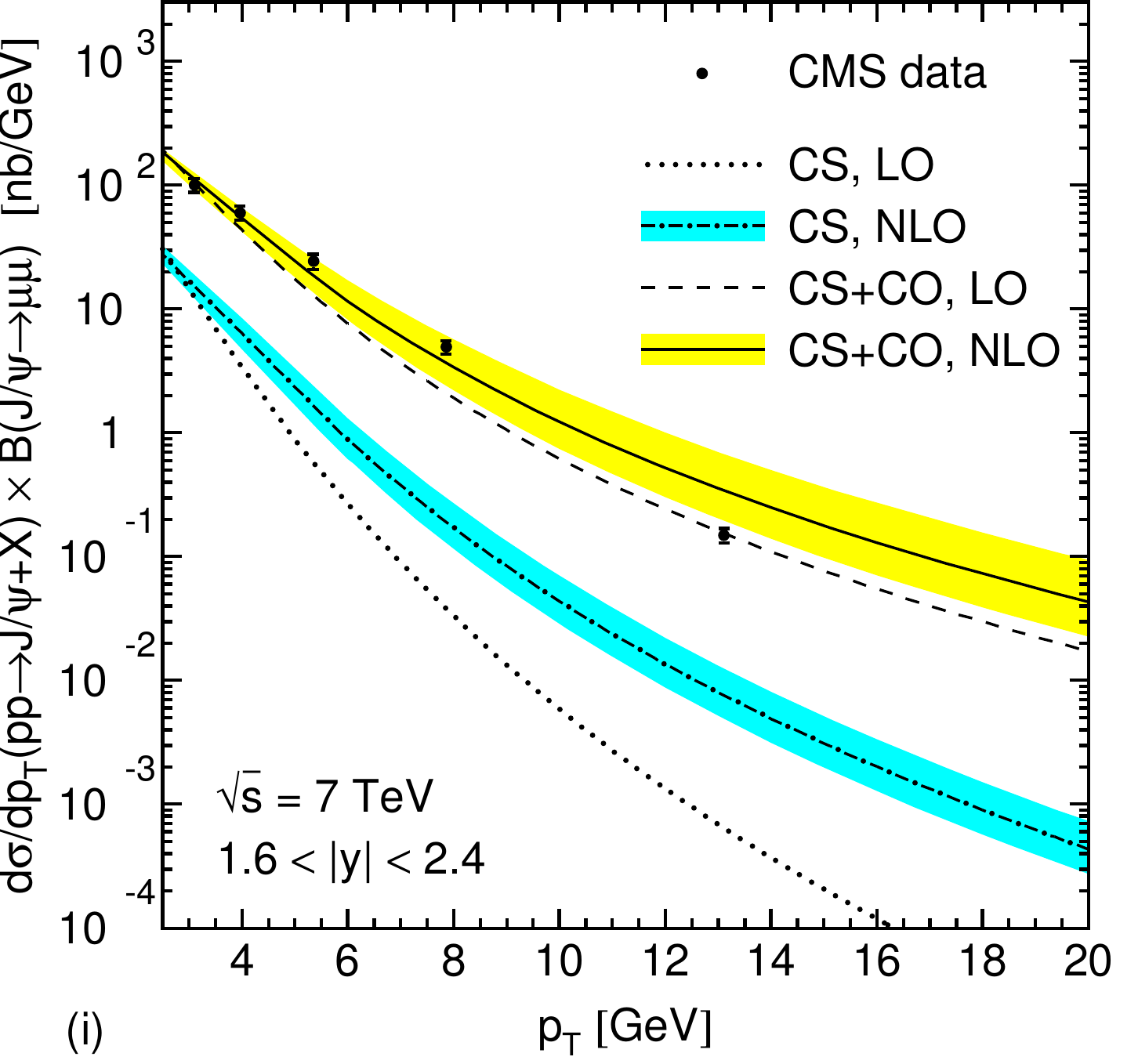}
\includegraphics[width=4cm]{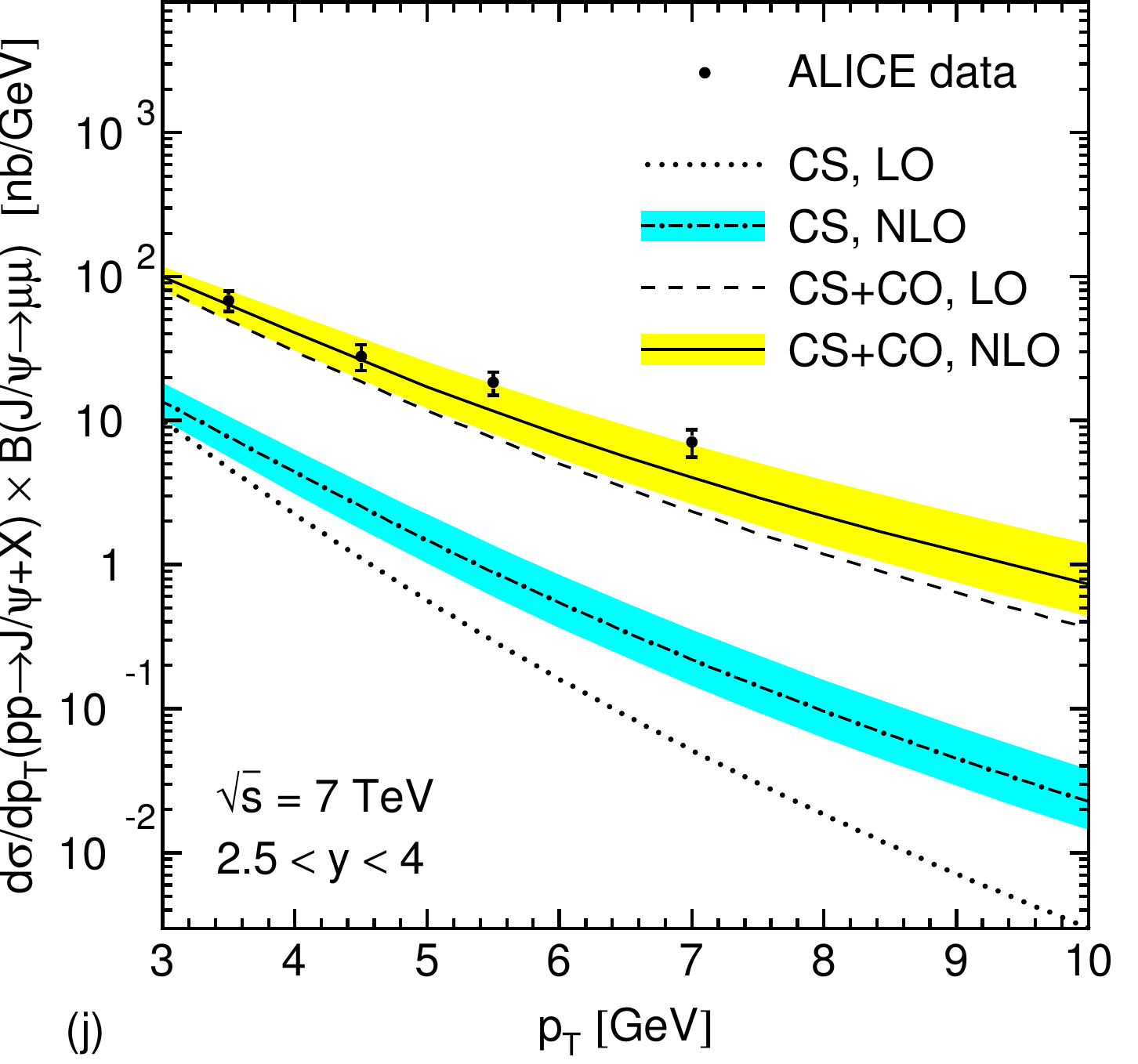}
\includegraphics[width=4cm]{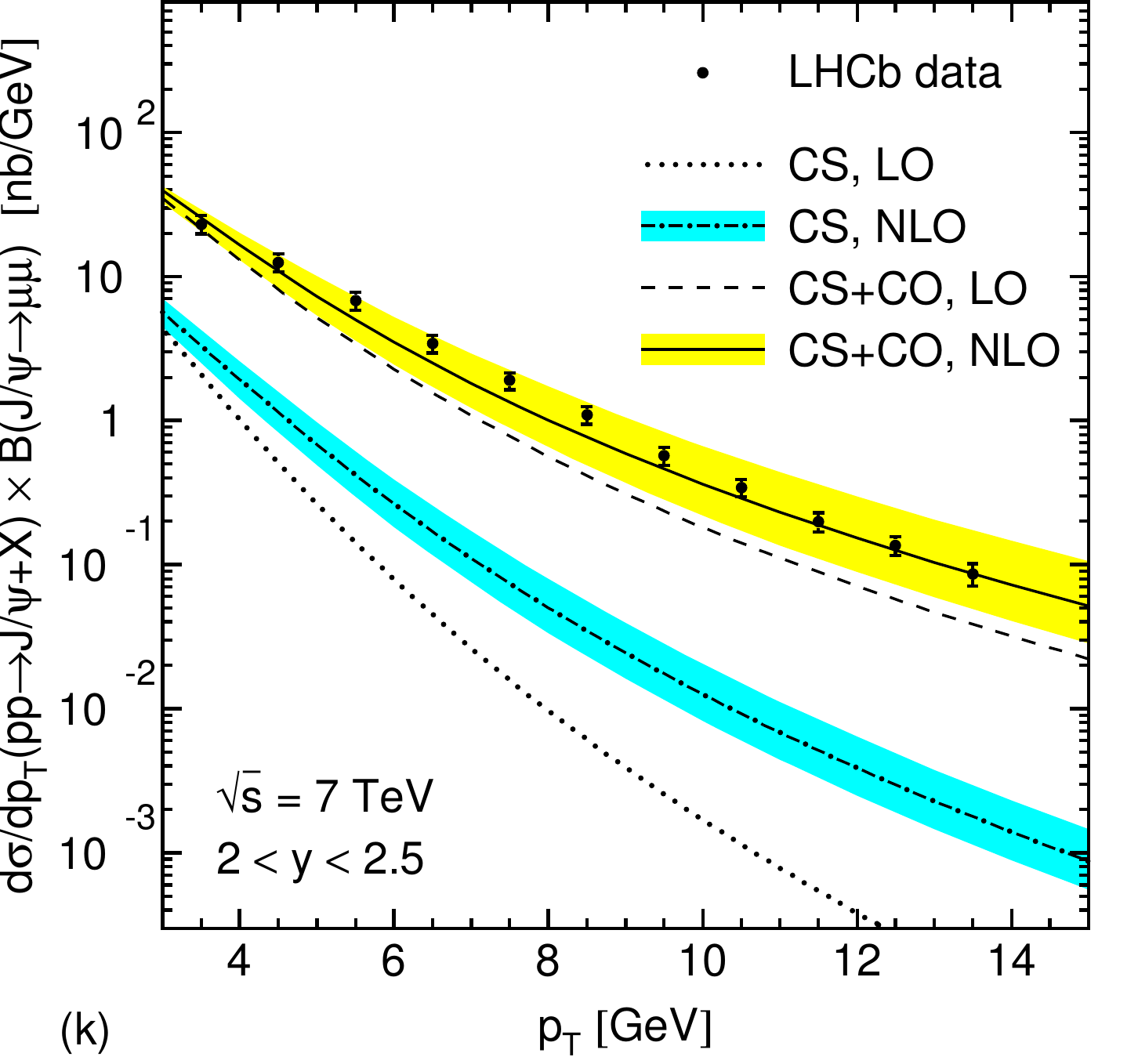}
\includegraphics[width=4cm]{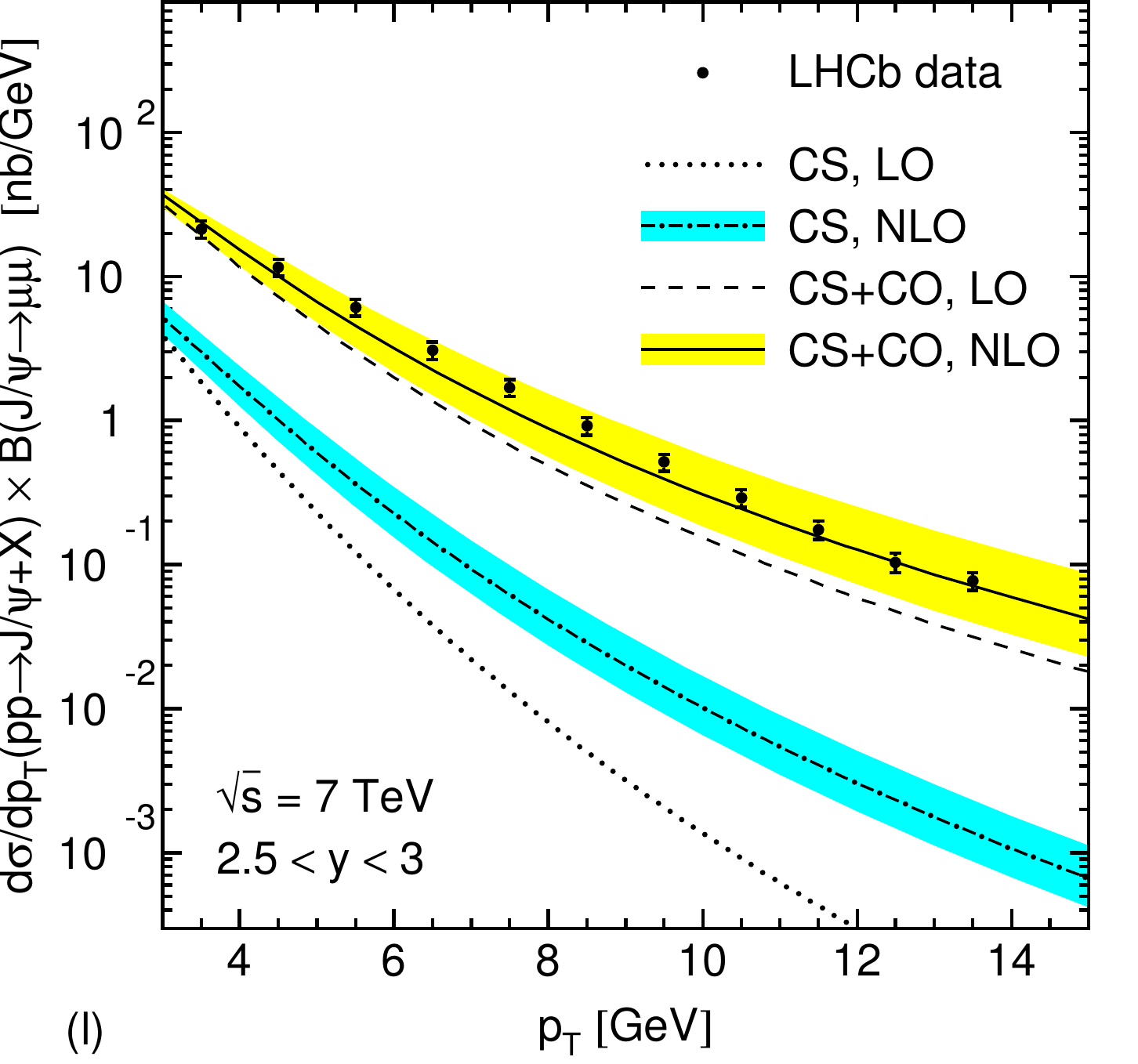}

\vspace{5pt}
\includegraphics[width=4cm]{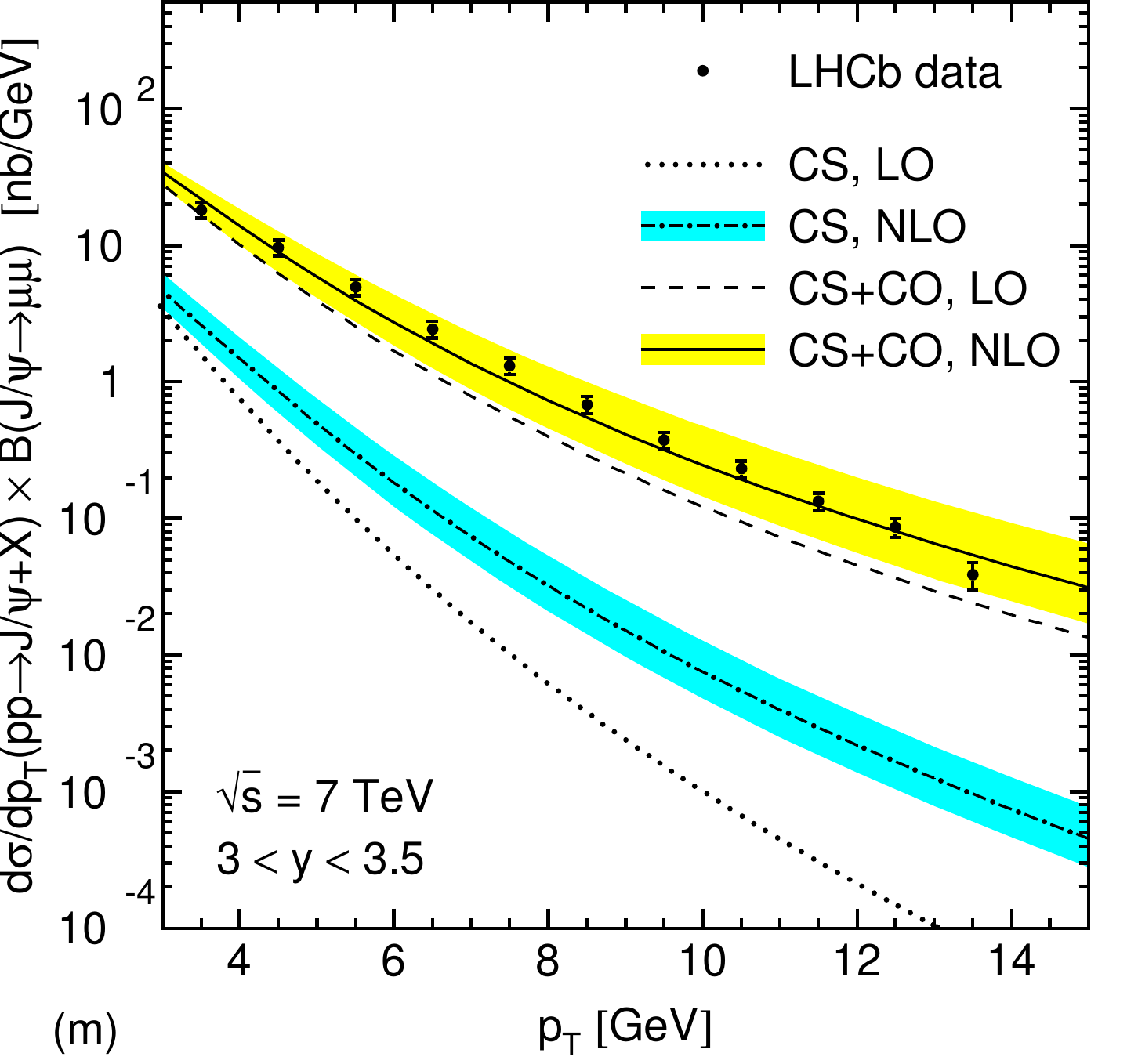}
\includegraphics[width=4cm]{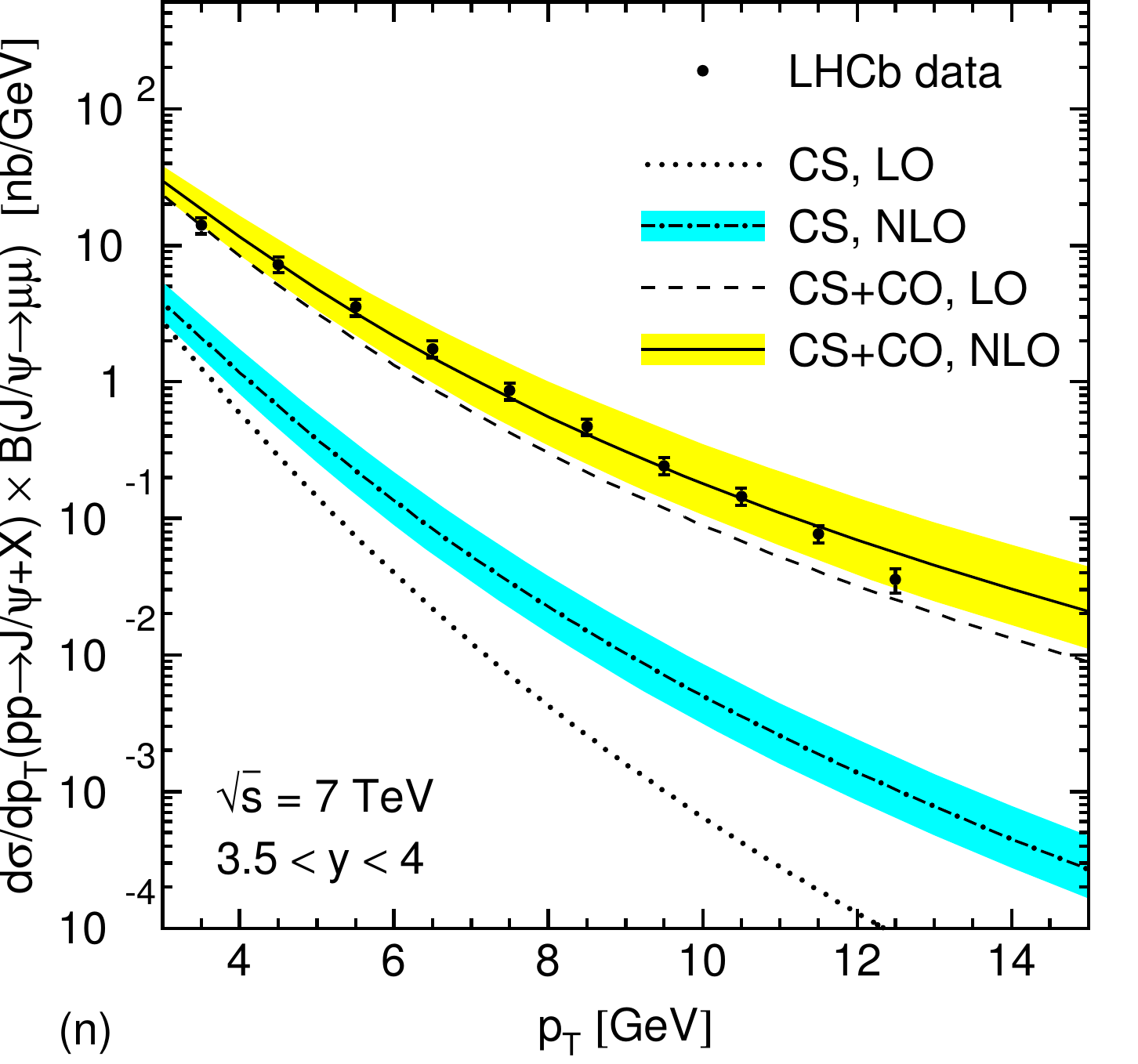}
\includegraphics[width=4cm]{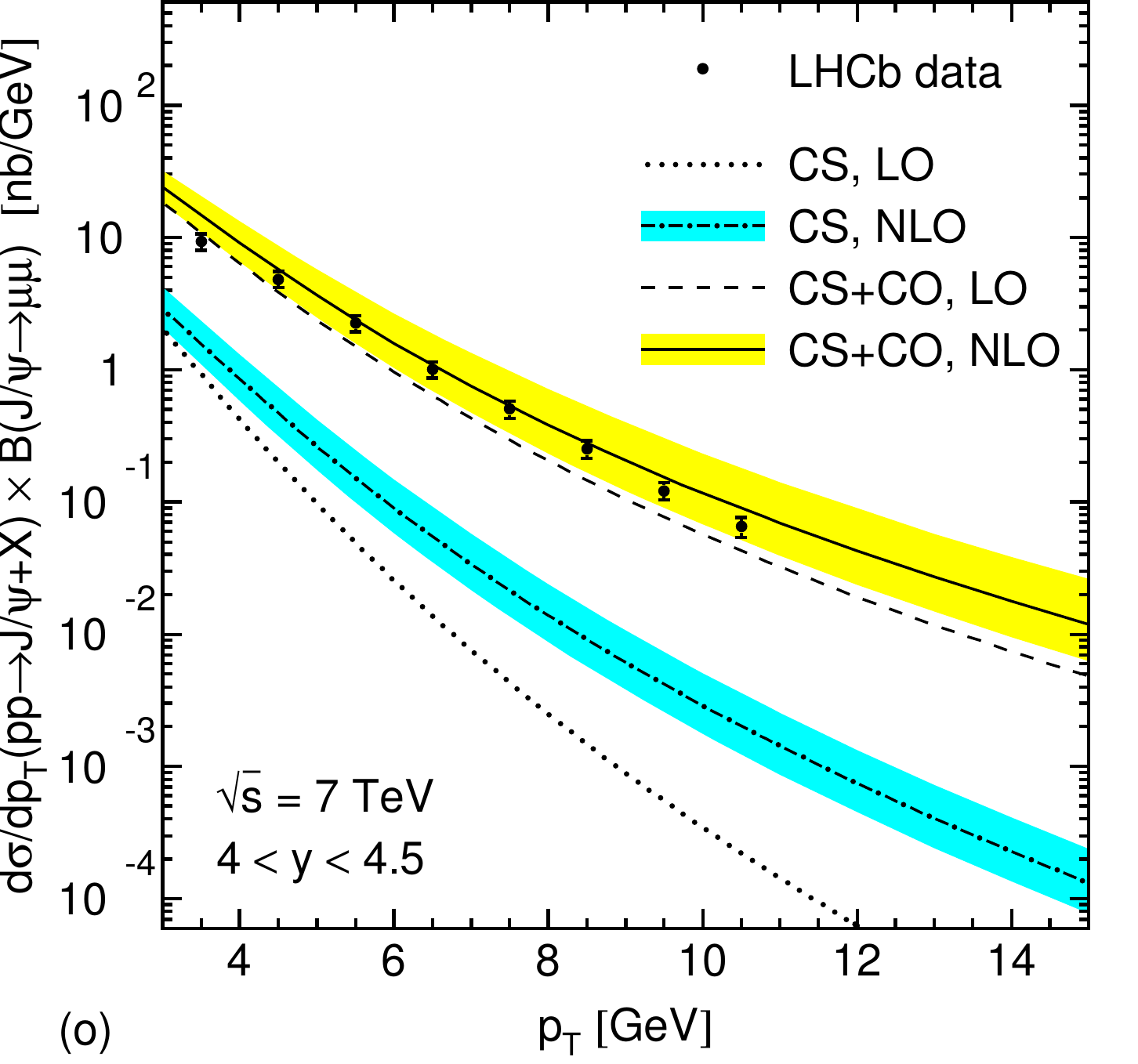}
\includegraphics[width=4cm]{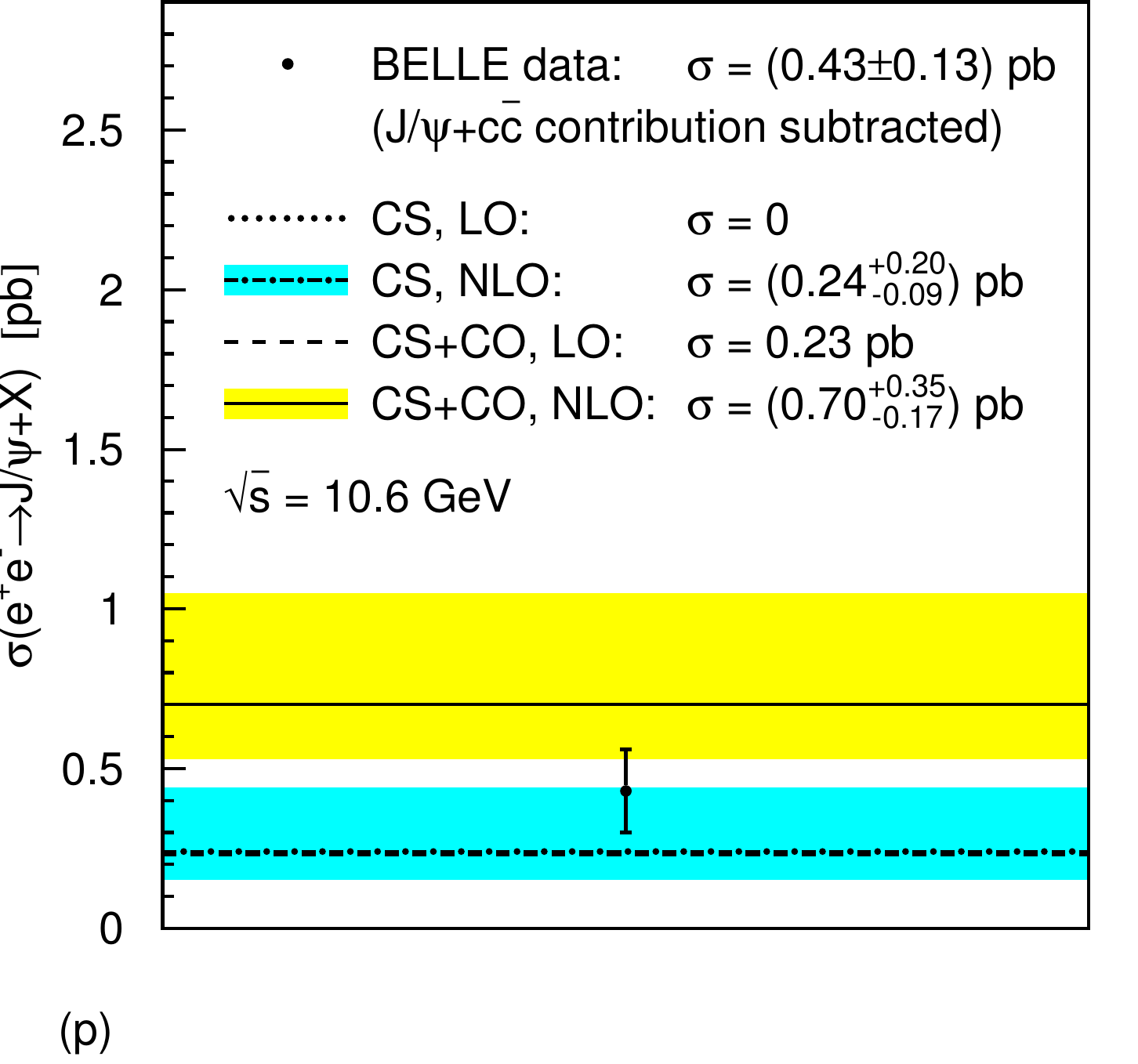}

\vspace{5pt}
\includegraphics[width=4cm]{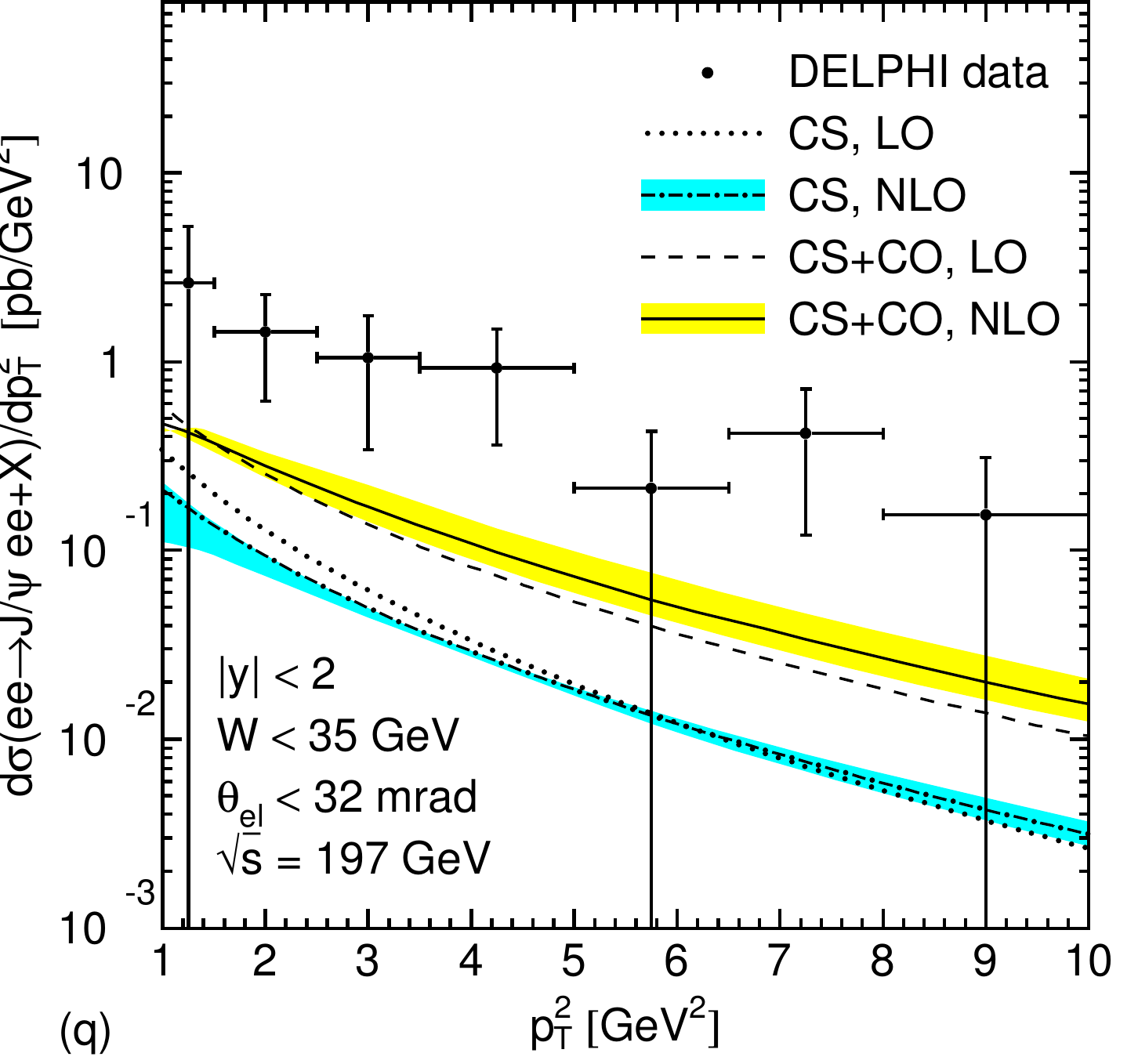}
\includegraphics[width=4cm]{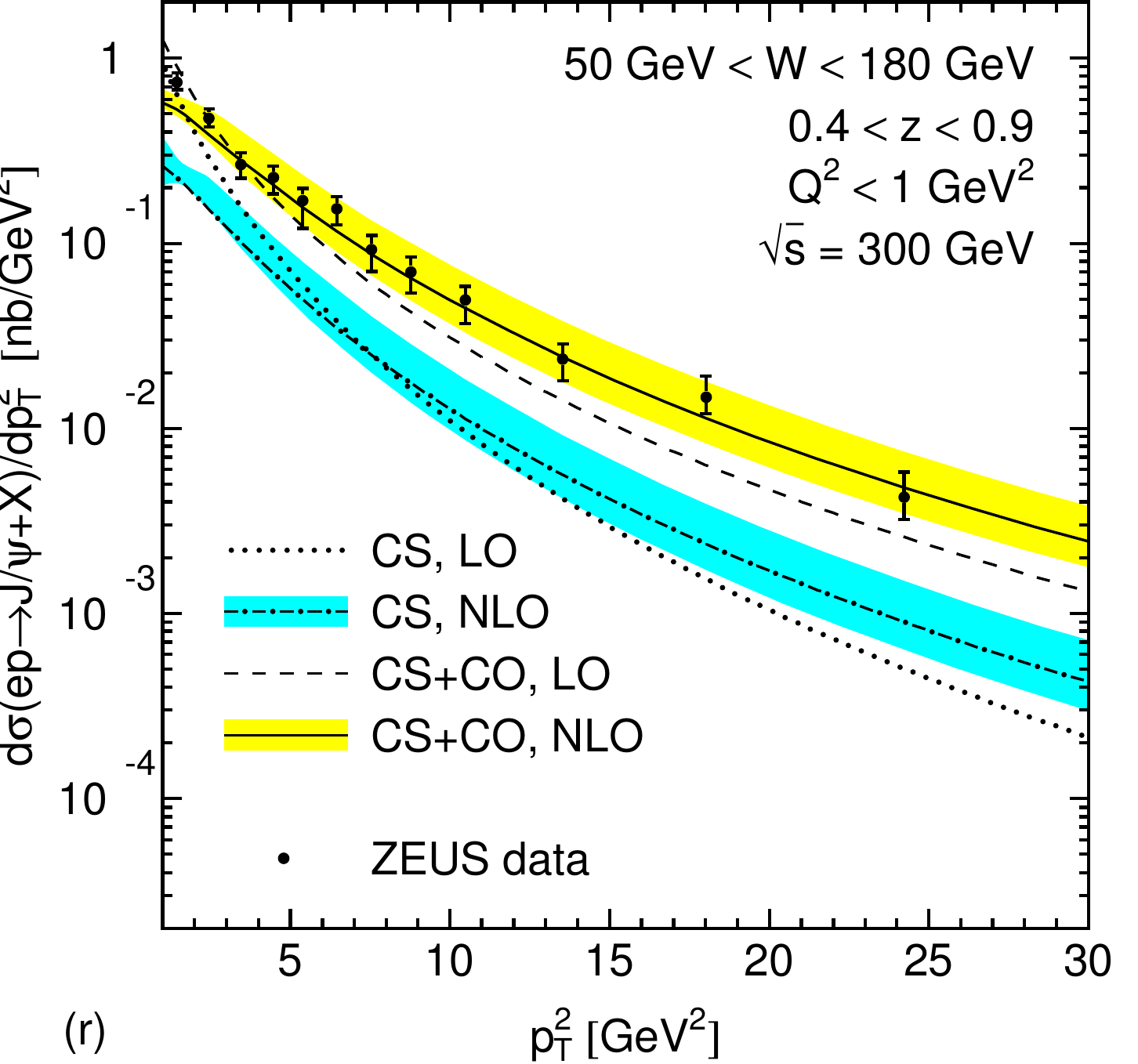}
\includegraphics[width=4cm]{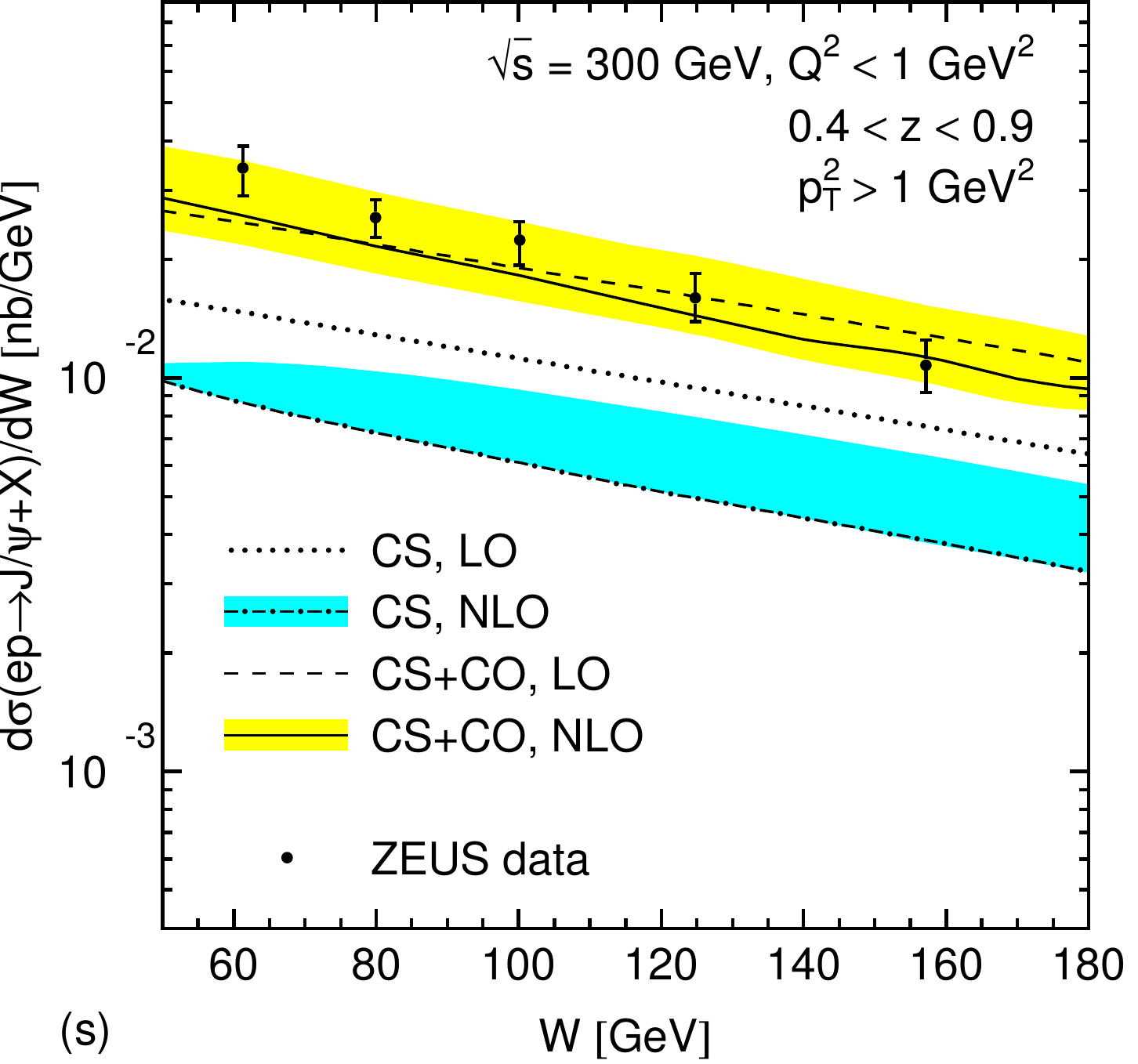}
\includegraphics[width=4cm]{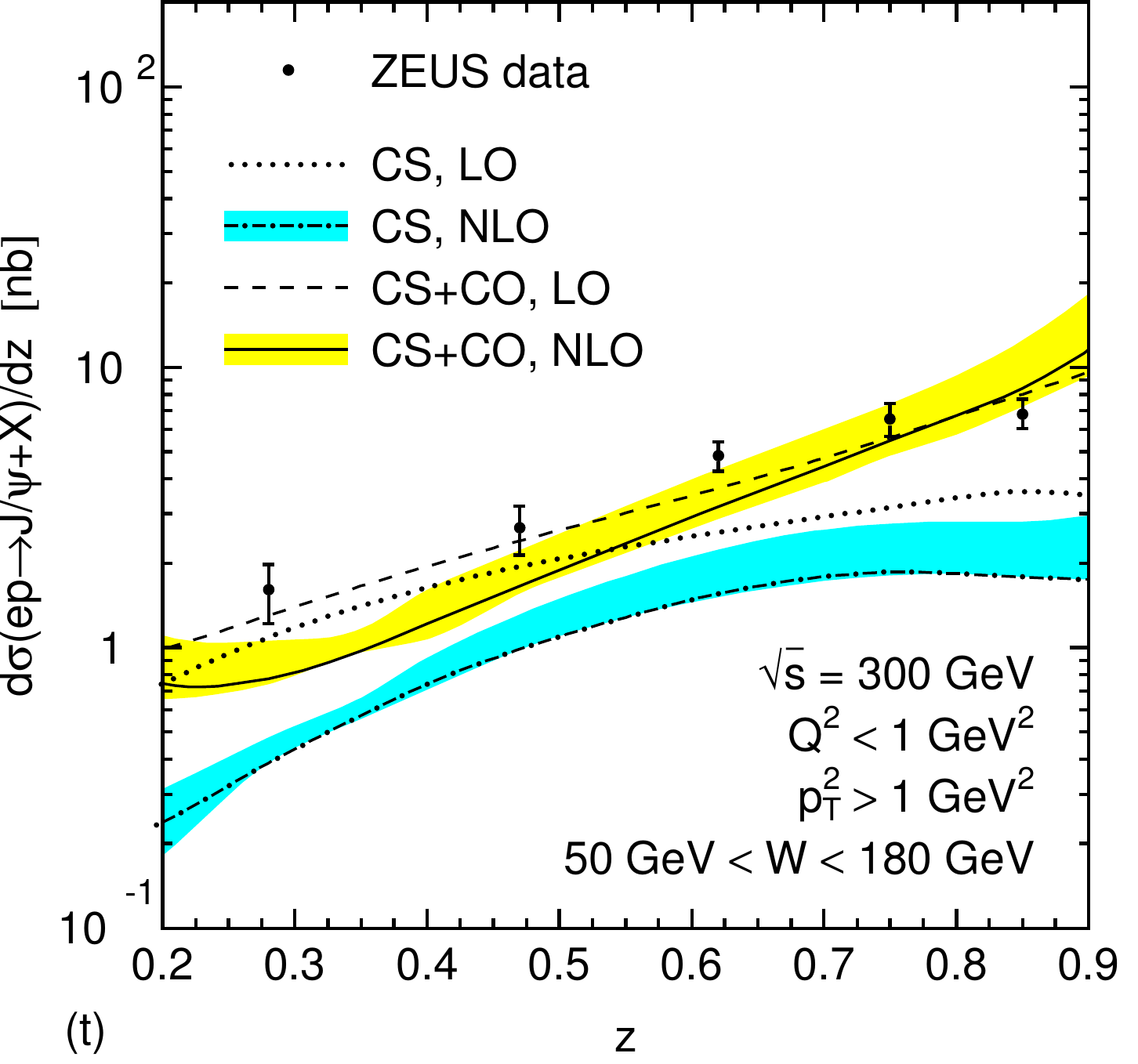}
\caption{\label{fig:fitgraphs1}
Plots a-t: Results of the global fit \cite{Butenschoen:2011yh} compared to ALICE \cite{ALICEdata}, ATLAS \cite{ATLASdata}, Belle \cite{:2009nj}, CDF \cite{Acosta:2004yw,Abe:1997jz}, CMS \cite{Khachatryan:2010yr}, DELPHI \cite{Abdallah:2003du}, LHCb \cite{Aaij:2011jh}, PHENIX \cite{Adare:2009js}, and ZEUS \cite{Chekanov:2002at} data. The blue bands are the CSM predictions, the yellow bands include the CO contributions. The bands are constructed by variation of the renormalization, factorization and NRQCD scales.}
\end{figure*}
\addtocounter{figure}{-1}
\begin{figure*}
\centering
\includegraphics[width=4cm]{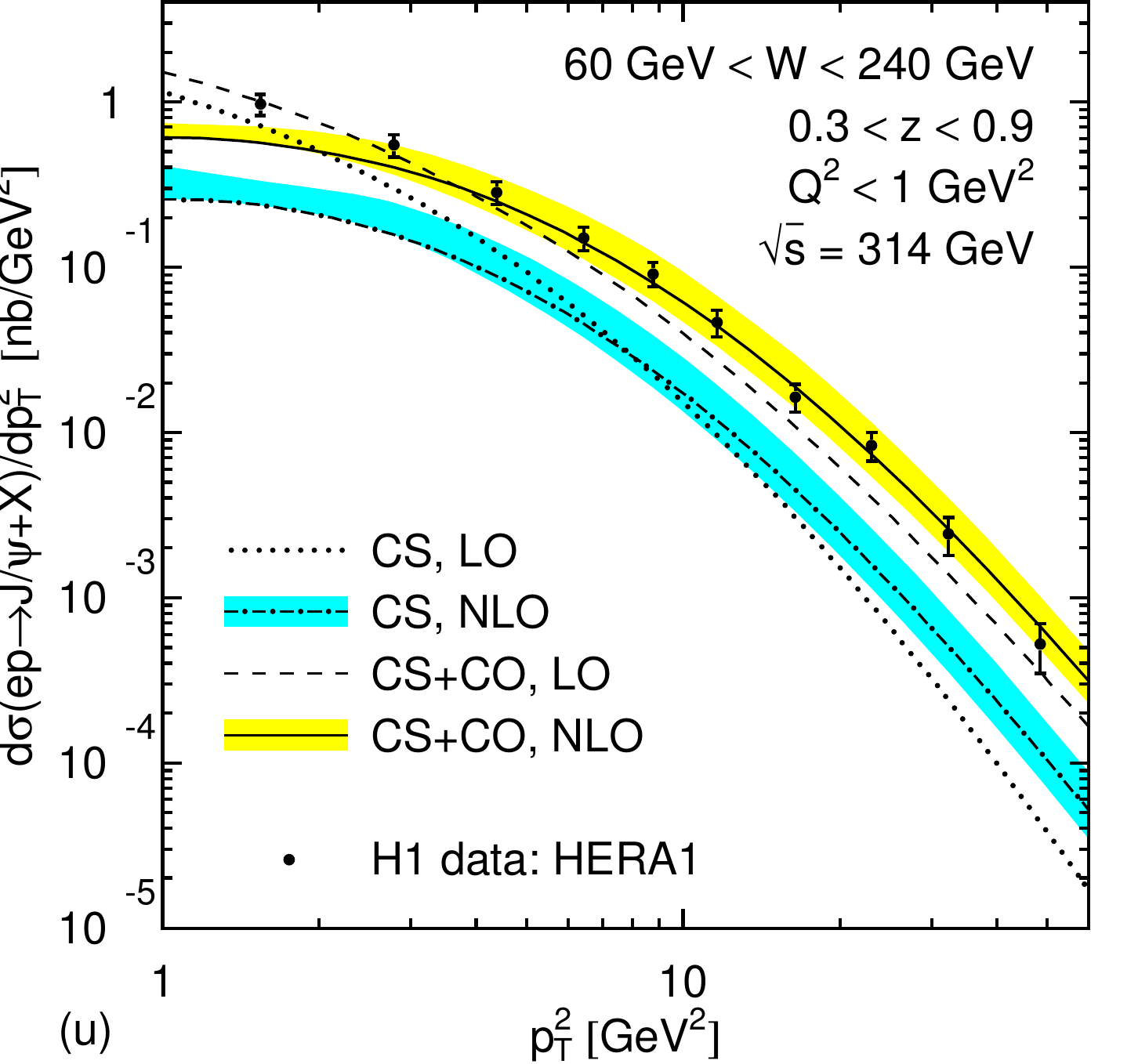}
\includegraphics[width=4cm]{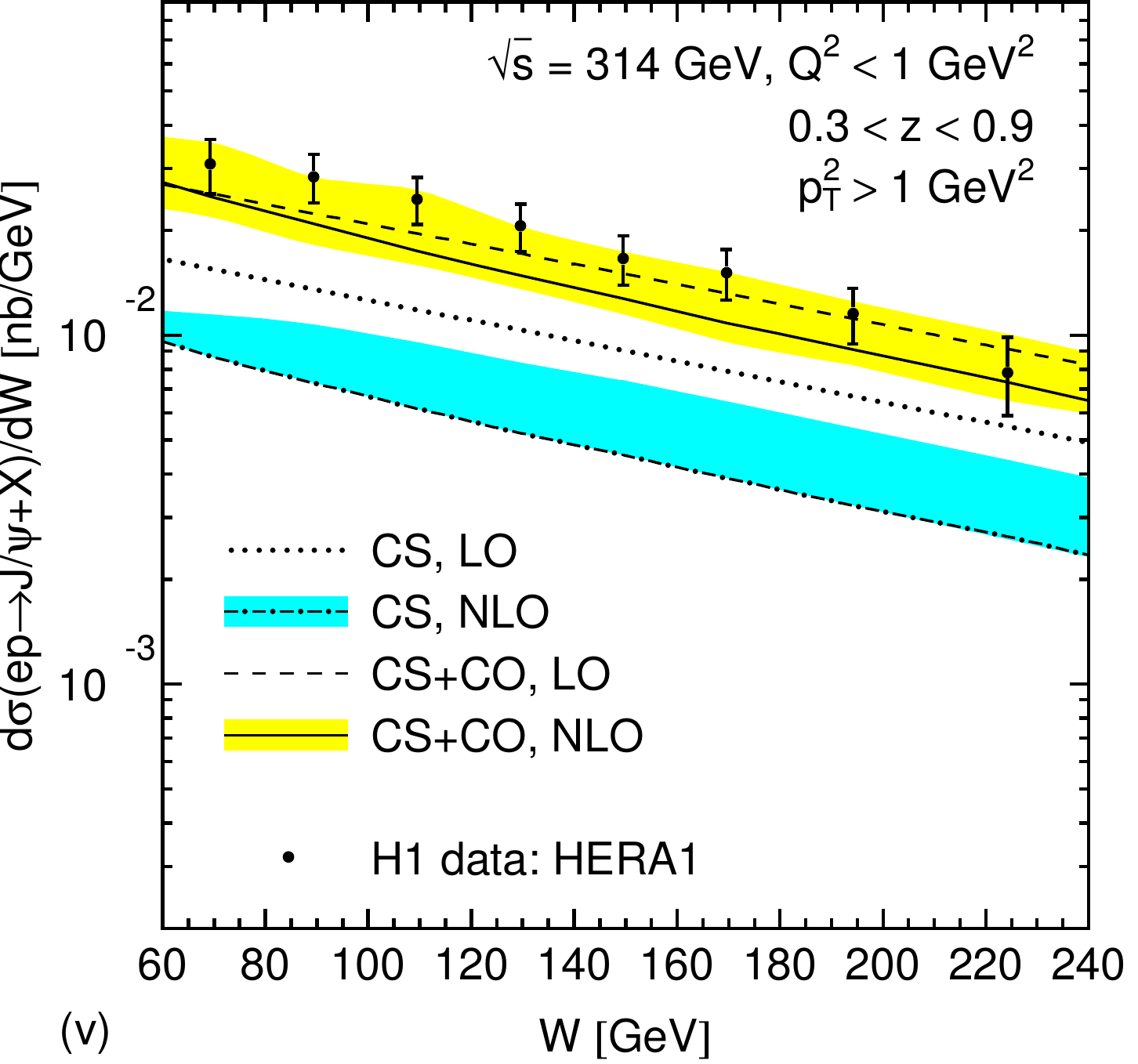}
\includegraphics[width=4cm]{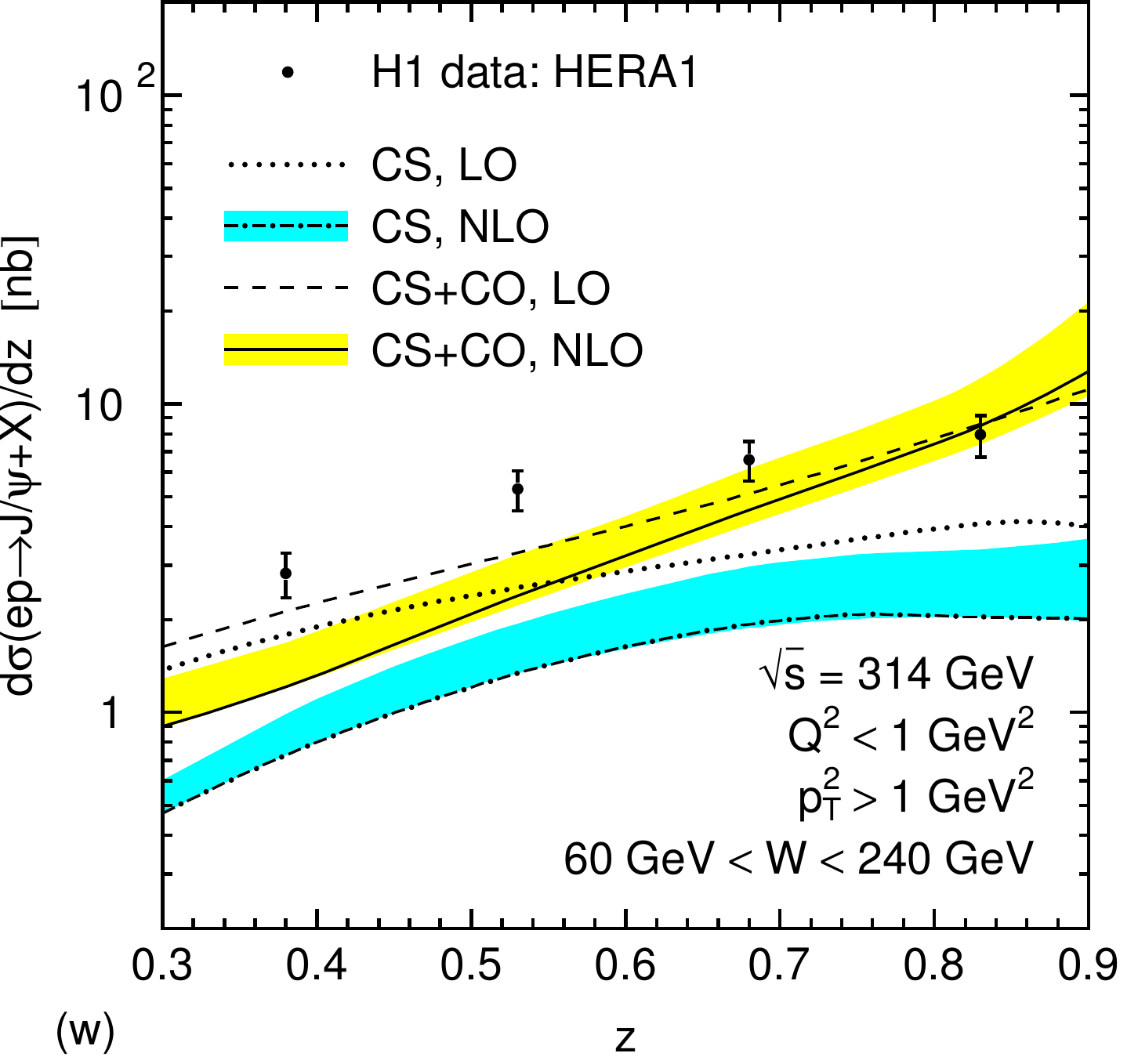}

\vspace{5pt}
\includegraphics[width=4cm]{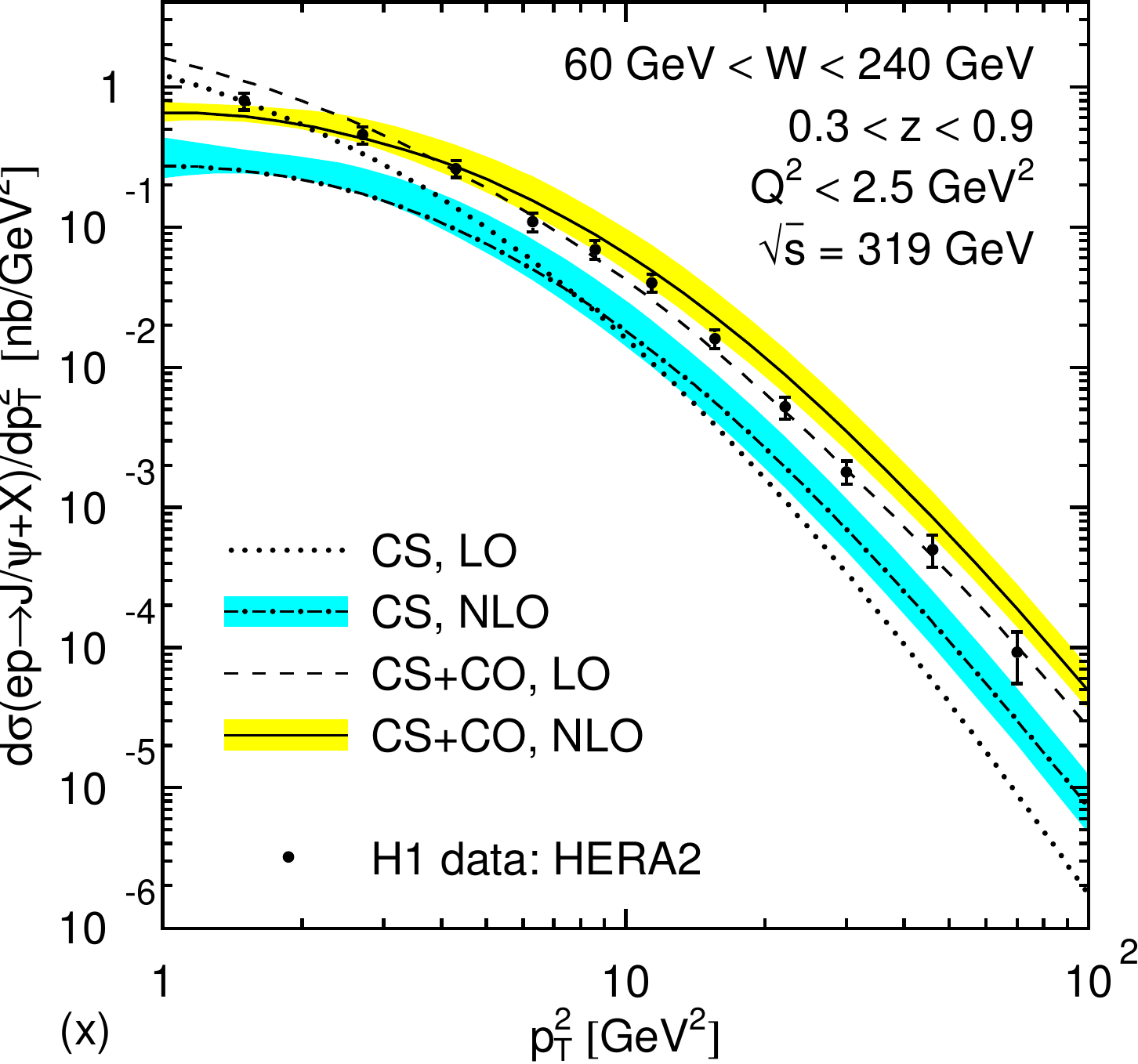}
\includegraphics[width=4cm]{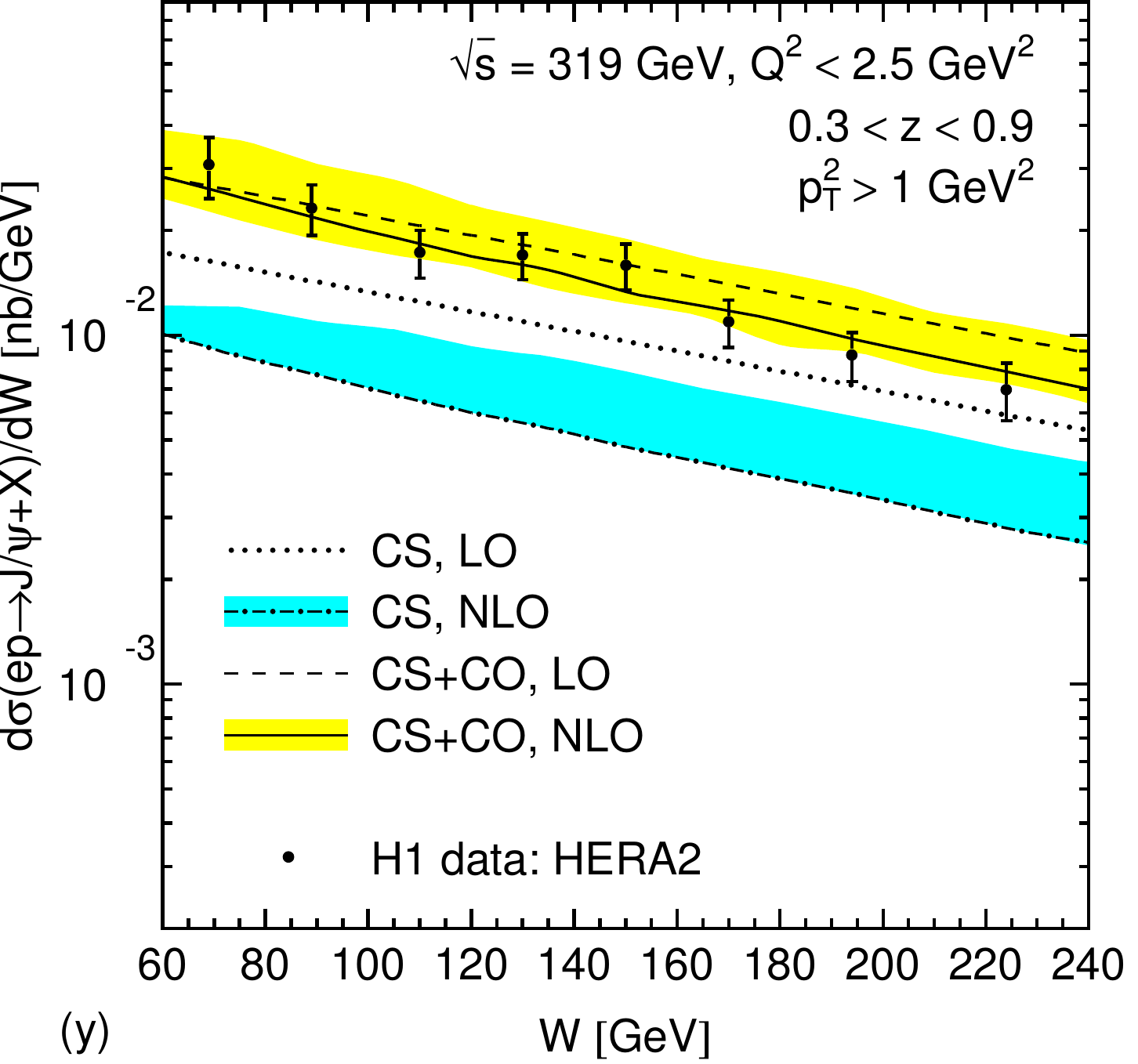}
\includegraphics[width=4cm]{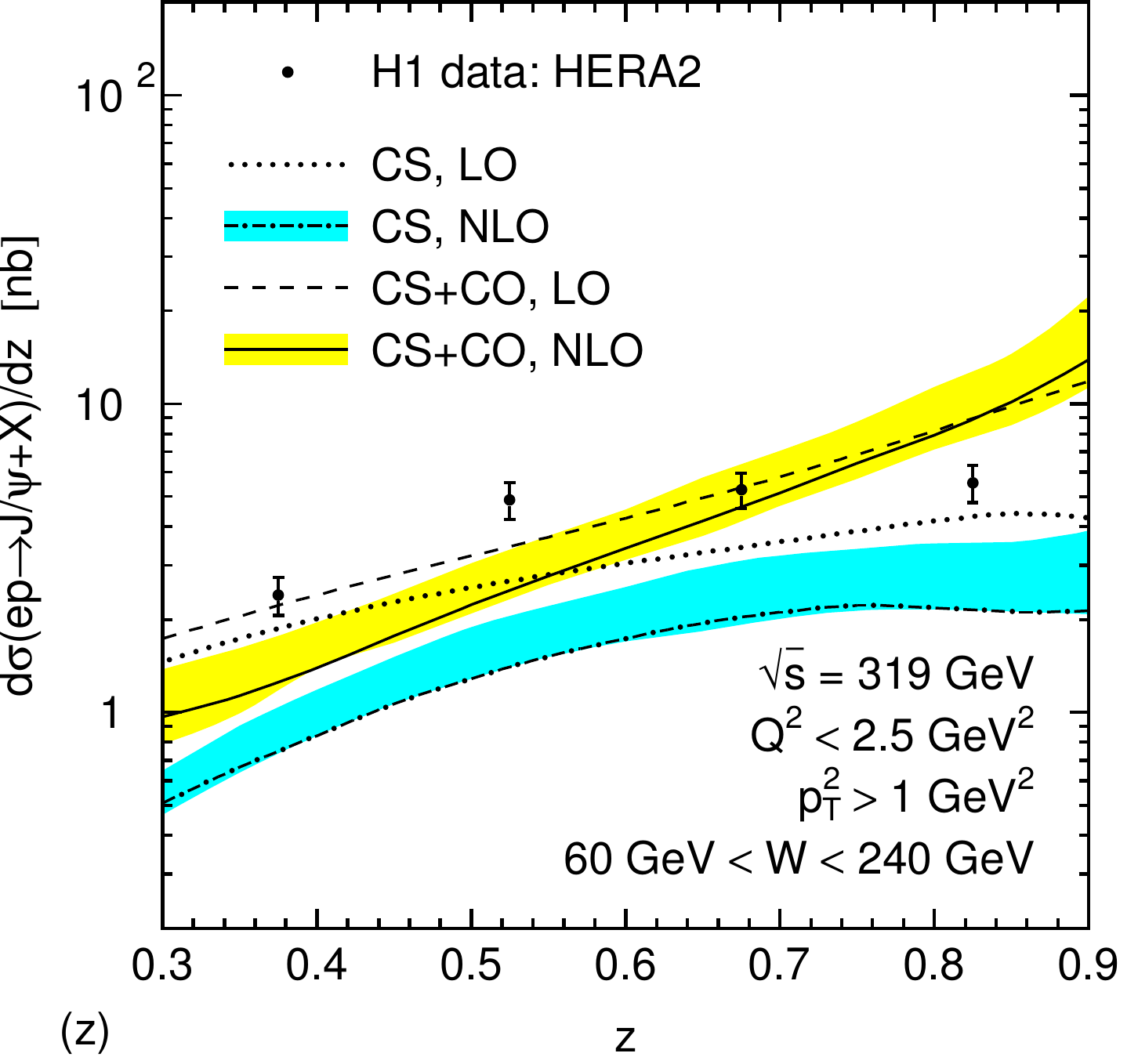}
\caption{Plots u-z (continuation): Results of the global fit \cite{Butenschoen:2011yh} compared to H1 \cite{Adloff:2002ex,Aaron:2010gz} data. The blue bands are the CSM predictions, the yellow bands include the CO contributions. The bands are constructed by variation of the renormalization, factorization and NRQCD scales.}
\end{figure*}

\begin{figure*}
\centering
\includegraphics[width=4cm]{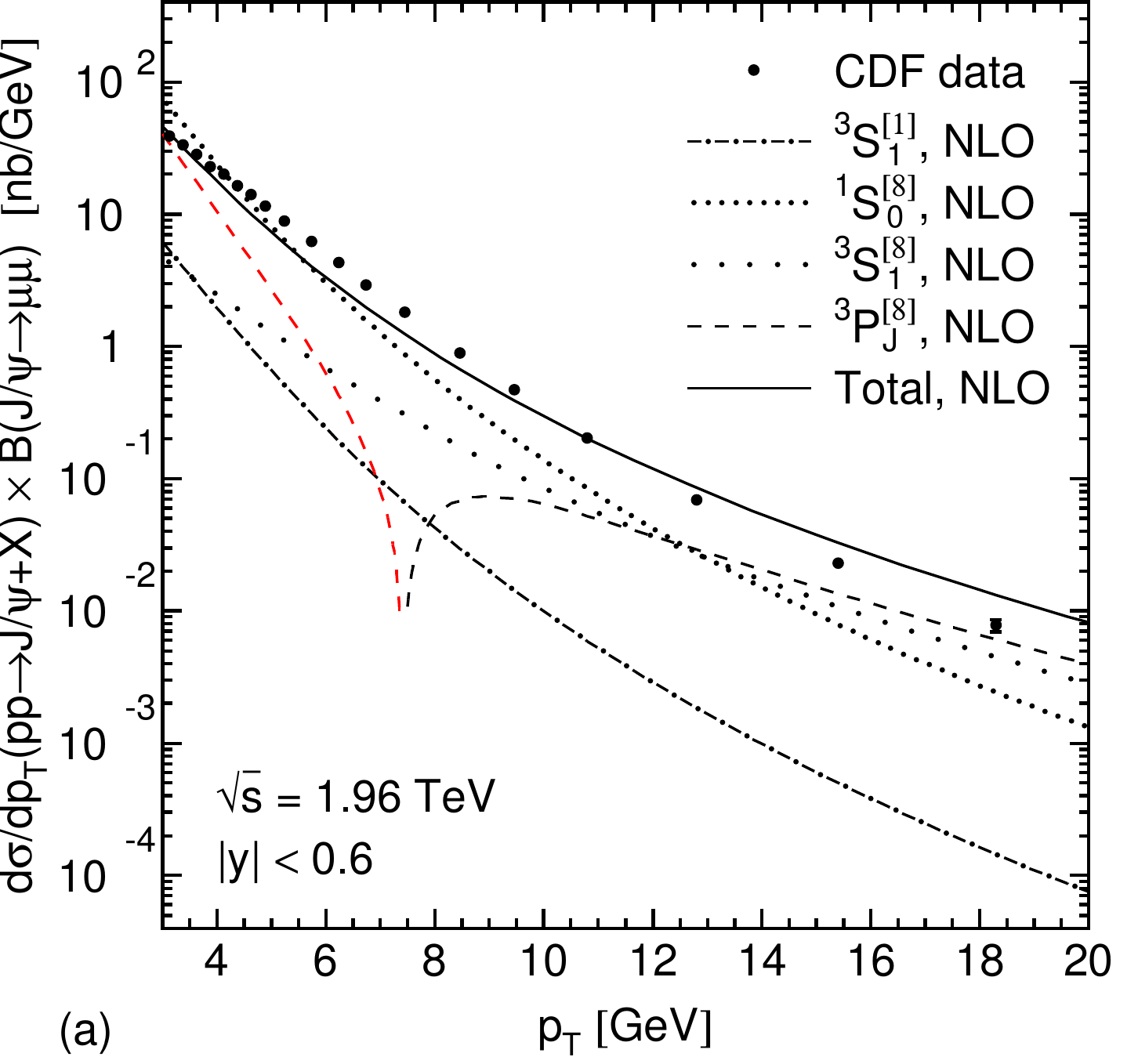}
\includegraphics[width=4cm]{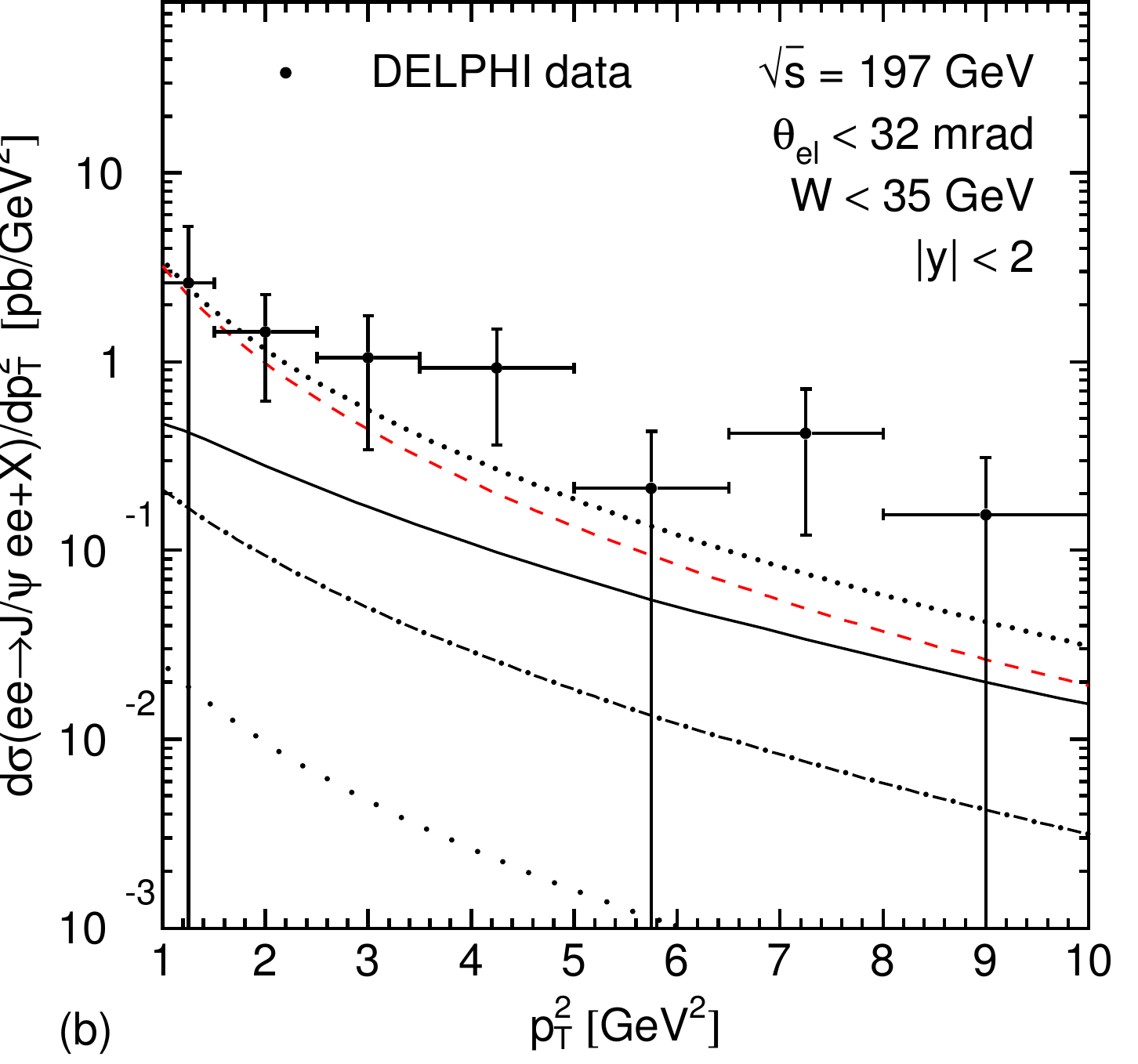}
\includegraphics[width=4cm]{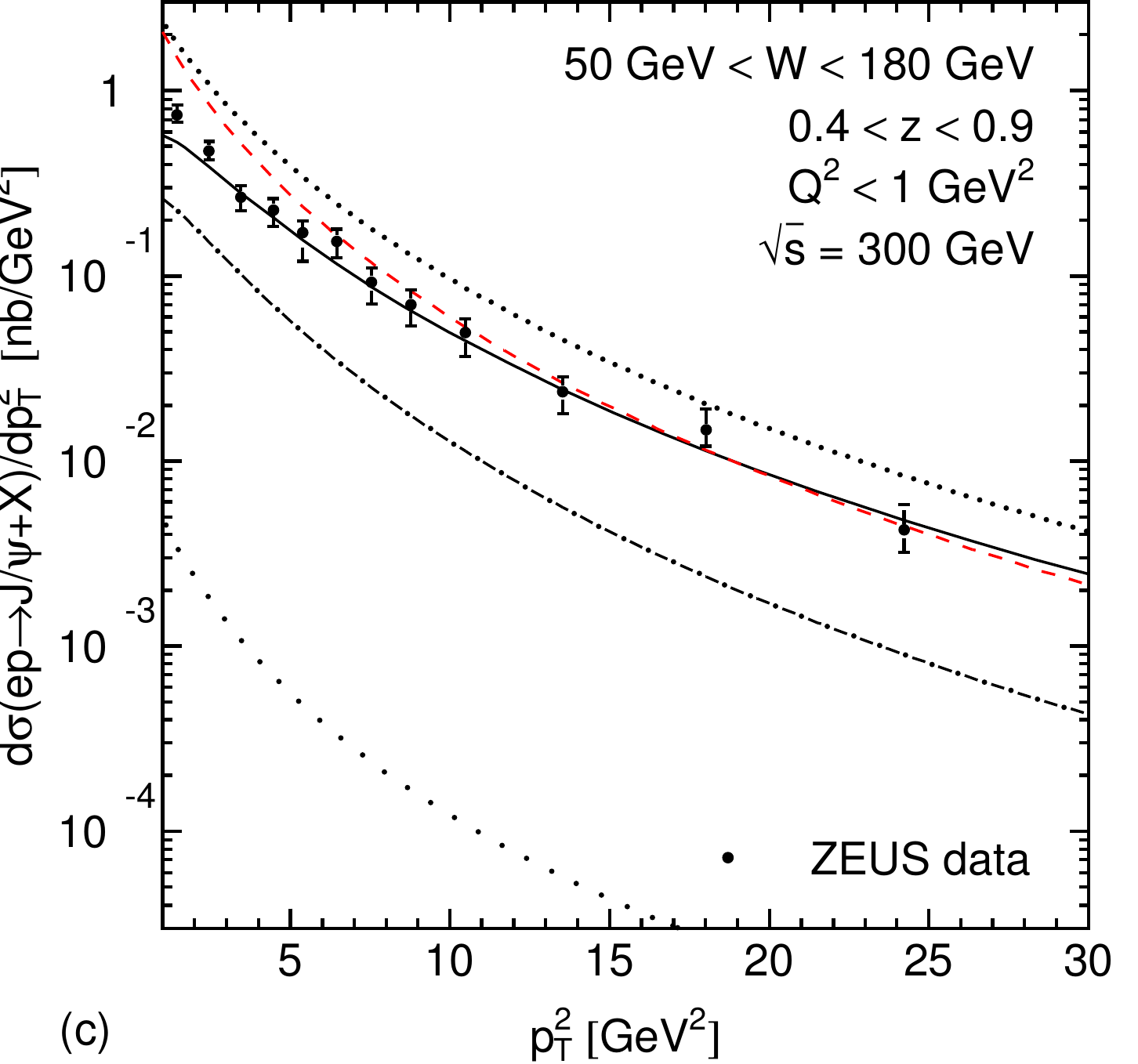}
\includegraphics[width=4cm]{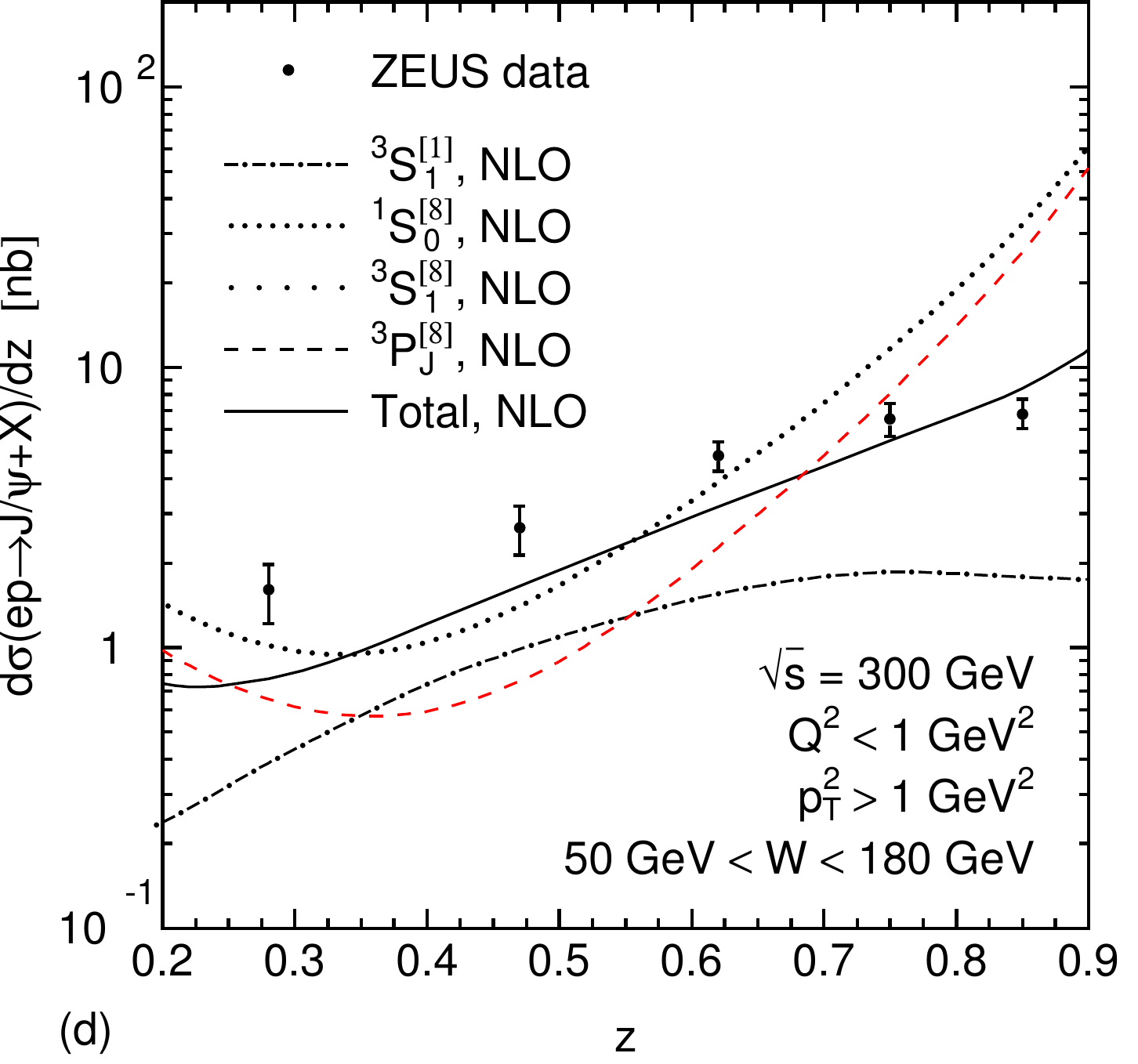}
\caption{\label{fig:decomposition}Decomposition of the NLO CS+CO curves from figures~\ref{fig:fitgraphs1}c, q, r, and t into the contributions of individual intermediate $c\overline{c}$ states. The line coding is the same for all four graphs. Red curves mean negative values. Please note that these curves are the short distance cross sections already multiplied by the corresponding LDMEs.}
\end{figure*}

\begin{figure*}
\centering
\includegraphics[width=4cm]{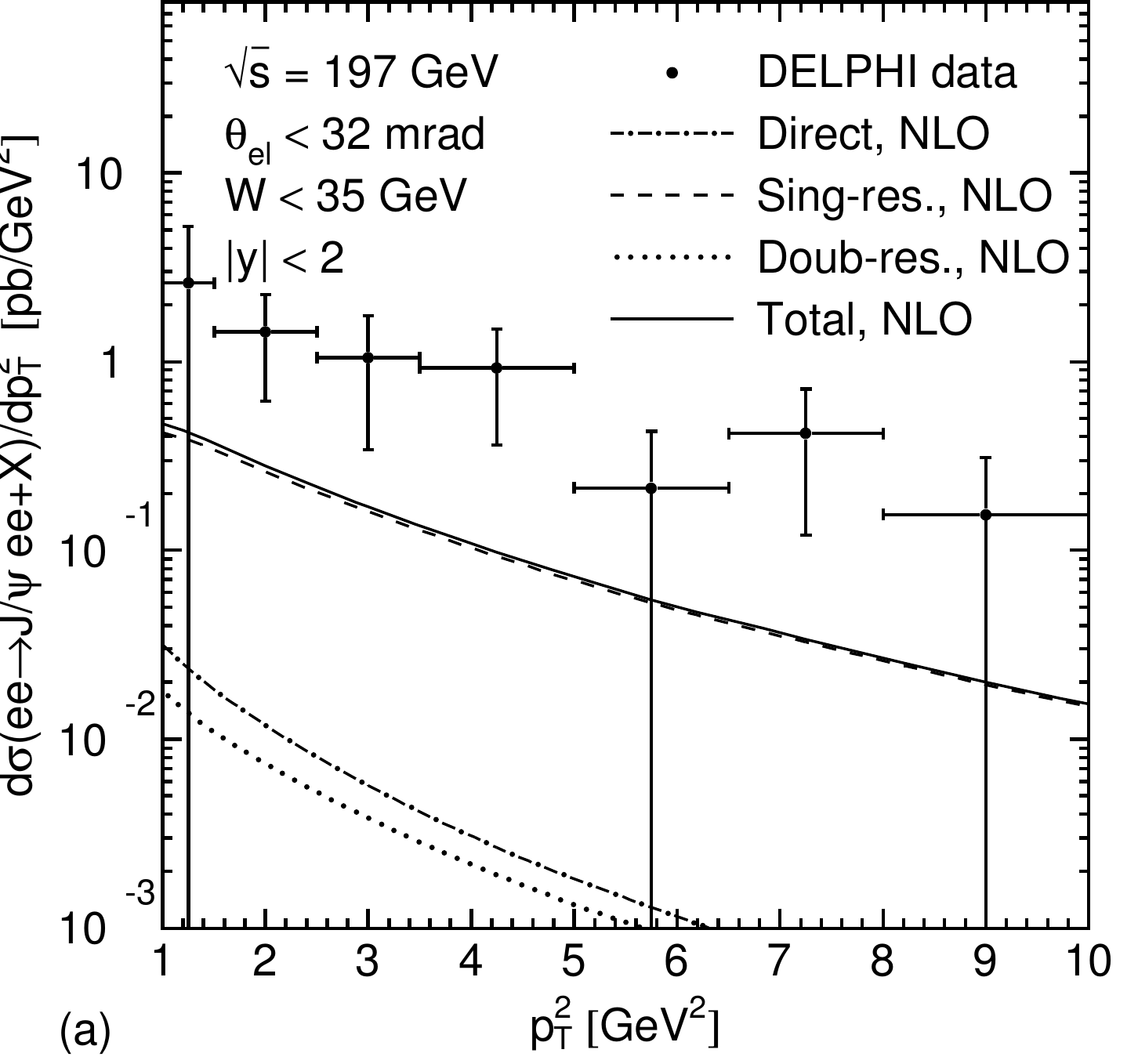}
\includegraphics[width=4cm]{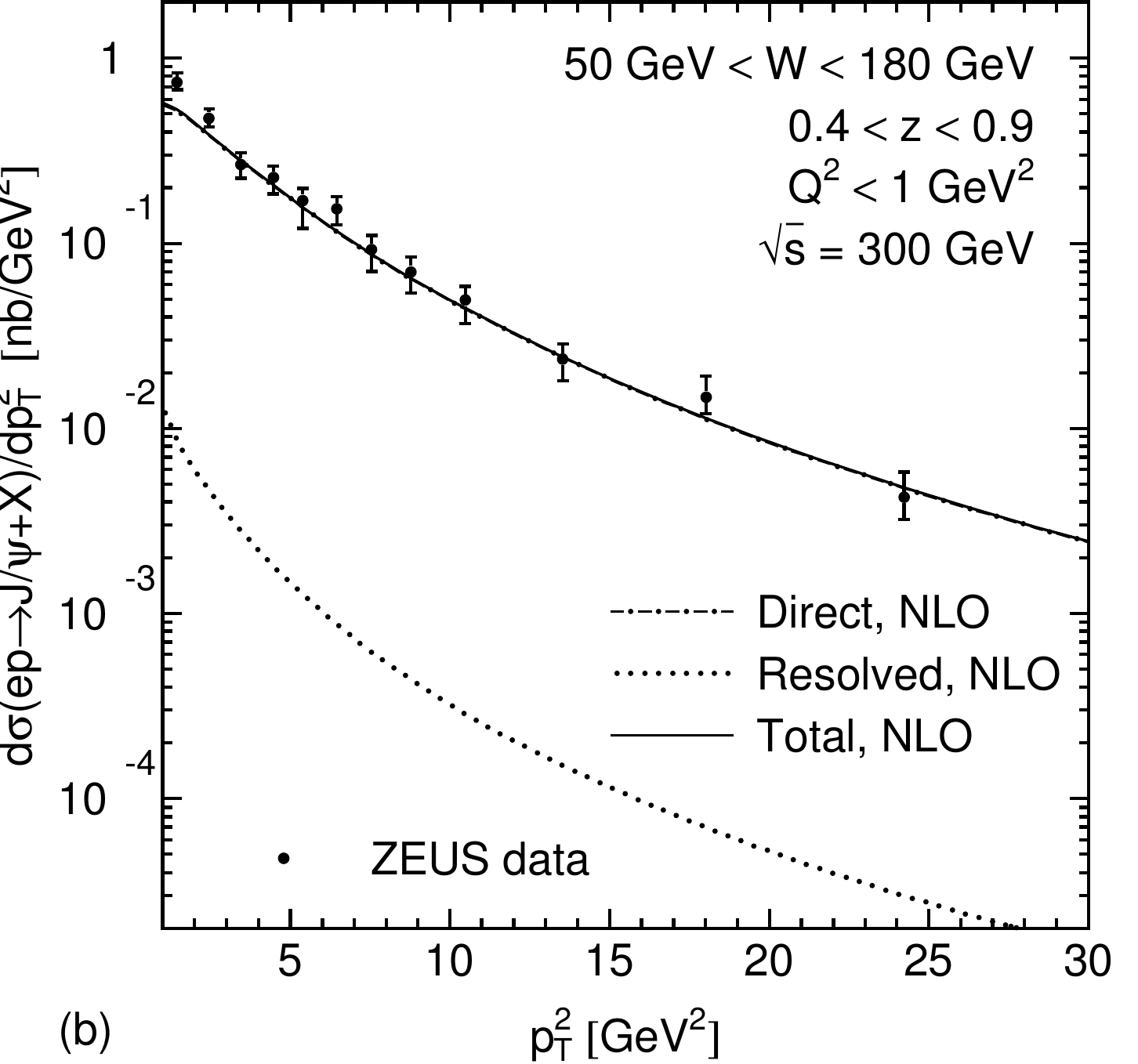}
\includegraphics[width=4cm]{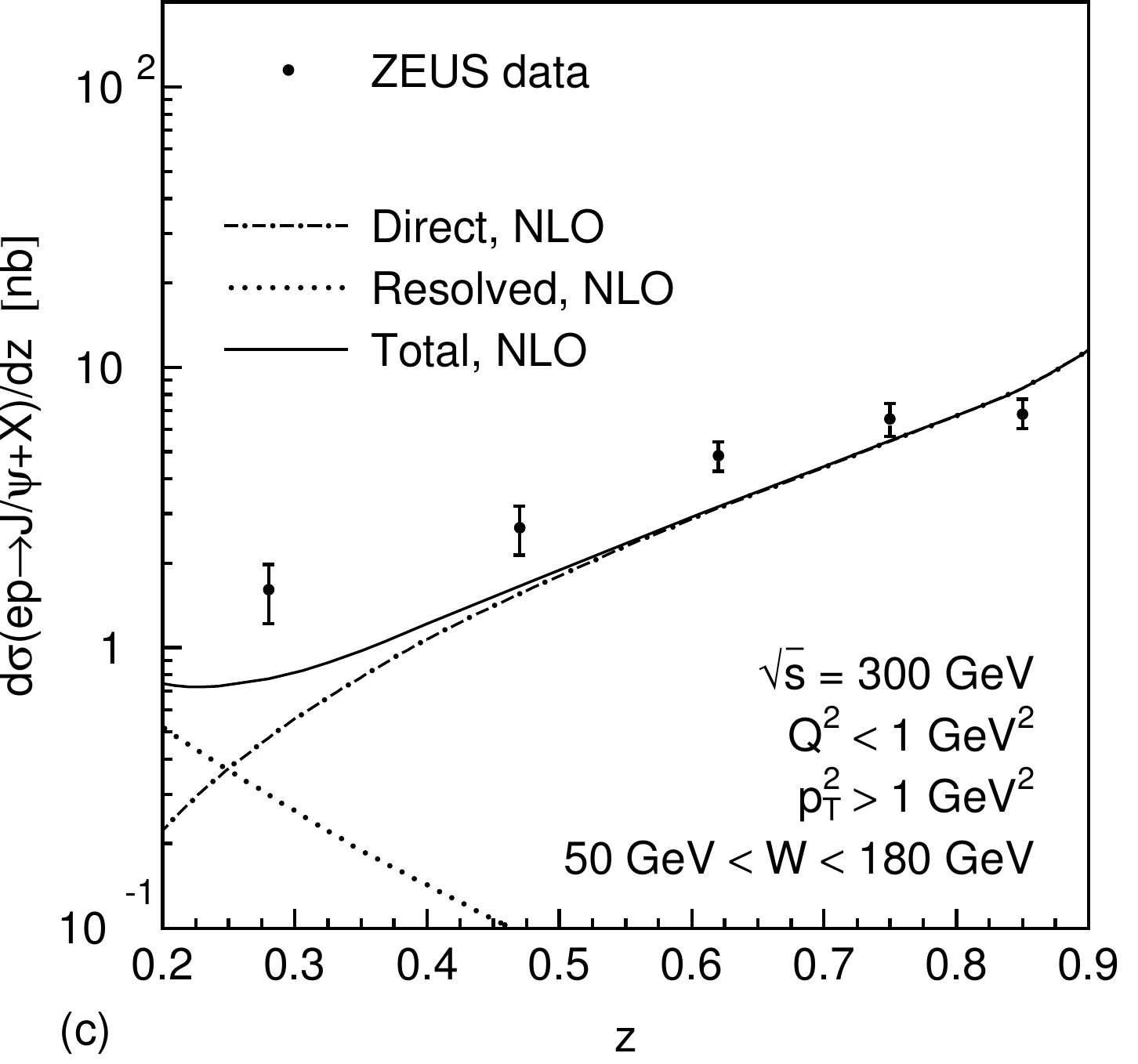}
\caption{\label{fig:fitdirres}Figure a shows a decomposition of the NLO CS+CO curve of the two-photon-scattering results from plot~\ref{fig:fitgraphs1}q into direct, single- and double-resolved photon contributions. Figures b and c show decompositions of the NLO CS+CO curves of the $p_T$ and $z$ distributions in photoproduction shown in figures~\ref{fig:fitgraphs1}r and t into direct and resolved contributions.}
\end{figure*}

\begin{table*}
\centering
\begin{tabular}{|c|c|c|}
\hline
 & Set A: Do not mind feed-downs & Set B: Subtract feed-downs first\\
\hline
$\langle {\cal O}^{J/\psi}(^1S_0^{[8]}) \rangle$ &
$(4.97\pm0.44)\times10^{-2}$~GeV$^3$ & $(3.04\pm0.35)\times10^{-2}$~GeV$^3$ \\
$\langle {\cal O}^{J/\psi}(^3S_1^{[8]}) \rangle$ &
$(2.24\pm0.59)\times10^{-3}$~GeV$^3$ & $(1.68\pm0.46)\times10^{-3}$~GeV$^3$ \\
$\langle {\cal O}^{J/\psi}(^3P_0^{[8]}) \rangle$ &
$(-1.61\pm0.20)\times10^{-2}$~GeV$^5$ & $(-9.08\pm1.61)\times10^{-3}$~GeV$^5$ \\
\hline
\end{tabular}
\caption{\label{tab:fit} Results of global fit \cite{Butenschoen:2011yh} for the $J/\psi$ CO LDMEs. Set A corresponds to the main fit shown in figure \ref{fig:fitgraphs1}. In set B, estimated feed-down contributions from higher charmonium states were subtracted from the prompt data prior to fitting (hadroproduction: 36\%, photoproduction: 15\%, $\gamma\gamma$ scattering: 9\%, $e^+e^-$ annihilation: 26\%).}
\end{table*}

\begin{figure}
\centering
\includegraphics[width=5.4cm]{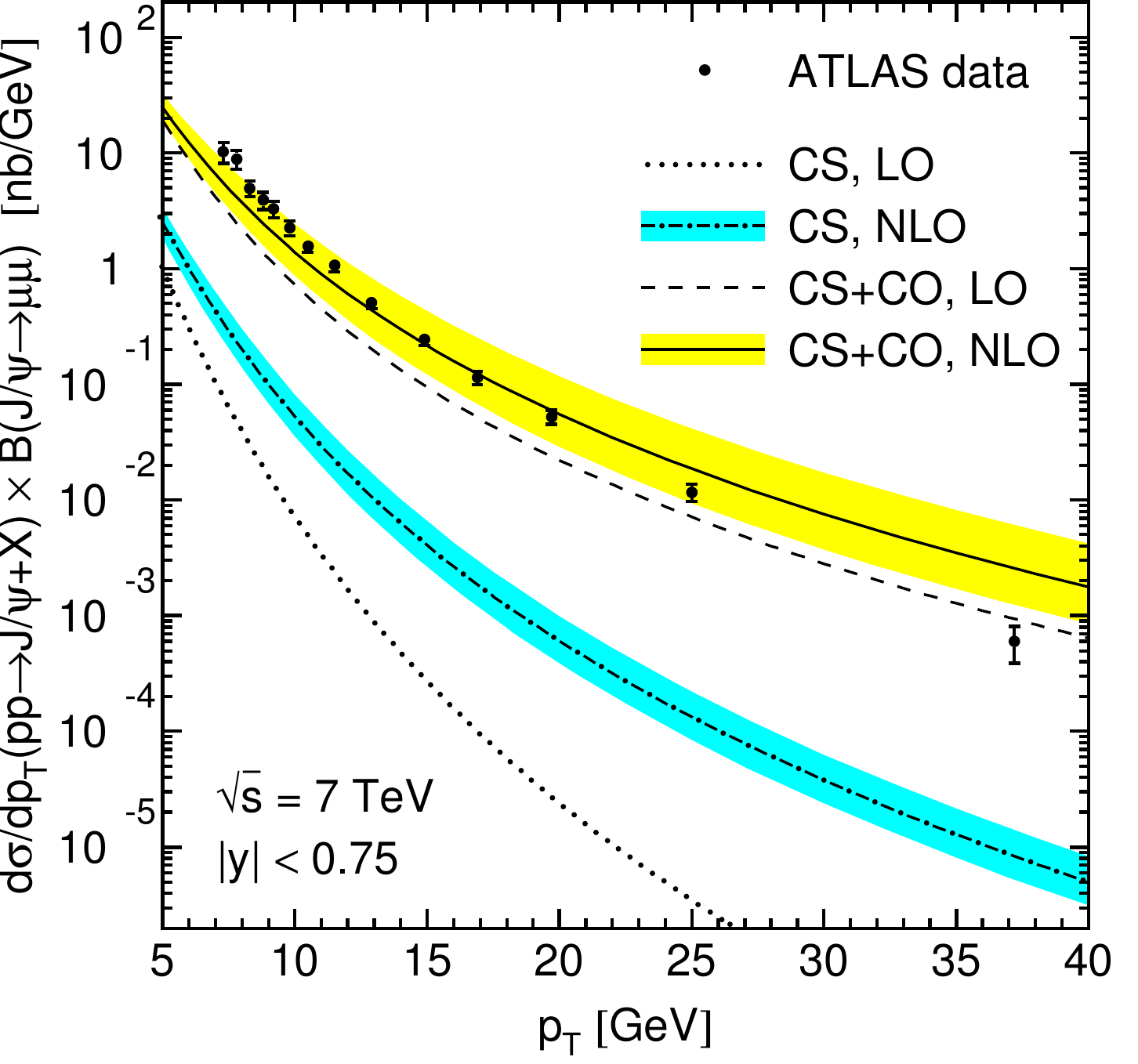}
\caption{\label{fig:atlasnew}Predictions for the ATLAS measurement \cite{Aad:2011sp}. Note that these data are not part of our global fit, since they became public after our global fit was finished. The bands are again constructed by variation of the renormalization, factorization and NRQCD scales. At very high~$p_T$, it will be necessary to resum large logarithms $\log(p_T^2/m_{J/\psi}^2)$. For instance, at $p_T=40$~GeV, $\alpha_s\log(p_T^2/M_{J/\psi}^2)\approx0.7$.}
\end{figure}

In \cite{Butenschoen:2011yh} we have published a global NLO fit of the three CO LDMEs to 194 data points of inclusive unpolarized $J/\psi$ production from 10 different experiments, see figure~\ref{fig:fitgraphs1} and table~\ref{tab:fit} for the fit results. These experiments include data from photoproduction at HERA, hadroproduction at RHIC, Tevatron and the LHC and additionally data from two-photon collisions measured at LEP and electron-positron collisions at KEKB. Most of the data fitted to is prompt, while none of our short distance cross sections does include feed-down contributions from higher charmonium states. Therefore, in table~\ref{tab:fit}, we list two sets of CO LDMEs: In set A, we ignore these contributions, this is our default fit, while in set B we take care of them by subtracting estimated feed-down contributions from the corresponding data prior to fitting. The values of set A are used for all curves of figures~\ref{fig:fitgraphs1}, \ref{fig:decomposition}, \ref{fig:fitdirres} and~\ref{fig:atlasnew}. In figure \ref{fig:decomposition} and \ref{fig:fitdirres} some diagrams of figure~\ref{fig:fitgraphs1} are decomposed into contributions of the different intermediate Fock states as well as into direct and resolved photon contributions. The global fit shows that at NLO all considered processes except perhaps the two-photon collisions can be described well with a single set of CO LDMEs. Furthermore, the CO LDMEs thus obtained do indeed exhibit the scaling behavior predicted by the scaling rules of NRQCD \cite{Bodwin:1994jh}: They are roughly of ${\cal O}(v^4)$ relative to $\langle {\cal O}^{J/\psi}(^3S_1^{[1]}) \rangle$, with $v^2\approx0.2$. The color-singlet (CS) contributions alone are on the other hand shown to fall clearly short of the data everywhere except for the BELLE total $e^+e^-$ cross section.

At this point let us look at the different production processes in more detail. Figures~\ref{fig:fitgraphs1}a--o show $J/\psi$ transverse momentum $p_T$ distributions of various hadroproduction measurements. Nearly all data points lie within the NRQCD NLO band, which estimates corrections due to even higher orders in $\alpha_s$. From a comparison of the LO and NLO curves the $\alpha_s$ expansion seems to converge much more rapidly when the CO contributions are included than in the CSM, where the slope drastically changes when going from LO to NLO. At higher values of transverse momentum than considered in the fit, resummations of large logarithms $\log(p_T^2/M_{J/\psi}^2)$ appearing in the short distance cross section will become necessary, as can be seen from our comparison to recent ATLAS data in figure~\ref{fig:atlasnew}. Such resummations are a standard tool to enhance the applicability of fixed-order calculations in multi-scale problems, see figure~8 of \cite{Kniehl:2008zza} as an example. In the case of quarkonium production, this might best be achieved within the framework outlined in \cite{Kang:2011zz}. In figure~\ref{fig:decomposition}a, the NLO CS+CO cross section of plot~\ref{fig:fitgraphs1}c, as an example, is decomposed into the various contributions of the different intermediate $c\overline{c}$ Fock states. These are the short distance cross sections already multiplied by their respective LDMEs. The fact that the overall $^3P_J^{[8]}$ contributions are negative for $p_T\lessapprox7$~GeV (here the short distance cross section is positive, but the LDME is negative, for $p_T\gtrapprox7$~GeV both are negative) is not worrying since individual contributions are unphysical, only the overall cross section has to be positive.

The global fit does also describe the photoproduction at HERA well. In the photoproduction limit, the incoming electron or positron interacts with the proton via a quasi-real bremsstrahlung photon. Figures~\ref{fig:fitgraphs1}r--z show distributions in the $J/\psi$ transverse momentum $p_T$, in the photon-proton invariant mass $W$ and in the inelasticity variable $z$, which in the proton rest frame is the fraction of the photon energy taken over by the $J/\psi$. Here, even the $z$ distribution is now much better described than in previous analyses, for two reasons: First, this analysis is the first one to include resolved photon contributions, in which the photon in turn interacts with the proton via its hadronic content. Resolved photoproduction dominates the cross section below $z\approx0.25$, see figure~\ref{fig:fitdirres}c. And secondly, the cross section is at high $z$ now much better described than in the older Born analyses, which predicted a steep rise in the cross section not found in the data. The reason is a strong cancellation between the $^1S_0^{[8]}$ and $^3P_J^{[8]}$ contributions, due to the negative value of $\langle {\cal O}^{J/\psi}(^3P_0^{[8]}) \rangle$, as shown in figure~\ref{fig:decomposition}d. We further note that the H1 HERA~2 data \cite{Aaron:2010gz} (fig.~\ref{fig:decomposition}x--z) lie systematically below the data measured by H1 at HERA~1 \cite{Adloff:2002ex} (fig.~\ref{fig:decomposition}u--w) and by the ZEUS collaboration both at HERA~1 \cite{Chekanov:2002at} (fig.~\ref{fig:decomposition}r--t) and HERA~2 \cite{ZEUStalk}.

Figure~\ref{fig:fitgraphs1}q shows the transverse momentum distribution for two-photon scattering as measured by the DELPHI collaboration at LEP \cite{Abdallah:2003du}. In this global analysis \cite{Butenschoen:2011yh}, for the first time also NLO contributions due to single and double resolved photons have been calculated. The single resolved contributions are in fact contributing up to 99\% to the overall cross section, see figure~\ref{fig:fitdirres}a. The fact that the DELPHI data overshoots the NRQCD prediction is not worrying since the experimental errors are huge with just 16 events entering the data of figure~\ref{fig:fitgraphs1}q.

As for $J/\psi$ production in electron-positron annihilation, the production in association with another charmonium or open $c\overline{c}$ state is dominating the inclusive cross section \cite{:2009nj}. Since these contributions are not included in our calculation, in figure~\ref{fig:fitgraphs1}p we compare our results for the total production cross section at KEKB with the measurement \cite{:2009nj} performed by Belle, in which the double charmonium and $J/\psi+c\overline{c}$ contributions are explicitly subtracted from the inclusive cross section. We see here that both the NLO CS and the NLO CS+CO results are compatible with the data. However, two things have to be noted here: First, in all our calculations, the partonic final states of the LO amplitudes consist of $J/\psi$ plus one light quark. But within the CSM, the Born process is $e^+e^-\to J/\psi+gg$, which in our terminology is an NLO processes. We do not consider NLO corrections to $e^+e^-\to J/\psi+gg$ for reasons of consistency, although they have been shown to further increase the CSM prediction \cite{Ma:2008gq}. The second caveat is that in the Belle analysis \cite{:2009nj} only events with more than 4 charged tracks have been measured. The correction factor making up for the missing events could not be determined, therefore the actual cross section is larger than the presented one by an unknown factor.

\subsection{Dependence on low $p_T$ cuts}

\begin{table*}
\centering
\begin{tabular}{c|ccccc}& $p_T > 1$ GeV & $p_T > 2$ GeV & $p_T > 3$ GeV & $p_T > 5$ GeV & $p_T > 7$ GeV \\
\small Hadroproduction data left & \small 148 points & \small 134 points & \small 119 points & \small 86 points & \small 60 points \\ \hline
$\langle {\cal O}^{J/\psi}(^1S_0^{[8]}) \rangle\; [10^{-2}\mbox{ GeV}^3]
$&$ 5.68 \pm 0.37  $&$ 4.25 \pm 0.43  $&$ 4.97 \pm 0.44  $&$ 4.92 \pm 0.49  $&$ 3.91 \pm 0.51 $\\
$\langle {\cal O}^{J/\psi}(^3S_1^{[8]}) \rangle\; [10^{-3}\mbox{ GeV}^3]
$&$ 0.90 \pm 0.50  $&$ 2.94 \pm 0.58  $&$ 2.24 \pm 0.59  $&$ 2.23 \pm 0.62  $&$ 2.96 \pm 0.64 $\\
$\langle {\cal O}^{J/\psi}(^3P_0^{[8]}) \rangle\; [10^{-2}\mbox{ GeV}^5]
$&$ -2.23 \pm 0.17 $&$ -1.38 \pm 0.20 $&$ -1.61 \pm 0.20 $&$ -1.59 \pm 0.22 $&$ -1.16 \pm 0.23 $
\end{tabular}
\caption{\label{tab:Cutdependence1} Global fit results for different low-$p_T$ cuts on the hadroproduction data. Besides the number of hadroproduction data points listed, the fits include 74 data points of photoproduction and two-photon scattering, and 1 data point of $e^+e^-$ annihilation. The fit with $p_T>3$ GeV is our default fit. One can see that the fit results are practically independent of the low-$p_T$ cut, unless the cut is taken below 2~GeV.}
\end{table*}

\begin{table*}
\centering
\begin{tabular}{c|ccccc}& $p_T > 1$ GeV & $p_T > 2$ GeV & $p_T > 3$ GeV & $p_T > 5$ GeV & $p_T > 7$ GeV \\
\small Photoproduction and $\gamma\gamma$ data left & \small 74 points & \small 30 points & \small 15 points & \small 5 points & \small 1 points \\ \hline
$\langle {\cal O}^{J/\psi}(^1S_0^{[8]}) \rangle\; [10^{-2}\mbox{ GeV}^3]
$&$ 4.97 \pm 0.44 $&$ 5.10 \pm 0.92 $&$ 4.05 \pm 1.17 $&$ 5.44 \pm 1.27 $&$ 9.56 \pm 1.59 $\\
$\langle {\cal O}^{J/\psi}(^3S_1^{[8]}) \rangle\; [10^{-3}\mbox{ GeV}^3]
$&$ 2.24 \pm 0.59 $&$ 2.11 \pm 1.22 $&$ 3.52 \pm 1.56 $&$ 1.73 \pm 1.68 $&$ -3.66 \pm 2.09 $\\
$\langle {\cal O}^{J/\psi}(^3P_0^{[8]}) \rangle\; [10^{-2}\mbox{ GeV}^5]
$&$ -1.61 \pm 0.20 $&$ -1.58 \pm 0.48 $&$ -0.97 \pm 0.63 $&$ -1.63 \pm 0.68 $&$ -3.73 \pm 0.83 $
\end{tabular}
\caption{\label{tab:Cutdependence2} Global fit results for different low-$p_T$ cuts on the photoproduction and two-photon scattering data. Besides the number of photoproduction and two-photon scattering data points listed, the fits include 119 data points of hadroproduction, and 1 data point of $e^+e^-$ annihilation. The fit with $p_T>1$ GeV is our default fit. The necessity of a critical amount of photoproduction data to stabilize the fit is clearly seen. In the region of a low-$p_T$ cut between 1~GeV and 5~GeV, where there is enough photoproduction data included, the fit results are practically independent of that cut.}
\end{table*}

\begin{table*}
\centering
\begin{tabular}{c|ccccc}
& $p_T > 1$ GeV & $p_T > 2$ GeV & $p_T > 3$ GeV & $p_T > 5$ GeV & $p_T > 7$ GeV \\
\small Hadroproduction data left & \small 148 points & \small 134 points & \small 119 points & \small 86 points & \small 60 points \\ \hline
$\langle {\cal O}^{J/\psi}(^1S_0^{[8]}) \rangle\; [10^{-2}\mbox{ GeV}^3]
$&$ 8.54 \pm 0.52  $&$ 16.85 \pm 1.23  $&$ 11.02 \pm 1.67 $&$ 1.68 \pm 2.20 $&$ 2.18 \pm 2.56 $\\
$\langle {\cal O}^{J/\psi}(^3S_1^{[8]}) \rangle\; [10^{-3}\mbox{ GeV}^3]
$&$ -2.66 \pm 0.69 $&$ -13.36 \pm 1.60 $&$ -5.56 \pm 2.19 $&$ 8.75 \pm 2.98 $&$ 10.34 \pm 3.55 $\\
$\langle {\cal O}^{J/\psi}(^3P_0^{[8]}) \rangle\; [10^{-2}\mbox{ GeV}^5]
$&$ -3.63 \pm 0.23 $&$ -7.70 \pm 0.61  $&$ -4.46 \pm 0.87 $&$ 2.20 \pm 1.23 $&$ 3.50 \pm 1.50 $ \\ \hline
$M_0 \; [10^{-2}\mbox{ GeV}^3]         $&$ 2.25 \pm 0.12  $&$ 3.51 \pm 0.19   $&$  3.29 \pm 0.20 $&$ 5.50 \pm 0.29 $&$ 8.24 \pm 0.58 $\\
$M_1\; [10^{-3}\mbox{ GeV}^3]          $&$ 6.37 \pm 0.19  $&$ 5.80 \pm 0.19   $&$  5.54 \pm 0.20 $&$ 3.27 \pm 0.29 $&$ 1.63 \pm 0.43 $\\
\end{tabular}
\caption{\label{tab:Cutdependence3} Fits restricted to only hadroproduction data: Fit results for different low-$p_T$ cuts. The fits include only the listed number of hadroproduction data points. The fits are underconstrained, and the fit results depend strongly on the low-$p_T$ cut chosen. For comparison, also the values for the linear combinations $M_0$ and $M_1$ used in references \cite{Ma:2010yw,Ma:2010jj} are given.}
\end{table*}

In our default global fit, we impose a low-$p_T$ cut on all hadroproduction data of 3~GeV, since data with lower $p_T$ can not be described in a fixed-order treatment due to soft gluon radiation: The data exhibits a flattening, which can not be successfully described. For a similar reason we consider only two-photon scattering data with $p_T>1$~GeV. In this section we show the stability of our global fit with respect to varying the low-$p_T$ cuts on the data. In that way we also show that our global fit is constrained, while a fit to hadroproduction data alone is not. We also successfully compare to the Tevatron-only fit published in \cite{Ma:2010yw,Ma:2010jj}.

In table~\ref{tab:Cutdependence1} we list our global fit results for different values of the low-$p_T$ hadroproduction cut. We see that the fit results are almost independent of this cut. The values for the fitted LDMEs vary little for low-$p_T$ cuts down to about 2~GeV. In table~\ref{tab:Cutdependence2} we then show results of our global fit with different cuts on the photoproduction and two-photon scattering data. We see that the fit is stable against varying this cut in the region between 1~GeV and 5~GeV. However, with even higher low-$p_T$ cuts, too little photoproduction data is left, so the fit becomes unstable.

The necessity to include data from different production mechanisms becomes even more obvious when we vary the low-$p_T$ cut in fits to hadroproduction data alone. In table~\ref{tab:Cutdependence3} we show the corresponding results when we restrict our global fit to hadroproduction data. The fit results show a very strong dependence on the low-$p_T$ cut, and the 3-parameter fit to only hadroproduction data is clearly underconstrained. Therefore, in \cite{Ma:2010yw,Ma:2010jj}, where a fit to CDF Tevatron run II data alone was performed, the authors fit the two linear combinations $M_0=\langle {\cal O}^{J/\psi}(^1S_0^{[8]})\rangle+3.9\, \langle {\cal O}^{J/\psi}(^3P_0^{[8]}) \rangle/m_c^2$ and $M_1=\langle {\cal O}^{J/\psi}(^3S_1^{[8]})\rangle-0.56\, \langle {\cal O}^{J/\psi}(^3P_0^{[8]}) \rangle/m_c^2$. For comparison we list our results for $M_0$ and $M_1$ in table \ref{tab:Cutdependence3} as well. We note that the results of our fit to only hadroproduction data with $p_T>7$~GeV agree well with their corresponding results
\begin{eqnarray*}
\lefteqn{M_0 = (8.54 \pm 1.02)\times 10^{-2}\; \mbox{GeV}^3,\mbox{ and}} \\
\lefteqn{M_1 = (1.67 \pm 1.05)\times 10^{-3}\; \mbox{GeV}^3,}
\end{eqnarray*}
which are listed in equation (18) of \cite{Ma:2010jj}.

\section{Predictions for polarization observables}

In order to scrutinize the universality of the LDMEs it is however not enough to consider only the inclusive $J/\psi$ yield. Therefore we now apply the set of CO LDMEs extracted in the last section to make predictions for $J/\psi$ polarization observables. More specifically, we use set ``B'' of table~\ref{tab:fit}. The results for photo- and hadroproduction are summarized in this section.

Measuring the polarization of $J/\psi$ means measuring the angular distribution of the two leptons, by which the $J/\psi$ is tagged. This distribution is parameterized via
\begin{eqnarray*}
W(\theta,\phi)&\propto&1+\lambda_\theta\cos^2\theta
+\lambda_\phi\sin^2\theta\cos(2\phi)
\nonumber\\
&&{}+\lambda_{\theta\phi}\sin(2\theta)\cos\phi,
\end{eqnarray*}
where $\theta$ and $\phi$ are the polar and azimuthal angles of the $\mu^+$ or $e^+$ in the $J/\psi$ rest frame. This definition does of course depend on 
the choice of the coordinate system axes. Among the frequently used coordinate frames are the helicity frame, in which the polar axis points in 
the direction of $-(\vec{p}_\gamma+\vec{p}_p)$, the Collins-Soper frame, in which it points to 
$\vec{p}_\gamma/|\vec{p}_\gamma|-\vec{p}_p/|\vec{p}_p|$ and the target frame, in which it points to $-\vec{p}_p$. Here, $\vec{p}_p$ is the three momentum of a colliding proton, and $\vec{p}_\gamma$ the three momentum of the colliding photon in case of photoproduction, and of the second proton or the antiproton in case of hadroproduction. $\lambda_\theta=0$ corresponds to unpolarized $J/\psi$, whereas $\lambda_\theta=+1$ (-1) stands for fully transversely (longitudinally) polarized $J/\psi$. Unfortunately, there are different naming conventions for the polarization parameters. In the photoproduction literature, the parameters $\lambda=\lambda_\theta$, $\mu=\lambda_{\theta\phi}$ and $\nu=2\lambda_\phi$ are widely used, and on top of that $\lambda_\theta$ is often called $\alpha$.

On the theoretical side, we calculate the parameters $\lambda_\theta$, $\lambda_\phi$ and $\lambda_{\theta\phi}$ via
\begin{eqnarray*}
\lefteqn{\lambda_\theta=\frac{d\sigma_{11}-d\sigma_{00}}{d\sigma_{11}+d\sigma_{00}},
\quad \lambda_\phi=\frac{d\sigma_{1,-1}}{d\sigma_{11}+d\sigma_{00}},}\\
\lefteqn{\lambda_{\theta\phi}=
\frac{\sqrt{2}\mbox{Re } d\sigma_{10}}{d\sigma_{11}+d\sigma_{00}},}
\end{eqnarray*}
where $d\sigma_{ij}$ are the differential $J/\psi$ production cross sections, calculated using the NRQCD factorization, but keeping the spin of the intermediate $c\overline{c}[n]$ pair fixed instead of summing over it. The spin polarization vectors $\epsilon^\ast(i)$ in the amplitude and $\epsilon(j)$ in the complex conjugated amplitude are instead replaced by their explicit expressions, as for example derived in \cite{Beneke:1998re}. In case of the Fock state $n={^3P_J^{[8]}}$, for which $S=L=1$, the orbital angular momentum is, however, still summed over. For the spin-zero Fock state $n={^1S_0^{[8]}}$, $d\sigma_{00}$ and $d\sigma_{11}$ are each set to one third of the unpolarized cross section and $d\sigma_{1,-1}=d\sigma_{10}=0$.

\subsection{Results for photoproduction}

\begin{figure*}
\centering
\begin{tabular}{|c|c|c|}
\hline
\includegraphics[width=5cm]{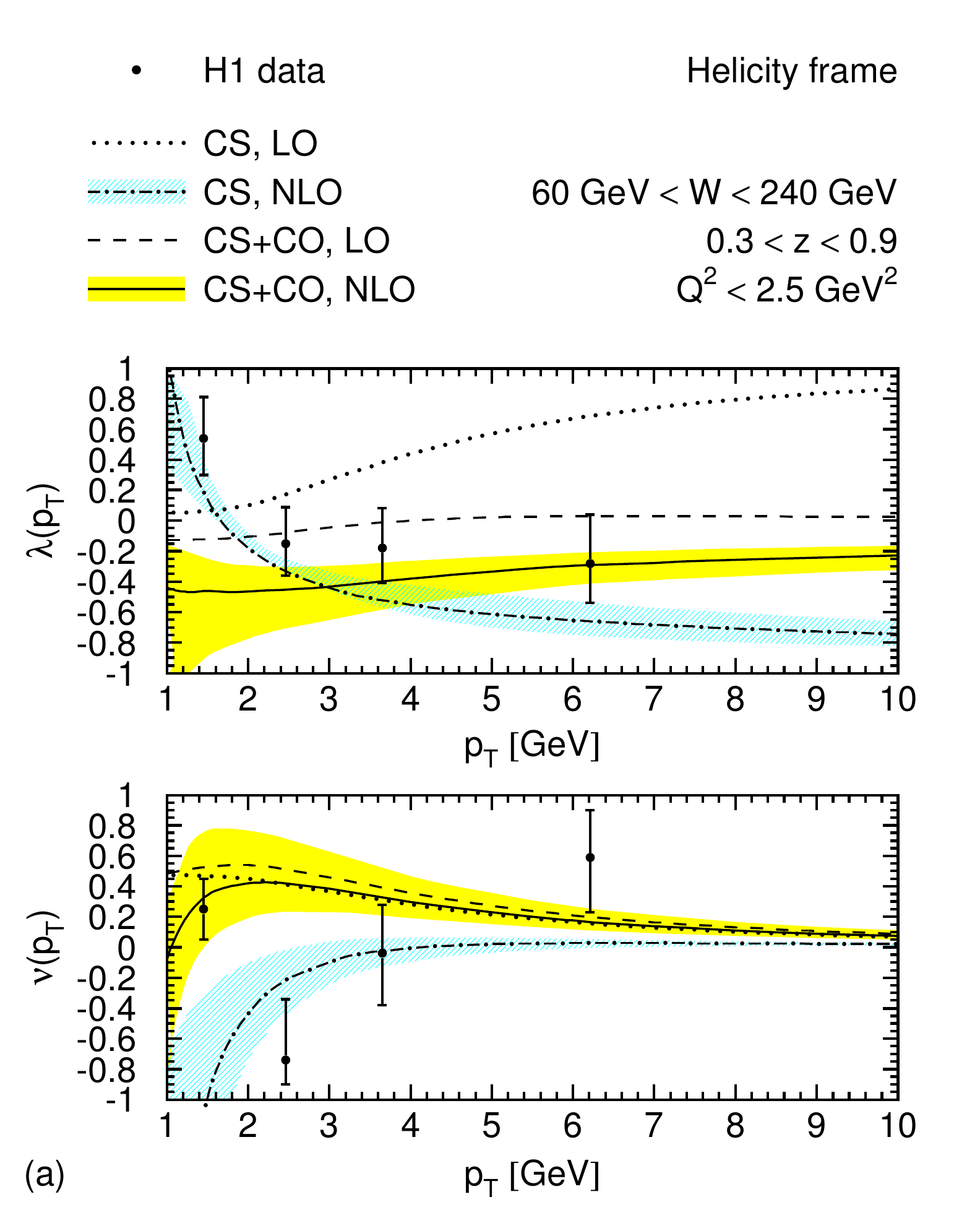}
&
\includegraphics[width=5cm]{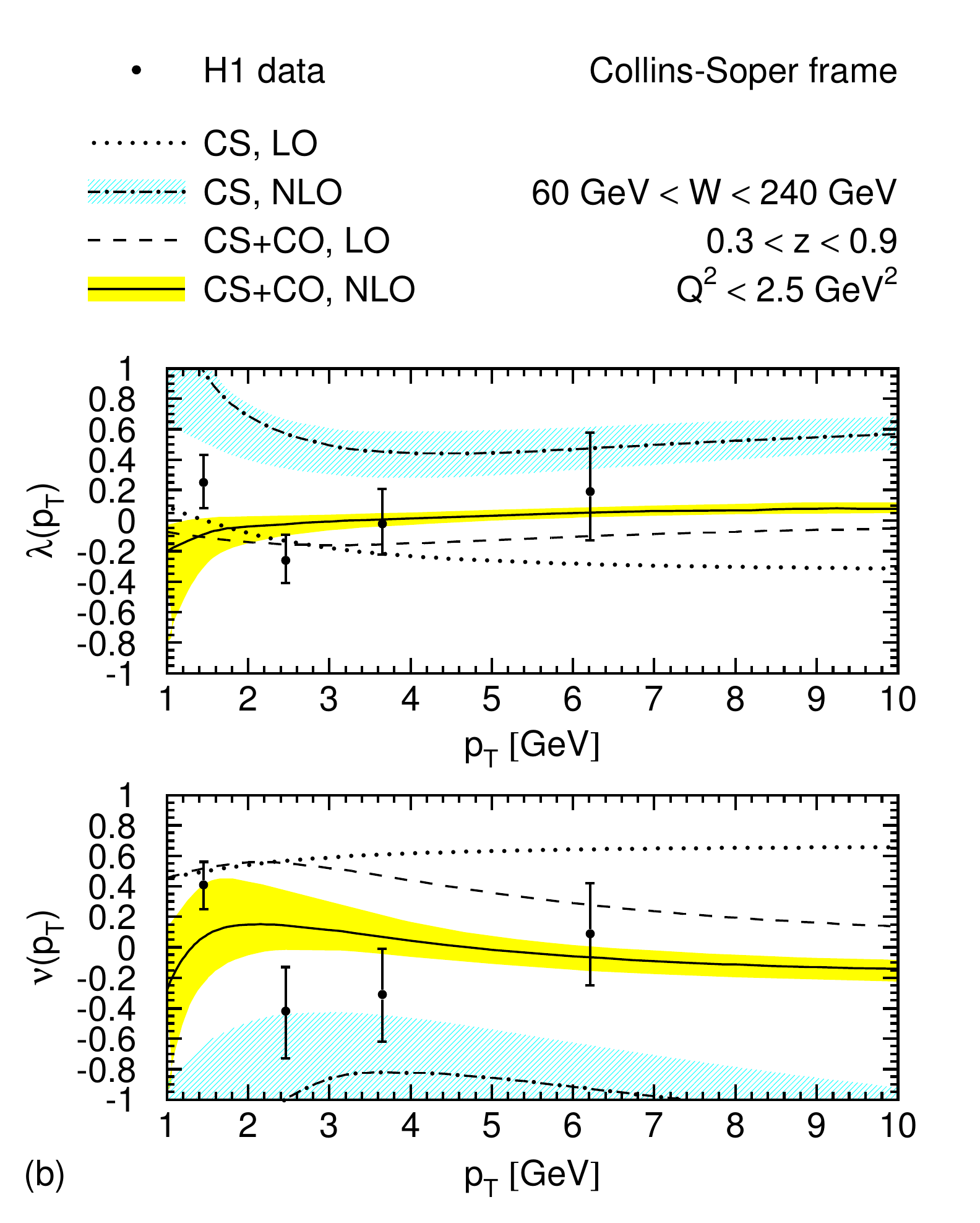}
&
\includegraphics[width=5cm]{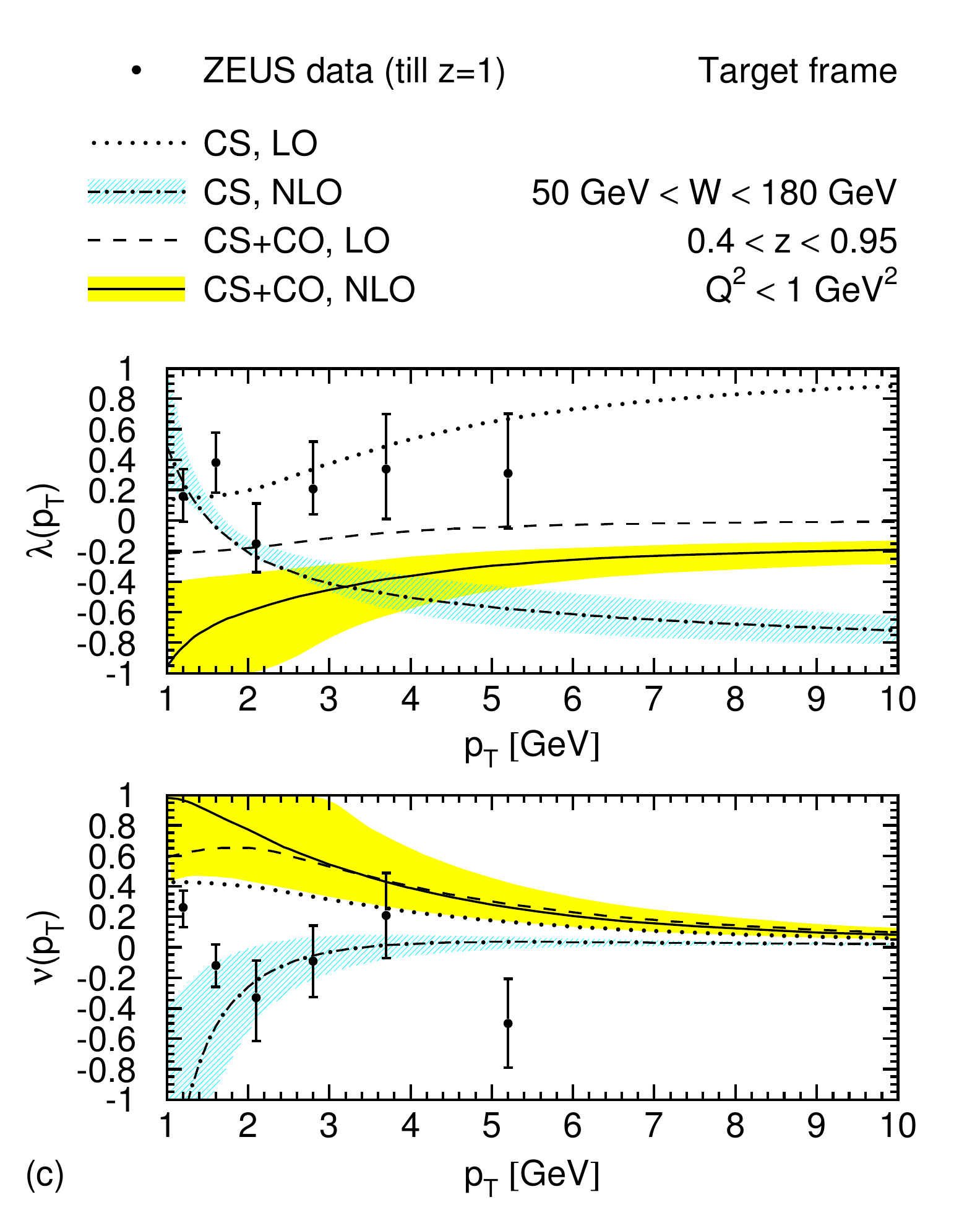}
\\ \hline
\includegraphics[width=5cm]{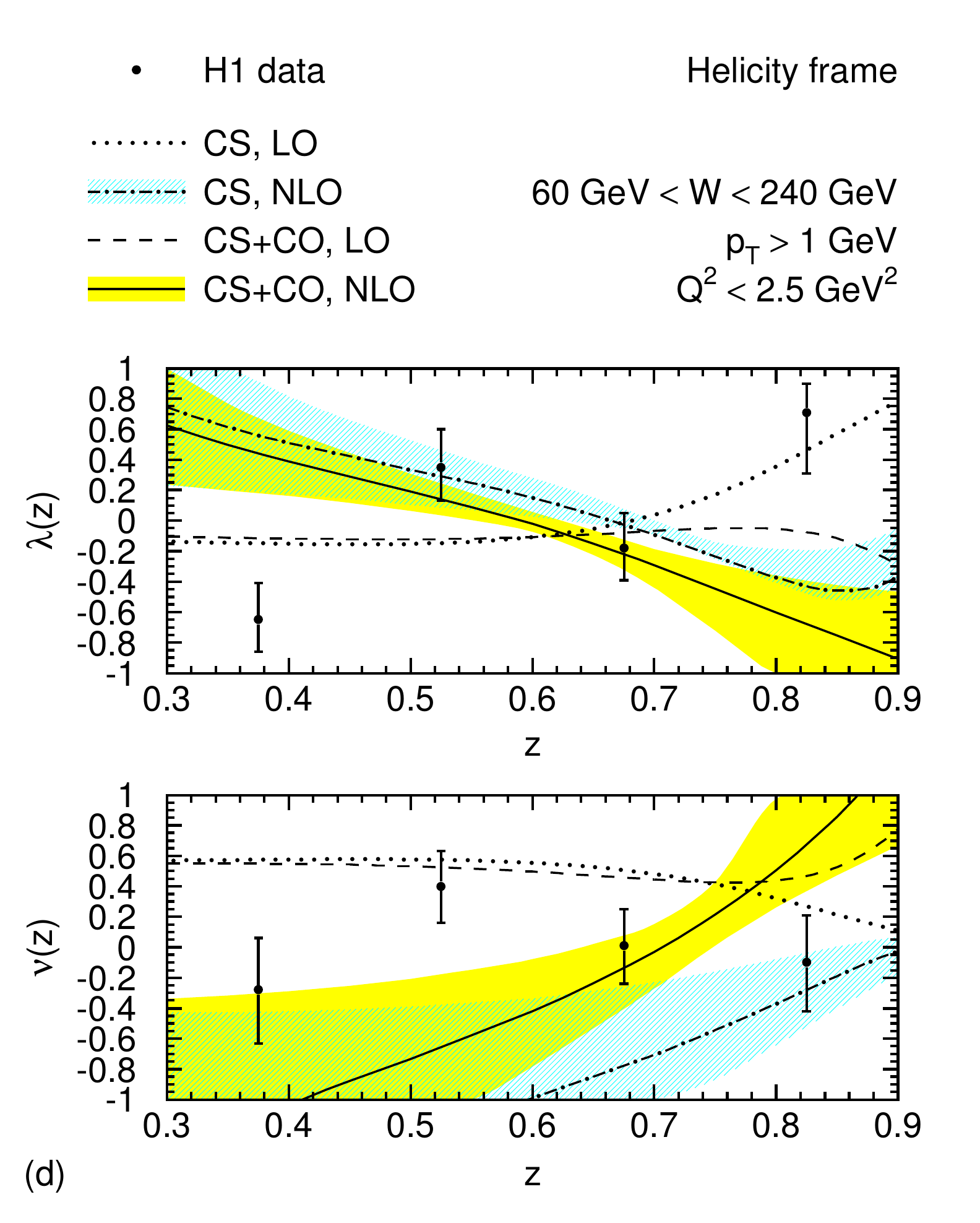}
&
\includegraphics[width=5cm]{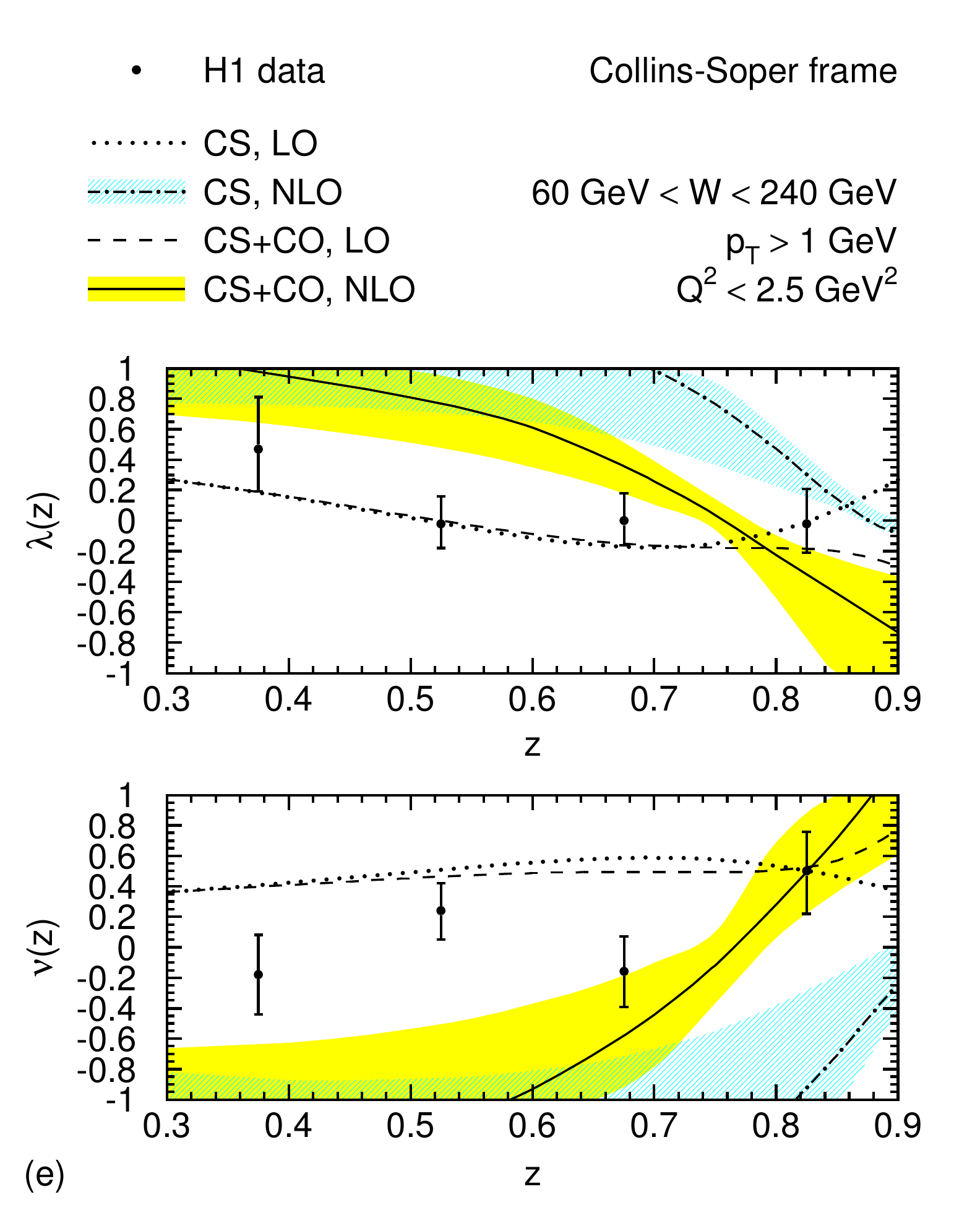}
&
\includegraphics[width=5cm]{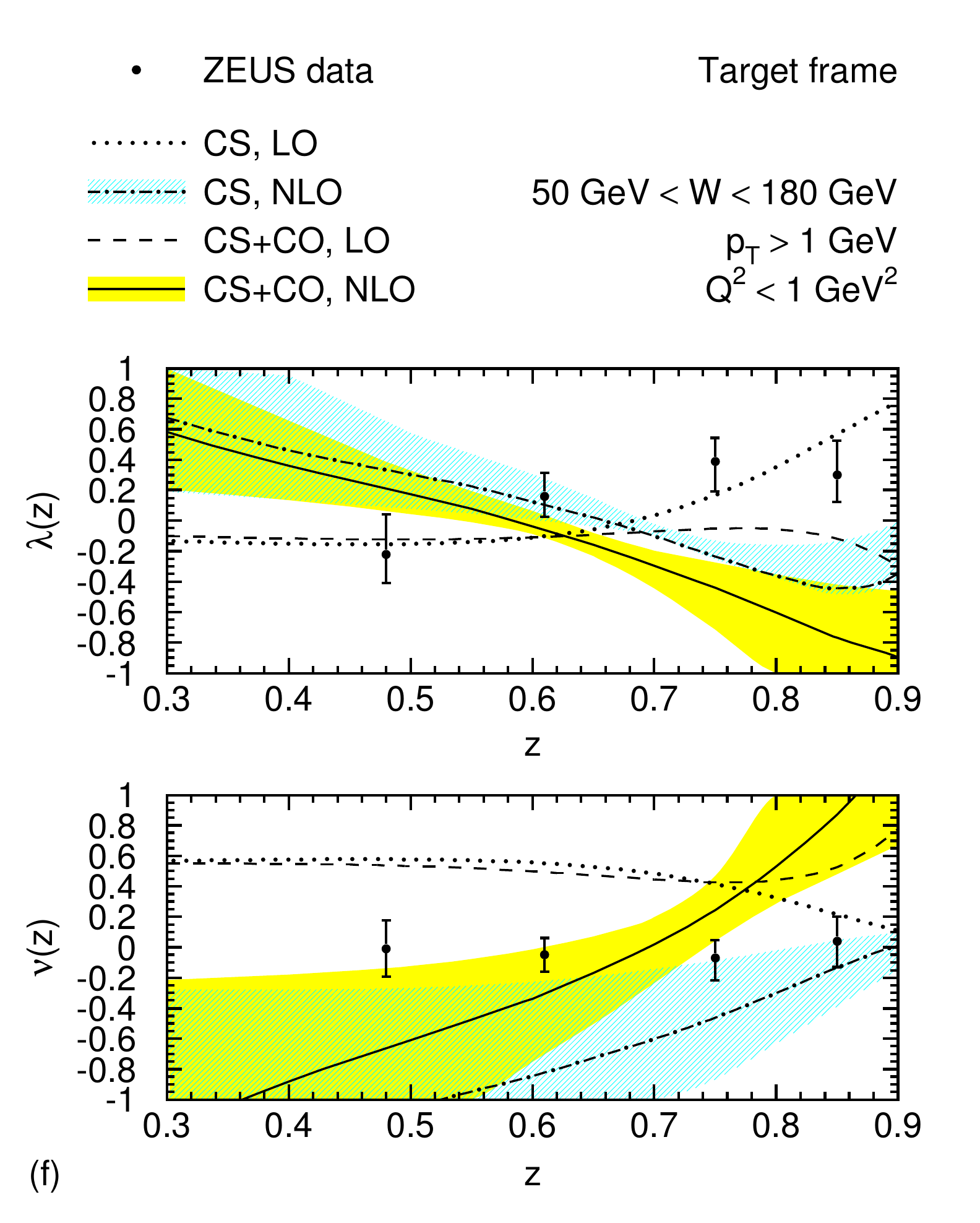}
\\ \hline
\end{tabular}
\caption{\label{fig:PhotoPol} Polarization parameters $\lambda$ and $\nu$ for direct photoproduction at HERA using CO LDME set B of table~\ref{tab:fit}, compared to H1 \cite{Aaron:2010gz} and ZEUS \cite{:2009br} data. Blue bands: Uncertainties of NLO CS curve due to scale variations. Yellow bands: Uncertainties of NLO CS+CO curve due to scale variations and uncertainties of the CO LDMEs. From \cite{Butenschoen:2011ks}.}
\end{figure*}

Our results for direct photoproduction \cite{Butenschoen:2011ks} are shown in figure~\ref{fig:PhotoPol}. We compare our predictions for the parameters $\lambda$ and $\nu$ as functions of $p_T$ and the inelasticity variable $z$ with data measured by the H1 \cite{Aaron:2010gz} collaboration in the helicity and Collins-Soper frames and by the ZEUS collaboration \cite{:2009br} in the target frame.

We note that the direct photoproduction mechanism is the dominant 
mechanism only in the intermediate $z$ region $0.3\lessapprox z \lessapprox 0.9$. At lower $z$, resolved photoproduction starts to dominate (see figure~\ref{fig:fitdirres}c), and at higher $z$ diffractive $J/\psi$ production overwhelms the cross section. Furthermore, close to the 
kinematical endpoint region $z\approx 1$, the NRQCD $v$ expansion is expected to break down.
Although the ZEUS $p_T$ distribution measurement does not impose an upper $z$ cut, we therefore nevertheless integrate only up to $z=0.95$ in figure~\ref{fig:PhotoPol}c. We note that the strong tendency towards
transversely polarized $J/\psi$ in plot~\ref{fig:PhotoPol}c could possibly be related to exactly that feature, since diffractively produced vector 
mesons are indeed predicted to be strongly transversely polarized in the endpoint region \cite{Brodsky:1994kf}.

Unfortunately, the H1 \cite{Aaron:2010gz} and ZEUS \cite{:2009br} data do not yet allow to distinguish the production mechanisms clearly, although the overall $\chi^2$ value of all the H1 and ZEUS data in figure~\ref{fig:PhotoPol} with regard to the default NLO predictions is reduced by more than 50\% as the CO contributions are included. But kinematical regions can be identified, in which a clear distinction could be possible in more precise experiments at a future $ep$ collider, like the LHeC: At higher $p_T$, NRQCD predicts the $J/\psi$ to be largely unpolarized in contrast to the CSM predictions. In the $z$ distributions, however, the scale uncertainties are sizeable and the error bands of the CSM and NRQCD predictions largely overlap.

\subsection{Results for hadroproduction}

\begin{figure*}
\centering
\begin{tabular}{|c|c|c|}
\hline
\includegraphics[width=5cm]{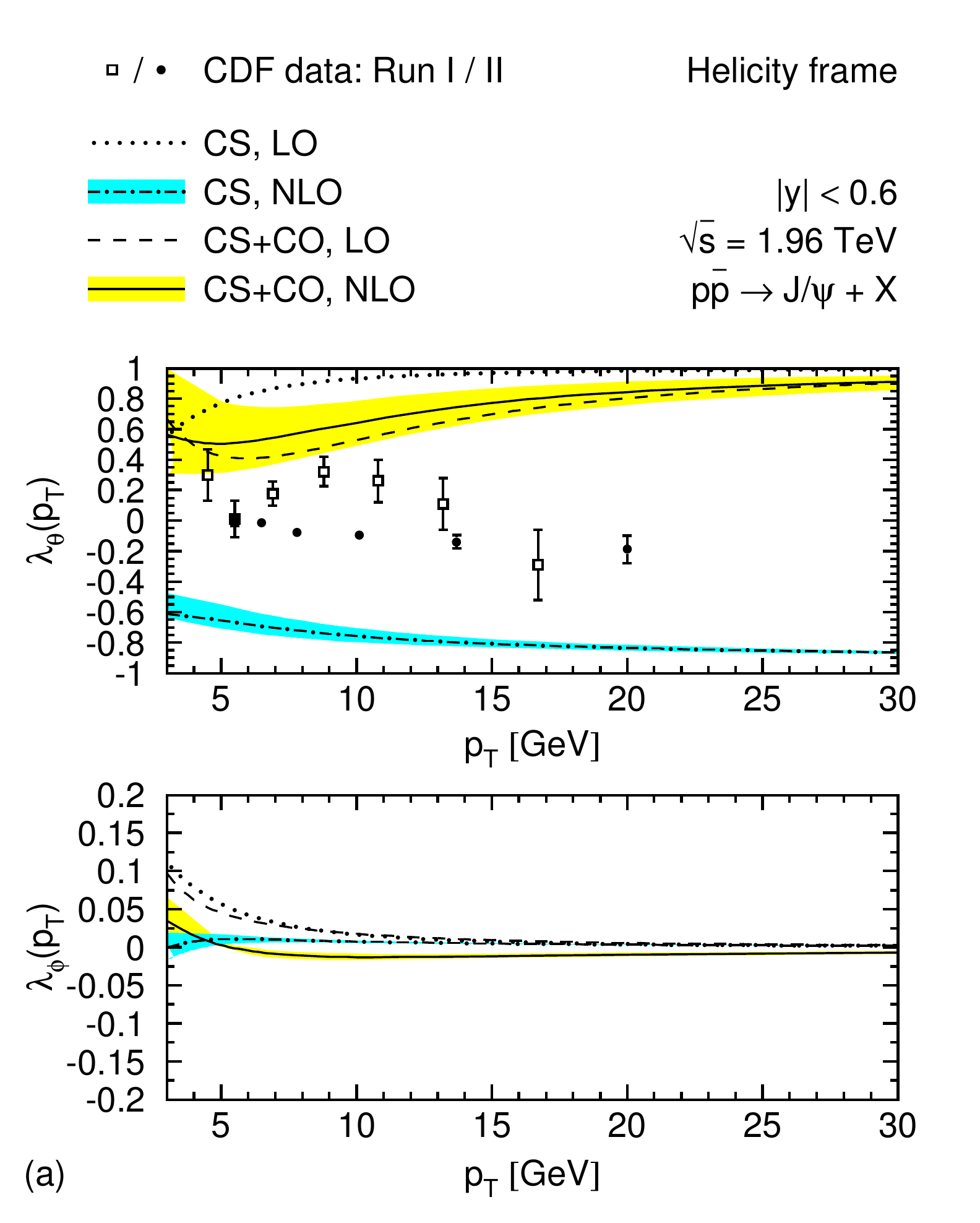}
&
\includegraphics[width=5cm]{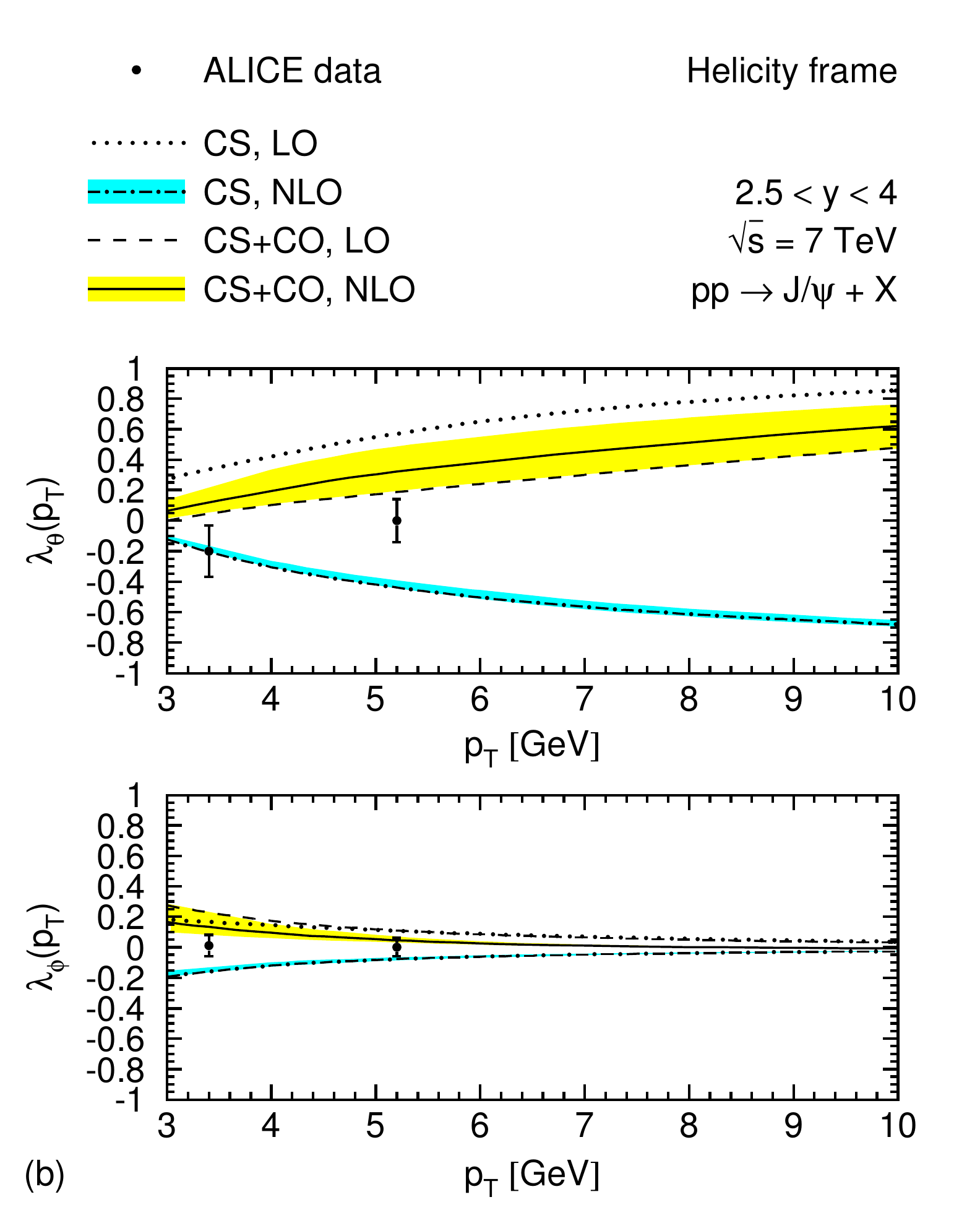}
&
\includegraphics[width=5cm]{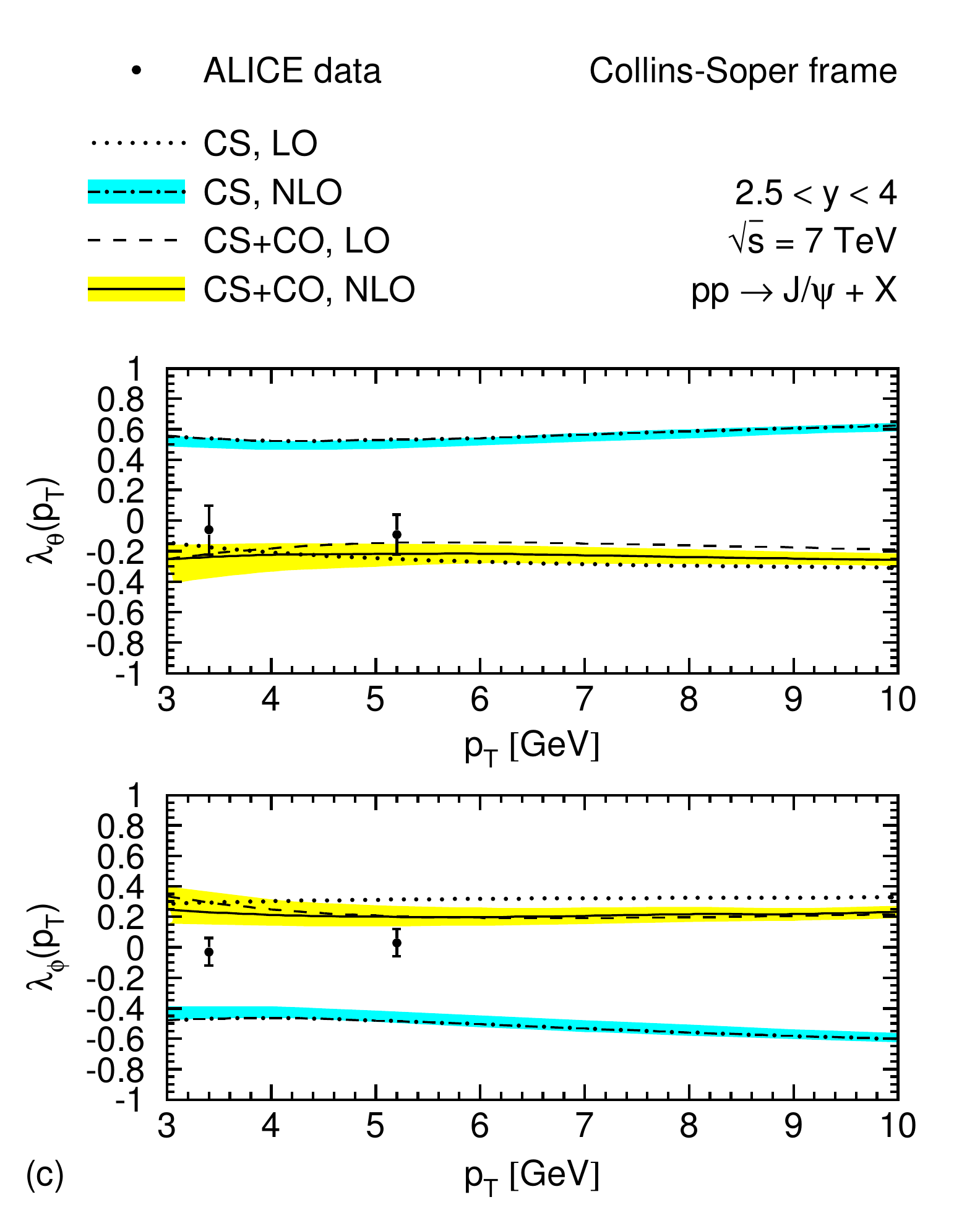}
\\ \hline
\end{tabular}
\caption{\label{fig:HadroPol} Polarization parameters $\lambda_\theta$ and $\lambda_\phi$ for hadroproduction using CO LDME set B of table~\ref{tab:fit}, compared to data measured by CDF at Tevatron in run I \cite{Affolder:2000nn} and II \cite{Abulencia:2007us} and by ALICE at the LHC \cite{Abelev:2011md}. Blue bands: Uncertainties of NLO CS curve due to scale variations. Yellow bands: Uncertainties of NLO CS+CO curve due to scale variations and uncertainties of the CO LDMEs. From \cite{HadroPolLetter}.}
\end{figure*}

\begin{figure*}
\centering
\includegraphics[width=5.4cm]{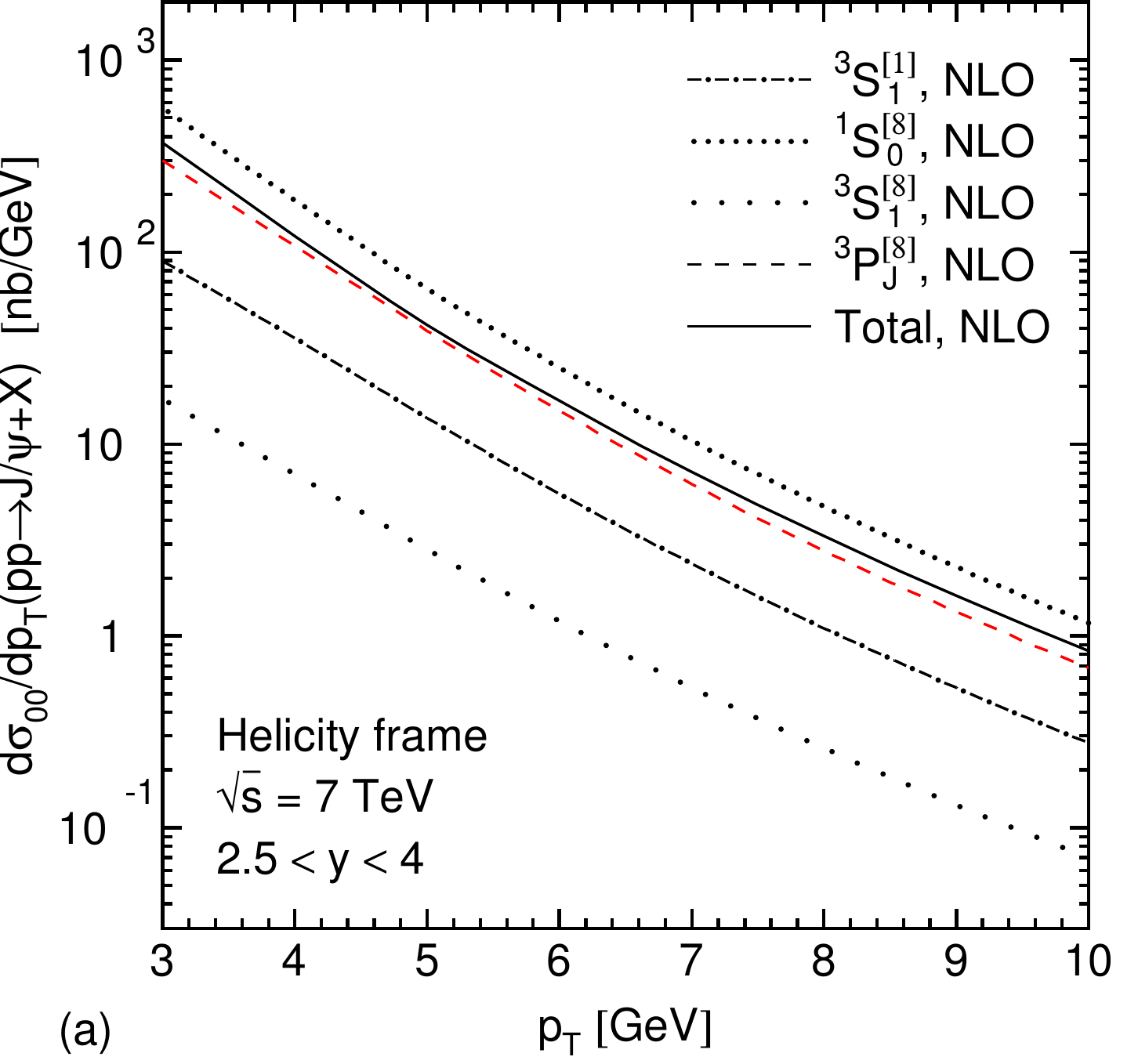}
\includegraphics[width=5.4cm]{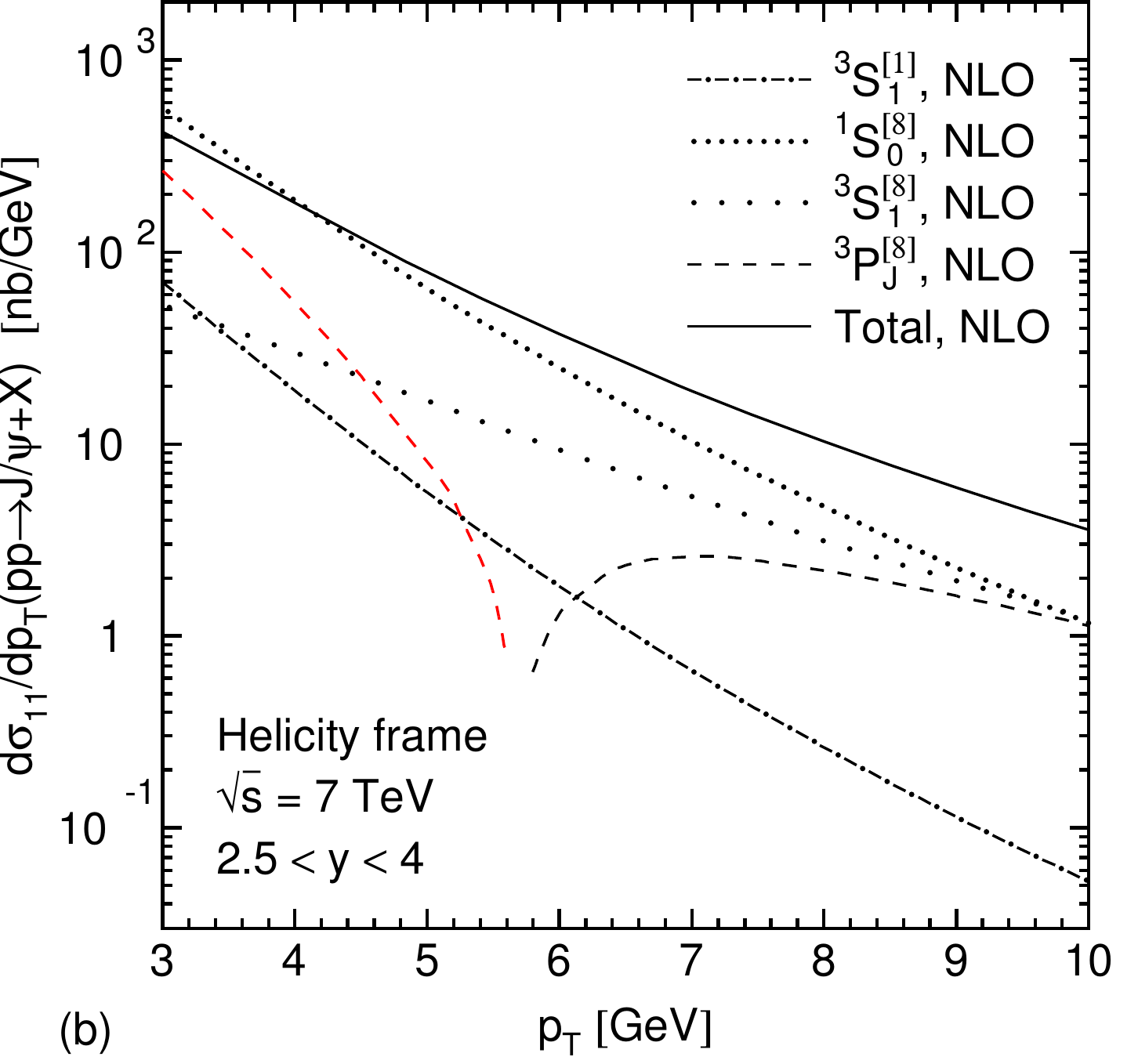}
\includegraphics[width=5.4cm]{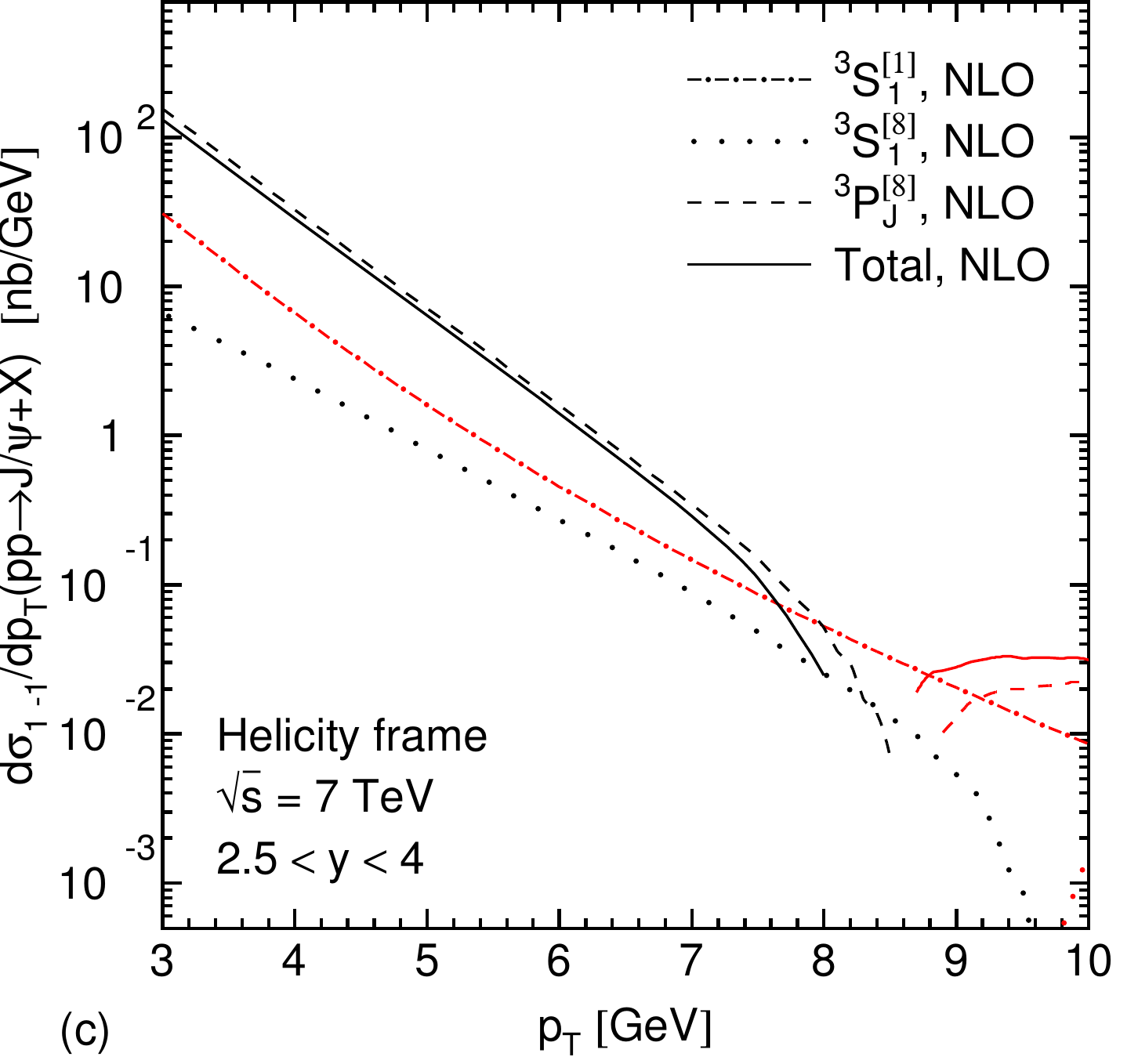}

\vspace{7pt}
\includegraphics[width=5.4cm]{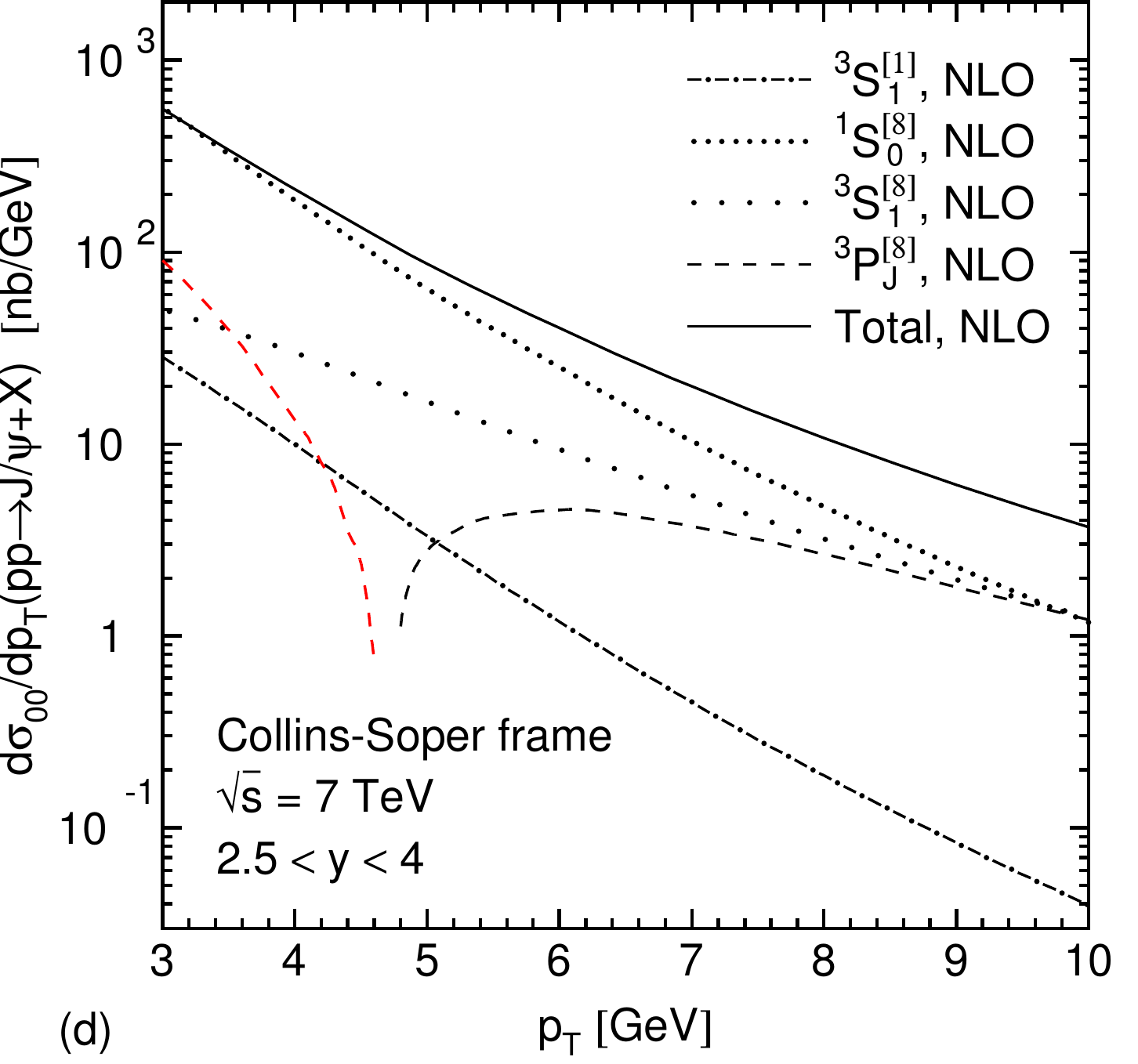}
\includegraphics[width=5.4cm]{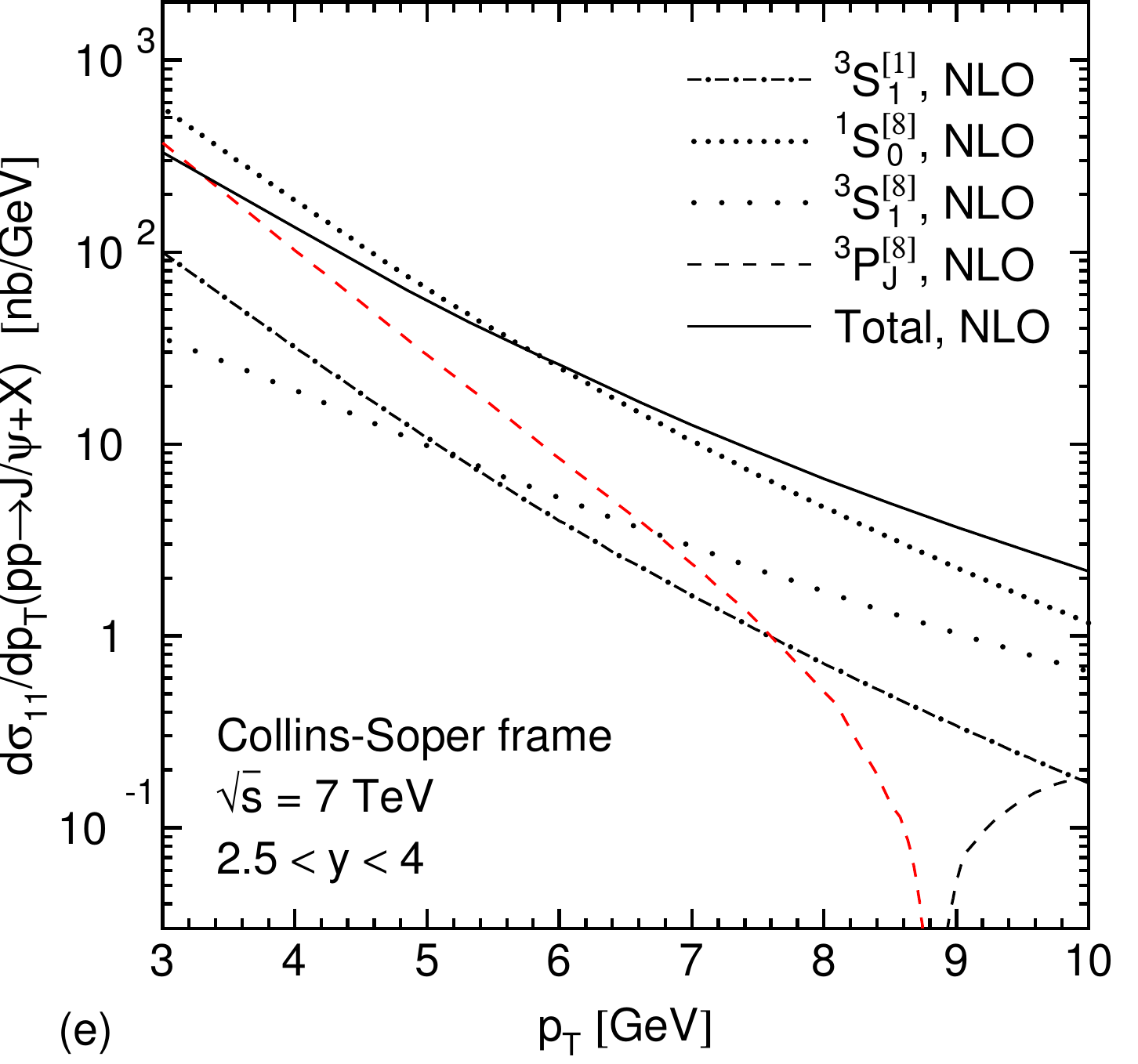}
\includegraphics[width=5.4cm]{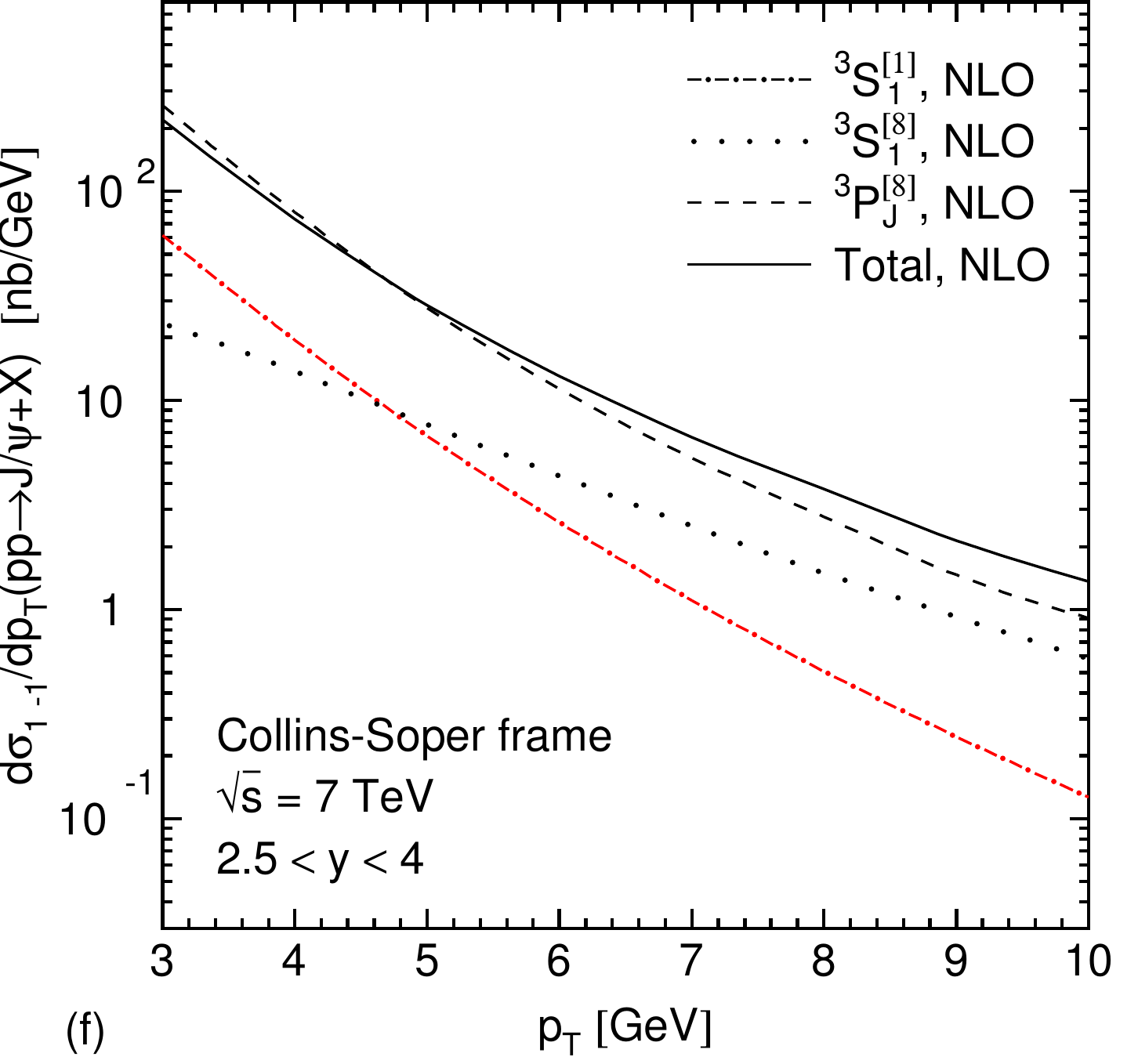}
\caption{\label{fig:HadroPolStates}The predictions for $d\sigma_{00}$, $d\sigma_{11}$ and $d\sigma_{1,-1}$ in the helicity and Collins-Soper frames decomposed into the contributions of the individual Fock states. Red curves mean negative values. Please note that these curves are the short distance cross sections already multiplied by the corresponding LDMEs (CO LDME set B of table~\ref{tab:fit}). Note also that $d\sigma_{1,-1}$ does not receive a contribution from the intermediate $^1S_0^{[8]}$ state.}
\end{figure*}

Our results for hadroproduction \cite{HadroPolLetter} are shown in figure~\ref{fig:HadroPol}. We compare our predictions for the parameters $\lambda_\theta$ and $\lambda_\phi$ as functions of $p_T$ in the helicity and Collins-Soper frames with the measurements by CDF  \cite{Affolder:2000nn,Abulencia:2007us} and ALICE \cite{Abelev:2011md}. Analogously to the unpolarized case we do not consider the $p_T$ range below 3~GeV, where nonperturbative effects are likely to dominate. In figure~\ref{fig:HadroPolStates} we decompose the contributions to $d\sigma_{00}$, $d\sigma_{11}$ and $d\sigma_{1,-1}$ used for the predictions of the ALICE measurements into the contributions of the various intermediate Fock states. The unpolarized cross section is recovered as $d\sigma_{00}+2d\sigma_{11}$. Our predictions do not yet include feed-down contributions from higher charmonium states, whereas the CDF data are prompt, and the ALICE data even non-prompt, but the leading-order (LO) NRQCD analysis \cite{Braaten:1999qk} has shown the impact of $\chi_{cJ}$ and $\psi^\prime$ feed-downs to be rather small. In the helicity frame, the CSM predicts strongly longitudinally polarized $J/\psi$ at NLO, while NRQCD predicts a strong transverse polarization. In the Collins-Soper frame the situation is inverted. The CDF measurement at Tevatron Run II \cite{Abulencia:2007us}, which is partially in disagreement with the measurement at run I \cite{Affolder:2000nn}, finds largely unpolarized $J/\psi$ in the helicity frame, which is in contradiction to both the CSM and NRQCD predictions. The early ALICE data \cite{Abelev:2011md} however favors NRQCD over the CSM. We look very much forward to the forthcoming, more precise polarization measurements at the LHC, which have the potential to clearly confirm or dismiss the various suggested $J/\psi$ production mechanisms.

We also note that in both photo- and hadroproduction the perturbative expansion in $\alpha_s$ seems to converge much more rapidly after inclusion of the CO states as compared to the CSM, where sign and shape of the various distributions are radically changed when going from LO to NLO.

We very much appreciate that very recently a second, independent, NLO NRQCD analysis involving the polarization parameter $\lambda_\theta$ in $J/\psi$ hadroproduction has been performed \cite{Chao:2012iv}. In that preprint it was shown that, neglecting data from all production mechanisms except hadroproduction, sets of CO LDMEs can be constructed that describe both the measured hadroproduction yield and values of $\lambda_\theta$ in the helicity frame close to zero. Unfortunately, as emphasized in that preprint, even after including the CDF polarization data it is still not possible to extract the three CO LDMEs independently in a hadroproduction-only fit, let alone to test their universality. The analysis \cite{Chao:2012iv} does however allow us to cross-check parts of our results, namely $d\sigma_{00}$ and $d\sigma_{11}$ in the helicity frame. Fortunately, both works agree there, since we can reproduce their figure~2, and when using the LDME set in the first row of their table~1, our prediction for $\lambda_\theta$ nicely lies within the ``NLO total'' band of their figure~1. Furthermore, we have verified that none of their proposed LDME sets is compatible with the observed photoproduction yield. Thus both works agree on that the observed hadro- and photoproduction yields and the CDF polarization measurement \cite{Abulencia:2007us} can not be simultaneously described by a single set of CO LDMEs.

\section{Summary}

We have presented a rigorous NLO analysis of the $J/\psi$ yield and polarization in various production mechanisms within the factorization theorem of NRQCD. We have extracted values for the three CO LDMEs by fitting to 194 points of inclusive $J/\psi$ production data from different hadroproduction, photoproduction, two-photon scattering and electron-positron annihilation measurements. We have shown that this fit is constrained and independent of the choice of possible lower cuts on the $J/\psi$ transverse momenta. Our global fit can describe all data well, except perhaps the two-photon scattering data by DELPHI. Besides that, the extracted LDMEs do exhibit the scaling behavior predicted by the NRQCD scaling rules.

In a second step we have used the CO LDMEs thus extracted to make predictions for $J/\psi$ polarization observables in photo- and hadroproduction and compare with the currently available experimental data. As for photoproduction, HERA data is not precise enough for definite conclusions, here we have to wait for future experiments, possibly at the LHeC. But as for hadroproduction, the precise results for $\lambda_\theta$ from the CDF collaboration measured at Tevatron run~II in the helicity frame are in stark conflict with the NRQCD predictions. CDF measured unpolarized $J/\psi$ here while NRQCD predicts a strong transverse polarization. However, the CDF measurements at Tevatron run~I and~II seem to be partly in conflict with each other, and the early data from the ALICE collaboration at low transverse momentum at the LHC is compatible with NRQCD. Future, more precise polarization measurements at the LHC, also at higher transverse momenta, will have the potential to clearly confirm or dismiss the LDME universality.

We stress that our present analysis still ignores feed-down contributions from higher charmonium states. Although they have been taken care of in our global fit through a subtraction of estimated contributions from the prompt experimental data, and their influence on the polarization observables is expected to be very minor, a rigorous inclusion of the feed-down contributions is still left for future work.





\end{document}